\documentclass{elsart}

\usepackage{amssymb}
\usepackage{amsmath}
\usepackage{graphicx}

\begin{document}
\begin{frontmatter}
\title{Effects of SO(10)-inspired scalar non-universality on the MSSM parameter space at large $\tan\beta$}
\author{M. R. Ramage}
\address{Rudolph Peierls Centre for Theoretical Physics, Department of Physics, University of Oxford, 1 Keble Road, Oxford OX1 3NP, U.K.}
\address{Astroparticle and High Energy Physics Group, Instituto de F\'isica Corpuscular, Edificio Institutos de Investigaci\'on, Universidad de Valencia, Valencia E-46071, Spain}
\address{email: ramage@thphys.ox.ac.uk}

\begin{abstract}
We analyze the parameter space of the ($\mu > 0$, $A_0 = 0$) CMSSM at large $\tan\beta$ with a small degree of non-universality originating from $D$-terms and Higgs-sfermion splitting inspired by ${\textrm{SO}}(10)$ GUT models. The effects of such non-universalities on the sparticle spectrum and observables such as $(g-2)_\mu$, $\mathcal{B}(b\rightarrow X_s \gamma)$, the SUSY threshold corrections to the bottom mass and $\Omega_{{\textrm{CDM}}}h^2$ are examined in detail and the consequences for the allowed parameter space of the model are investigated. We find that even small deviations to universality can result in large qualitative differences compared to the universal case; for certain values of the parameters, we find, even at low $m_{1/2}$ and $m_{16}$, that radiative electroweak symmetry breaking fails as a consequence of either $|\mu|^2 < 0$ or $m_{A^0}^2 < 0$. We find particularly large departures from the mSugra case for the neutralino relic density, which is sensitive to significant changes in the position and shape of the $A^0$ resonance and a substantial increase in the Higgsino component of the LSP. However, we find that the corrections to the bottom mass are not sufficient to allow for Yukawa unification.    
\end{abstract}
\end{frontmatter}

\section{Introduction}
The simplest supersymmetric extension of the Standard Model, the MSSM, contains well over a hundred free parameters after supersymmetry breaking is taken into account. In the most general case, a large number of mixing angles and complex phases are present and this tends to result in predictions of flavour changing neutral current (FCNC) and CP-violating processes in excess of the current strict experimental bounds. Moreover, to analyze a parameter space of such magnitude in any detail would require an enormous amount of computing power and the results would not be particularly illuminating. To avoid these problems, simplifying theoretical constraints are usually imposed on the model in order to restrict the form of the soft supersymmetry breaking mass and coupling matrices; i.e. they are taken to be proportional to the identity matrix in flavour space. Many analyses of recent years, for example~\cite{Kane:1994td,Baer:2003yh,Ellis:2003cw,Roszkowski:2001sb,Djouadi:2001yk} to cite but a few, have been based on the ``Constrained'' MSSM (CMSSM) or mSugra scenario in which a universal form for the mass and coupling matrices is assumed. In this model, gravity is presumed to be responsible for mediating the breaking of supersymmetry from a hidden sector of the theory sharing none of the Standard Model gauge interactions, to the visible sector~\cite{Ibanez:1982ee,Inoue:1982pi,Chamseddine:1982jx,Barbieri:1982eh,Nilles:1982ik,Ohta:1982wn,Hall:1983iz,Soni:1983rm,Nilles:1983xx,Ellis:1983wr,Alvarez-Gaume:1983gj,Nilles:1984ge,Haber:1985rc}. With certain simplifying assumptions it is possible to construct models of supergravity that lead to this preferred form for the soft breaking parameters. However, supergravity theories by no means inevitably predict this universality and, in any case, universality is readily violated below the supergravity scale by corrections deriving from, for example, renormalization group running between the supergravity/Planck scale and the GUT scale~\cite{Polonsky:1994sr,Polonsky:1995rz} or from the breaking of GUT~\cite{Polonsky:1995rz,Drees:1986vd,Hagelin:1990ta,Faraggi:1992bb,Lleyda:1993xf,Kawamura:1994yv,Kawamura:1994uf,Kawamura:1995ys,Rattazzi:1994bm,Kolda:1996iw} and/or family symmetries~\cite{Pomarol:1996xc,Babu:1999js,Murakami:2001hk,King:2001uz,Ross:2002fb,Kobayashi:2002mx,Maekawa:2002eh,King:2003rf,Ramage:2003pf}. 

In a previous paper~\cite{Ramage:2003pf} we explored the low energy constraints on an ${\textrm{SO}}(10)$ SUSY GUT model with an additional ${\textrm{SU}}(3)_{\textrm{F}}$ family symmetry~\cite{King:2001uz,Ross:2002fb,King:2003rf} spontaneously broken in such a way that a phenomenologically acceptable set of Yukawa matrices can be obtained, accounting for the observed fermion masses and mixings including the neutrino sector. When both symmetries remain unbroken the universal form for the mass and coupling matrices for the gauginos and sfermions is ensured regardless of the supersymmetry breaking mechanism whether it be gravity, gauge, or anomaly mediation. However, this sfermion universality is spoilt by $D$-terms arising from the breaking of the ${\textrm{SU}}(3)_{\textrm{F}}$ family symmetry. In this instance the mass squared of the third family of sfermions is split from the first two by around $20\%$ with the sign of the splitting undetermined. We found that, in the case of decreased third family sfermion masses, the boundary of correct electroweak symmetry breaking (EWSB), where $\mu$ vanishes, occurs at a substantially lower value of $m_0$ than for the universal case. Correspondingly, a new area of parameter space allowed by the various constraints appears for large $\tan\beta$. A general conclusion we reached was the allowed parameter space is very sensitive to the universality assumption at least for larger values of $\tan\beta$. Here we pursue this idea and consider the case where the dominant additional contributions to the sfermion mass matrices originate from the breaking of ${\textrm{SO}}(10)$ and its subgroups rather than from the breaking of the family symmetry. To isolate the effects of ${\textrm{SO}}(10)$ breaking we shall ignore the effects of the family symmetry breaking. The impact of scalar soft mass non-universality, either deriving from ${\textrm{SO}}(10)$ breaking or otherwise, on low energy phenomenology has been studied before~\cite{Matalliotakis:1995ft,Olechowski:1995gm,Cheng:1995bi,Berezinsky:1996cj,Nath:1997qm,Kawamura:1998ek,Baer:1999mc,Baer:2000jj,Baer:2001vw,Blazek:2002ta,Dermisek:2003vn,Ellis:2002iu,Ellis:2002wv,Auto:2003ys,Pallis:2003aw,Profumo:2003em,Ananthanarayan:2003ca,Cerdeno:2004zj,Baer:2004xx,Auto:2004km,Baer:2004fu}, but our analysis differs somewhat in perspective and should complement previous studies. In what follows, we will assume that ${\textrm{SO}}(10)$ is broken directly to the Standard Model or, in the case that there exists a secondary stage of breaking with an intermediate gauge group, for example the Pati-Salam group~\cite{Pati:1974yy}:  
\begin{displaymath}
{\textrm{SO}}(10) \longrightarrow {\textrm{SU}}(4)_{\textrm{PS}} \times {\textrm{SU}}(2)_{\textrm{L}} \times {\textrm{SU}}(2)_{\textrm{R}} \longrightarrow {\textrm{SU}}(3)_{\textrm{C}} \times {\textrm{SU}}(2)_{\textrm{L}} \times {\textrm{U}}(1)_{\textrm{Y}},
\end{displaymath}
or ${\textrm{SU}}(5)$~\cite{Georgi:1974sy}:
\begin{displaymath}
{\textrm{SO}}(10) \longrightarrow {\textrm{SU}}(5) \times {\textrm{U}}(1)_{\textrm{X}} \longrightarrow {\textrm{SU}}(3)_{\textrm{C}} \times {\textrm{SU}}(2)_{\textrm{L}} \times {\textrm{U}}(1)_{\textrm{Y}},
\end{displaymath}
for simplicity we will assume that the secondary breaking scale is sufficiently close to the scale of ${\textrm{SO}}(10)$ breaking that we can neglect the effects of RG running between those two scales. 

Additional $D$-term contributions to the soft mass matrices arise from the reduction of rank of the gauge group, associated with a broken ${\textrm{U}}(1)$ generator proportional to $2I_R + 3(B-L)/2$ where $I_R$ is the 3rd component of weak isospin pertaining to the ${\textrm{SU}}(2)_{\textrm{R}}$ subgroup of ${\textrm{SO}}(10)$ and $(B-L)$ is the difference between baryon and lepton number~\cite{Hagelin:1990ta,Faraggi:1992bb,Kawamura:1994uf,Kawamura:1995ys,King:2000vp}. The sfermion and Higgs soft masses squared each receive a contribution proportional to a quantity $D^2$, which can be positive or negative, that parametrizes the $D$-term contribution from the breaking of the GUT group. Besides the $D$-terms, there is no reason deriving from standard GUT scenarios why the EWSB Higgs soft masses should be related to the sfermion soft masses since they belong to different representations of the gauge group. Therefore we assume that the Higgs soft masses are independent of the sfermion masses in general, with $m_{H_1}^2$ and $m_{H_2}^2$\footnote{In our notation, as usual, $m_{H_1}^2$ and $m_{H_2}^2$ are the soft mass parameters for the Higgs fields that give mass to the down-type quarks and up-type quarks respectively.} taking the common value $m_{10}^2$ at the GUT scale (since they are both contained in the $\mathbf{\underline{10}}$ representation of ${\textrm{SO}}(10)$). Rather than allow $m_{10}$ and $D^2$ to vary arbitrarily, we take the ratio of GUT scale Higgs masses to sfermion masses, $m_{10}/m_{16}$, and similarly $ \mathcal{D} \equiv {\mathrm{sign}}(D^2)\sqrt{|D^2|}/m_{16}$ as the independent variables. Therefore the soft SUSY breaking masses and couplings are defined by the following free parameters: 
\begin{displaymath}
m_{16}, \quad m_{10}/m_{16}, \quad m_{1/2}, \quad \mathcal{D}, \quad A_0, \quad \tan\beta, \quad {\mathrm {sign}}(\mu). 
\end{displaymath}
Throughout this paper we will always take ${\mathrm {sign}}(\mu) > 0$  in accordance with the present experimental deviation of $(g-2)_\mu$ from its Standard Model value and the value favoured by the branching ratio $\mathcal{B}(b \rightarrow X_s \gamma)$, $\tan\beta = 50$ because for much lower $\tan\beta$ no new allowed regions of parameter space were found that had not already been discovered in the CMSSM case, and $A_0 = 0$ since it doesn't have a particularly large effect on the allowed regions (except when it takes on large values) and also to keep the analysis simple so we can focus on the additional parameters introduced by the ${\textrm{SO}}(10)$ breaking effects\footnote{Additional motivation for $A_0 = 0$ comes from the family symmetry. If the matter superfields $Q$, $u_R^c$, etc. are triplets of ${\textrm{SU}}(3)_{\textrm{F}}$, as is the case in~\cite{King:2003rf}, then, at least in the limit of unbroken family symmetry, $A_0 = 0$.}. 

The structure of the paper is as follows: in Section~\ref{breaking} we review the various sources of non-universality present in a typical ${\textrm{SO}}(10)$ SUSY-GUT model focusing on the origin of the $D$-term contributions to the soft SUSY breaking masses. In Section~\ref{spectrum} we comment in detail on the effects on the sparticle spectrum of introducing the $D$-terms and splitting the Higgs from the sfermion soft masses. In particular this can have significant implications for the mass of the pseudo-scalar Higgs boson $A^0$ and for electroweak symmetry breaking which in turn has consequences for the gaugino masses and mixing angles, all this in addition to the mass splittings and shifts directly or indirectly induced in the sfermion spectrum. These changes will feed through to affect the results of calculations of the various observables we use to constrain the theory. We discuss the nature of these effects on $\tan\beta$-enhanced amplitudes, e.g. $(g-2)_\mu$, in Section~\ref{tanb}. We go on in Section~\ref{method} to give details of the numerical calculation of the sparticle spectrum, mixing angles and corrections to the couplings, and to describe the constraints on the parameter space such as the branching ratio of the inclusive b-decay $\mathcal{B}(b \rightarrow X_s \gamma)$, the WMAP bound on the cold dark matter density $\Omega_{{\textrm{CDM}}}h^2$, and the discrepancy between the theoretical prediction of the anomalous magnetic moment of the muon and its measured value. In Section~\ref{results} we present the results of the constrained fit and discuss the main features. Finally, in Section~\ref{summary} we summarize and conclude.

\section{{\textrm{SO}}(10) Breaking and Scalar Masses\label{breaking}}
As we touched on briefly in the introduction, even assuming universality at the supergravity/Planck scale, $M_P \sim \mathcal{O}$($10^{19}$ GeV), non-universality at the GUT scale, $M_G \sim \mathcal{O}$($10^{16}$ GeV) can arise either from renormalization group running between the GUT scale and the Planck scale, from GUT threshold effects or from $D$-terms arising from the breaking of ${\textrm{SO}}(10)$ or the ${\textrm{SU}}(3)_{\textrm{F}}$ family symmetry. In this analysis we can ignore the possibility of RGE effects since the full ${\textrm{SO}}(10) \times {\textrm{SU}}(3)_{\textrm{F}}$ group remains unbroken down to the GUT scale and hence sfermion universality is maintained. We also neglect GUT threshold effects originating from the discrepancy between $M_G$ and the masses of the heavy particles, $M_H$, that obtain their masses when the GUT group is broken. There are also potentially important corrections due to the splitting between the masses of the fermions, $M_F$, and bosons, $M_B$ in the heavy superfields (see~\cite{Polonsky:1995rz} for more details). This leads to corrections proportional to the logarithms $\ln\frac{M_H^2}{M_G^2}$ and $\ln\frac{M_B^2}{M_F^2}$, and finite terms. These corrections depend on the details of the GUT scale particle spectrum and since we don't pin ourselves down to a particular ${\textrm{SO}}(10)$ model, and although they may be significant, we will ignore them for simplicity. Therefore, we choose only to include the $D$-term contributions to the sfermion masses that are expected to appear when this symmetry group is broken. 

Here we will outline the origin of the $D$-terms. For a detailed derivation, see~\cite{Kolda:1996iw,Martin:1997ns}. Consider a ${\textrm{U}}(1)$ subgroup of ${\textrm{SO}}(10)$, ${\textrm{U}}(1)_{\textrm{X}}$, which is broken when the rank of the gauge group is reduced from 5 to 4. We assume that there exists two heavy scalars $S_{(\pm)}$ with opposite ${\textrm{U}}(1)_{\textrm{X}}$ charges, for example $\pm 1$, where in this case the charge $X$ is given by $2I_R + 3(B-L)/2$. The $D$-term contribution to the scalar potential corresponding to the ${\textrm{U}}(1)_{\textrm{X}}$ generator is given by 
\begin{displaymath}
\frac{1}{2} g_{10}^2 D^2 = \frac{1}{2} g_{10}^2 \left( |S_{(+)}|^2 - |S_{(-)}|^2 + \sum_{i} x_{i} |\phi_{i}|^2 \right)^2,
\end{displaymath}
where $g_{10}$ is the unified coupling and $x_i$ are the ${\textrm{U}}(1)_{\textrm{X}}$ charges of the MSSM fields $\phi_{i}$. When the fields $S_{(\pm)}$ obtain VEVS, this leads to corrections to the mass terms for the MSSM scalar fields: 
\begin{displaymath}
\delta m_{\phi_{i}}^2 |\phi_{i}|^2 = x_{i} g_{10}^2 \left(\langle S_{(+)} \rangle^2 - \langle S_{(-)} \rangle^2 \right) |\phi_{i}|^2. 
\end{displaymath}
However, the magnitudes of the $D$-terms are not of the order of the large $S_{(\pm)}$ VEVS as would be naively expected. A more detailed analysis shows that to leading order
\begin{displaymath}
D^2 \equiv \left(\langle S_{(+)} \rangle^2 - \langle S_{(-)} \rangle^2 \right) \simeq \frac{1}{2 g_{10}^2}\left(m_{S_{(-)}}^2 - m_{S_{(+)}}^2\right),  
\end{displaymath}
where $m_{S_{(\pm)}}^2$  are the soft SUSY breaking masses squared for the $S$ scalars. Therefore the new $D$-term contributions are of order the SUSY breaking scale. Note that in the case of universal soft SUSY breaking masses these $D$-term contributions vanish. Since the $S_{(\pm)}$ scalars are expected to belong to different representations of ${\textrm{SO}}(10)$ (for example, potential $\mathbf{16_H}$ and $\mathbf{\overline{16}_H}$ heavy Higgs representations) and indeed any family symmetry that may also exist at that scale, they are not expected a-priori to have equal masses at the GUT scale. Even if they are equal at $M_P$ this equality can easily be wiped out by differences in the RG running down to $M_G$. Drawing an analogy with $m^2_{H_1}$ and $m^2_{H_2}$, a significant difference in Yukawa couplings could drive $m_{S_{(\pm)}}^2$ apart. Since we are otherwise ignorant of the size of $D^2$ we will take it to be a free parameter in our subsequent analysis, although we will take it to be relatively small. For ${\textrm{SO}}(10)$, including the $D$-term contributions to the sfermion and Higgs masses, we obtain the well-known GUT scale inputs for the sfermion and Higgs soft SUSY breaking masses~\cite{Hagelin:1990ta,Faraggi:1992bb,Kawamura:1994uf,Kawamura:1995ys,King:2000vp}:
\begin{eqnarray}
m^2_{\tilde{Q}_{\phantom{L}}} &=& m_{16}^2 + g_{10}^2 D^2 \nonumber\\
m^2_{\tilde{u}_R} &=& m_{16}^2 + g_{10}^2 D^2 \nonumber\\
m^2_{\tilde{e}_R} &=& m_{16}^2 + g_{10}^2 D^2 \nonumber\\
m^2_{\tilde{L}_{\phantom{L}}} &=& m_{16}^2 - 3 g_{10}^2 D^2 \nonumber\\
m^2_{\tilde{d}_R} &=& m_{16}^2 - 3 g_{10}^2 D^2 \nonumber\\
m^2_{H_1} &=& m_{10}^2 + 2 g_{10}^2 D^2 \nonumber\\
m^2_{H_2} &=& m_{10}^2 - 2 g_{10}^2 D^2. \label{Deqs}
\end{eqnarray} 
We use these GUT scale boundary conditions in our analysis, varying $m_{10}/m_{16}$ from 0.75 to 1.25 and $\mathcal{D}$ from -0.4 to 0.4. These represent reasonably small perturbations about universality, but as we shall see, the effects on the allowed parameter space can be significant.  

In ${\textrm{SO}}(10)$, a right-handed neutrino superfield completes the $\underline{\mathbf{16}}$ representation. The presence of any neutrino Yukawa couplings and associated soft SUSY breaking parameters would, in principle, feed into the renormalization group equations for the remaining parameters above the scale of the right-handed neutrinos. However, we choose to ignore such effects in light of our ignorance of the details of the neutrino sector and our desire to keep the analysis simple.      

\section{The Sparticle Spectrum}\label{spectrum}
The impact of the $D$-terms and the Higgs-sfermion mass splitting on the low scale phenomenology can be understood in terms of how the soft mass differences at the GUT scale, either directly or through their influence on the RGEs, affect EWSB and the sparticle and Higgs masses. Here we will discuss aspects of the sparticle spectrum and mixings and EWSB. This section and the next are meant to provide the reader with an intuitive guide to how the $D$-terms and Higgs-sfermion splitting feed through the different soft parameters to affect the masses, mixings and observables. We will explain these effects through various approximations valid in certain circumstances before proceeding to discuss the more exact numerical results. Readers only interested in the results may wish to skip to Section~\ref{method}. We start by considering how the changes to the universal boundary conditions feed through the RGEs to create differences at the EWSB scale.
\subsection{Renormalization Group Evolution}\label{rge}
To begin with, we will implicitly assume that $m_{10}/m_{16} = 1$ to isolate the results of varying $D^2$. It turns out that the effect of the $D$-terms on the phenomenology is almost entirely due to their tree-level contribution to the boundary conditions at $M_G$ since they almost completely cancel out in the RGEs~\cite{Kolda:1996iw}. We will not write down all of the soft mass RGEs here, as they are well known and can be found in~\cite{Martin:1994zk}  to 2-loop order. However, we will reproduce the RGE for $m_{H_2}^2$ in the third family approximation as an example of this cancellation and because it will be useful to refer to it later\footnote{The third family 1-loop approximation is sufficient to obtain a qualitative understanding of the results which is our intention here. The numerical work, on the other hand, was carried out with the full two-loop RGEs.}. The $m_{H_2}^2$ RGE is given by
\begin{equation}
16\pi^2 \frac{dm_{H_2}^2}{dt} = 6|y_t|^2 \left(m^2_{\tilde{t}_{L}} + m^2_{\tilde{t}_{R}} + m^2_{H_2}\right) + 6|a_t|^2 - 6g_2^2|M_2|^2 - \frac{6}{5}g_1^2|M_1|^2 + \frac{3}{5}g_1^2\mathcal{S},
\end{equation}
where $t = \ln(Q/M_G)$ and 
\begin{eqnarray}
\mathcal{S} &=& Tr(Ym^2) \nonumber \\
&=& m^2_{H_2} - m^2_{H_1} + \sum_{\textrm{generations}} \left(m^2_{\tilde{Q}} - 2m^2_{\tilde{u}_R} + m^2_{\tilde{d}_R} - m^2_{\tilde{L}} + m^2_{\tilde{e}_R} \right).
\end{eqnarray}
Here, $Y$ is the weak hypercharge generator. So at $M_G$,
\begin{displaymath}
\mathcal{S} = -4g_{10}^2 D^2,
\end{displaymath}
but the $D$-terms cancel out of the soft masses in the term proportional to $|y_t|^2$. This cancellation also occurs for the analogous terms in the RGEs for the other scalar soft masses at the one-loop level since, aside from the $\mathcal{S}$ term, the sfermion soft masses only enter into the RGEs in the following combinations (see for example~\cite{Martin:1997ns}):
\begin{eqnarray}
  X_t &=& 2|y_t|^2 \left(m^2_{\tilde{t}_{L}} + m^2_{\tilde{t}_{R}} + m^2_{H_2}\right) + 2|a_t|^2 \nonumber \\
  X_b &=& 2|y_b|^2 \left(m^2_{\tilde{b}_{L}} + m^2_{\tilde{b}_{R}} + m^2_{H_1}\right) + 2|a_b|^2 \nonumber \\
  X_\tau &=& 2|y_\tau|^2 \left(m^2_{\tilde{\tau}_{L}} + m^2_{\tilde{\tau}_{R}} + m^2_{H_1}\right) + 2|a_\tau|^2,
\end{eqnarray}
and the only remaining trace of the $D$-terms in the RGEs lies within the terms proportional to $\mathcal{S}$. Therefore, in the running from $M_G$ to the EWSB scale, the $D$-terms only enter the RGEs multiplied by the (GUT normalised) weak hypercharge coupling $g_1^2$ and since we are not considering large deviations from universality, we find this to be a very small effect. In the rest of this analysis we neglect such RG effects due to $D$-terms although of course we retain them in our numerical calculations. For more details, see~\cite{Lleyda:1993xf,Kolda:1996iw}. Our main point here is that the splitting induced at the GUT scale {\em by the $D$-terms} does not change appreciably in magnitude as we evolve the RGEs down to the electroweak scale. 

Switching the $D$-terms off, we now allow $m_{10}/m_{16}$ to vary. This time we need not worry about $\mathcal{S}$ since $m_{10}$ and $m_{16}$ cancel out. However the splitting directly affects the $X_i$ terms ($i=t,b,\tau$). If we take $m_{10}/m_{16} > 1$ the $X_i$ are initially larger. This persists throughout the RG evolution and leads to an overall suppression of the 3rd family sfermion masses at low scales compared to the mSugra case since the effect of the $X_i$ is to reduce the masses as we evolve from high to low scales. In the case of the Higgs soft masses, on the other hand, the tendency to be driven to lower values by the larger $X_i$ is countered by the greater effect of the tree-level increase and so if we increase $m_{10}/m_{16}$ at $M_G$ we increase $m^2_{H_1}$ and $m^2_{H_2}$ at the EWSB scale.

\subsection{The Heavy Higgs Masses}
The mass of the pseudo-scalar Higgs, $A^0$, deriving from the radiatively corrected Higgs potential in the tadpole formalism~\cite{Barger:1994gh,Chankowski:1994er,Pierce:1997zz}, is given by
\begin{eqnarray}
m_{A^0}^2 &=& \frac{1}{\cos 2\beta}\left(\overline{m}_{H_2}^2 - \overline{m}_{H_1}^2\right) - M_Z^2 - \mathcal{R}e \Pi^T_{ZZ}(M_Z^2)  \notag \\
&& - \mathcal{R}e \Pi_{AA}(m_{A^0}^2) + \frac{t_1}{v_1}\sin^2\beta + \frac{t_2}{v_2}\cos^2\beta. 
\end{eqnarray}
Here, $\overline{m}_{H_{i}}^{2}=m_{H_{i}}^{2}-t_{i}/v_{i}$ where $t_{i}/v_{i}$ are the corresponding tadpole contributions from loop diagrams, $\Pi _{ZZ}^{T}$ is the transverse part of the $Z$ self-energy and $\Pi_{AA}$ is the $A^0$ self-energy. At large $\tan\beta$, $\cos 2\beta \simeq -1$ and one can see that $m_{A^0}^2$ depends overwhelmingly on the difference in the soft Higgs mass parameters assuming that $|m_{H_i}^2| \gg M_Z^2$ 
\begin{displaymath} 
m_{A^0}^2 \simeq m_{H_1}^2 - m_{H_2}^2.
\end{displaymath} 
Of importance is the RGE for $m^2_{H_1} - m^2_{H_2}$, which is given by
\begin{equation}
16\pi^2 \frac{d(m_{H_1}^2 - m_{H_2}^2)}{dt} = 3X_b + X_\tau - 3X_t.
\end{equation}
In general, $3X_t > 3X_b + X_\tau$ due to the large top Yukawa coupling, causing $m_{H_2}^2$ to run to low values faster than $m_{H_1}^2$ and increasing the difference between the Higgs soft masses. Increasing $m_{10}/m_{16}$ will increase $X_i$ by a factor $2(m^2_{10}-m^2_{16})|y_i|^2$ at $M_G$. Since  $3|y_t|^2 > 3|y_b|^2 + |y_\tau|^2$, one would think at first sight that the additional contribution would tend to increase the mass difference as we run the parameters down to the EWSB scale compared to the CMSSM. However we have neglected an additional effect. In our parameter space the above inequality regarding the Yukawa couplings always holds because $\mu > 0$ suppresses $y_b$ through the SUSY threshold corrections involving $\tilde{b} \tilde{g}$ and $\tilde{\chi}^{\pm} \tilde{t}$ loops which contribute with opposite signs with the gluino loop dominating. The main contributions come from terms enhanced by $\tan\beta$ which arise from helicity-flipping mass insertions in the sparticle propagators. In this case both the gluino and chargino loops are proportional to $\mu$. The decrease in $\mu$ caused by an increase in $m_{10}/m_{16}$ means that the net size of the correction to $y_b$ is reduced leaving us with a larger $y_b$. It turns out that this more than counters the increase in $X_t$ and for this reason we can expect the soft Higgs mass squared difference and thus $m_{A^0}$ to be lighter in the case of increased $m_{10}/m_{16}$. As an aside we should note here that this is a small effect and, unfortunately from the point of view of ${\textrm{SO}}(10)$, we do not find Yukawa unification to better than about 20\% in any part of the parameter space probed in this paper because $y_b$ is always too small for any reasonable range of $m_{10}/m_{16}$ that yields allowed points in the parameter space. 

\begin{figure}[tbp]
\begin{center}
\includegraphics[width=0.47\textwidth]{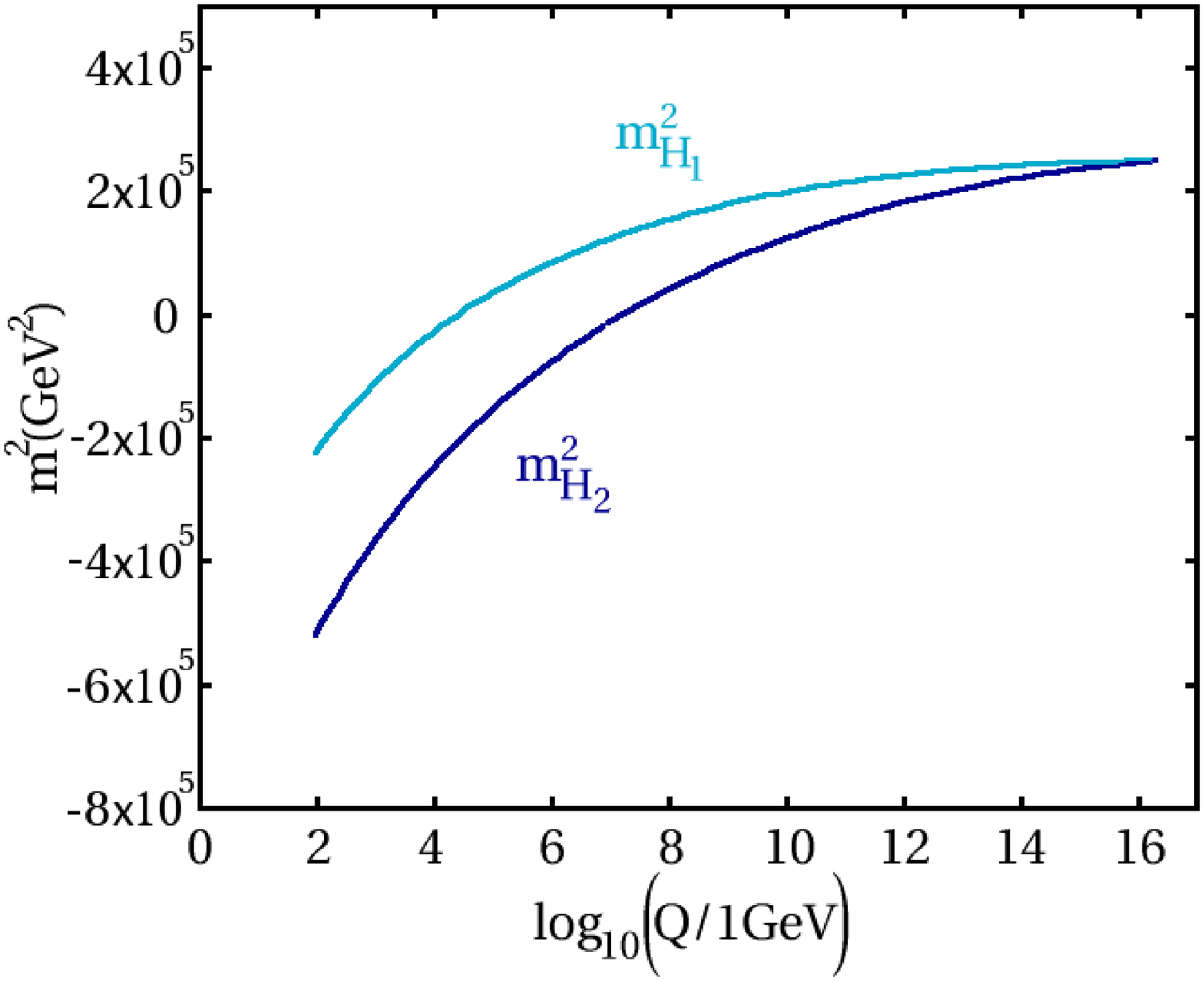} 
\includegraphics[width=0.47\textwidth]{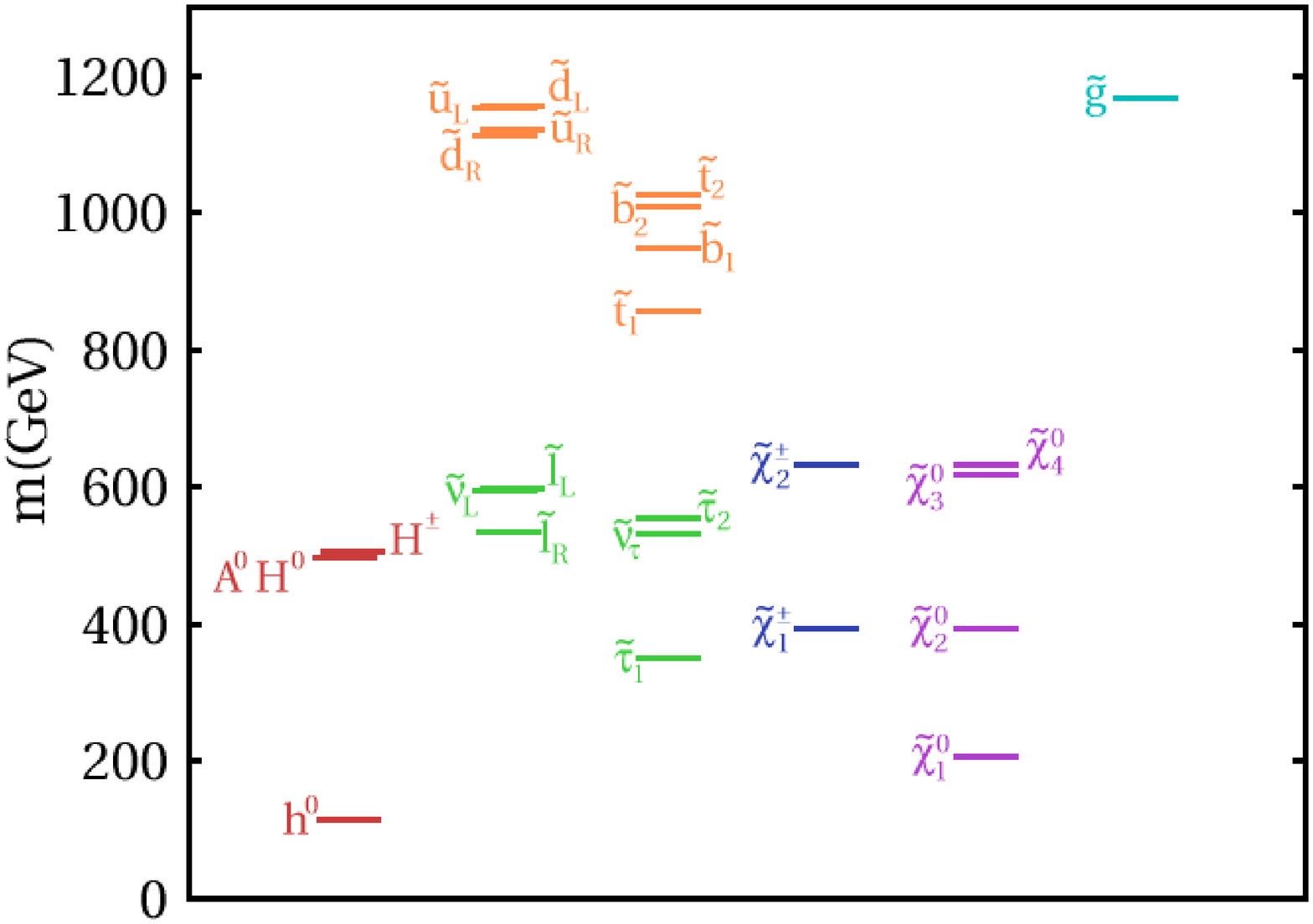}(a)
\includegraphics[width=0.47\textwidth]{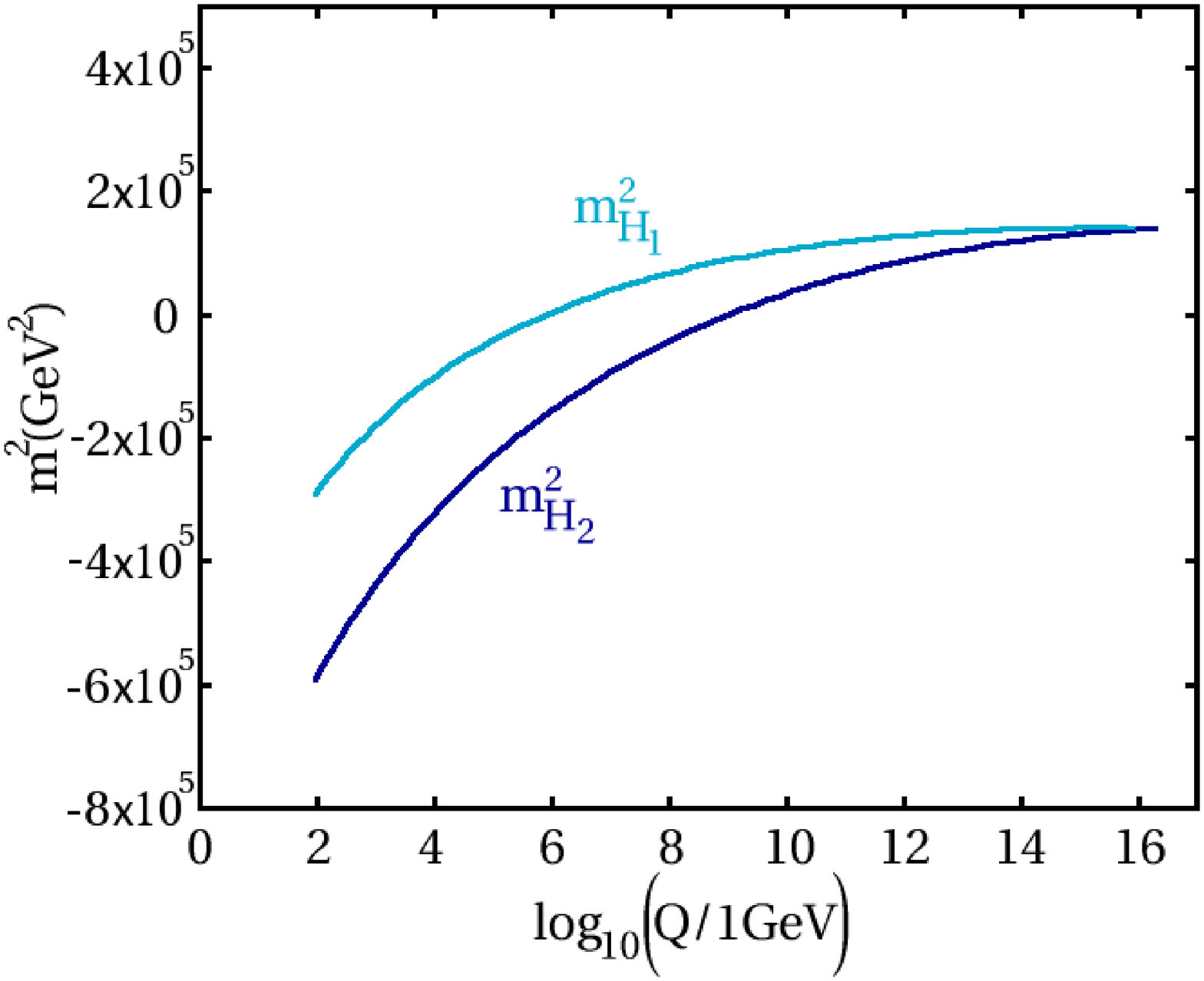}  
\includegraphics[width=0.47\textwidth]{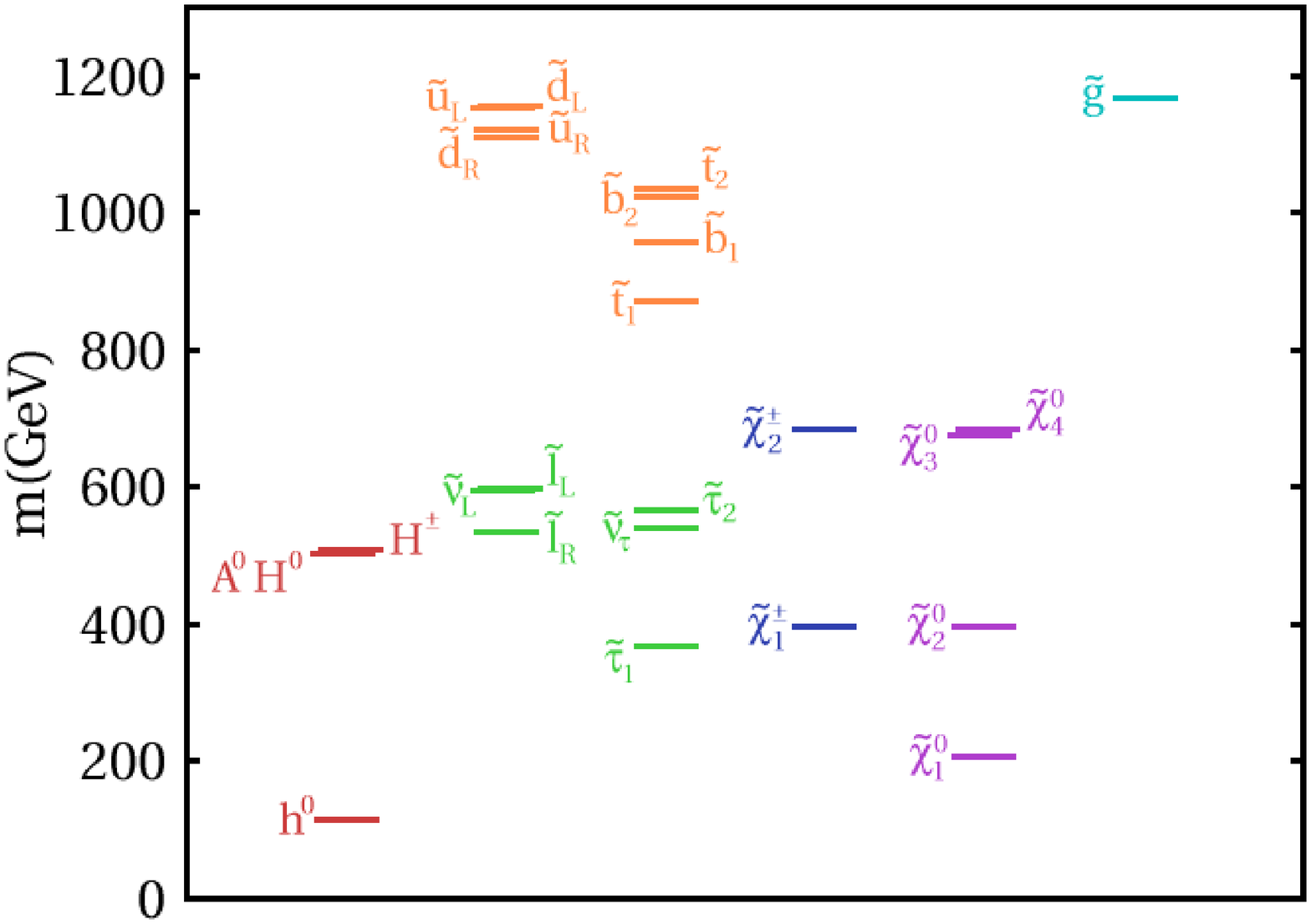}(b)
\includegraphics[width=0.47\textwidth]{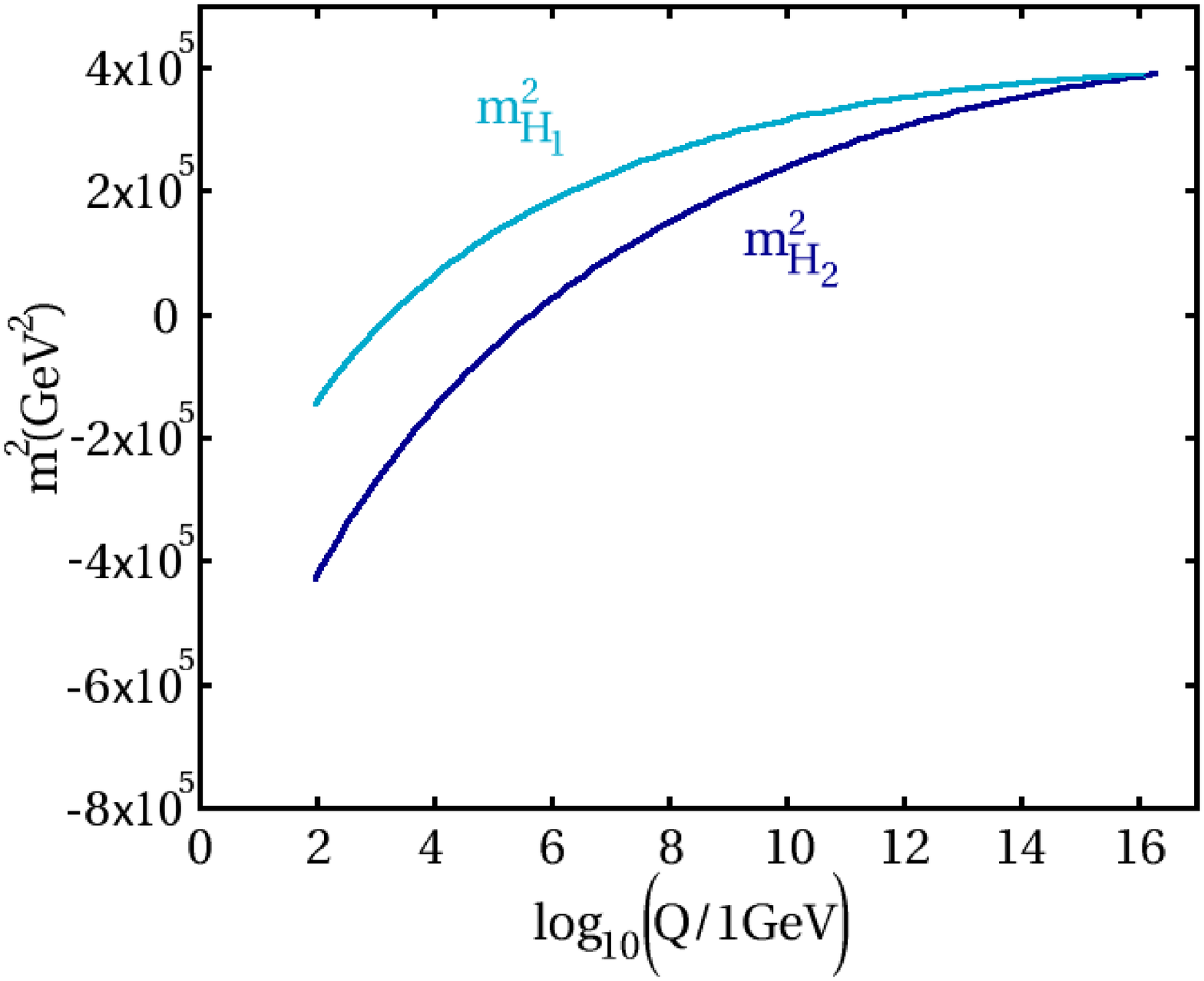} 
\includegraphics[width=0.47\textwidth]{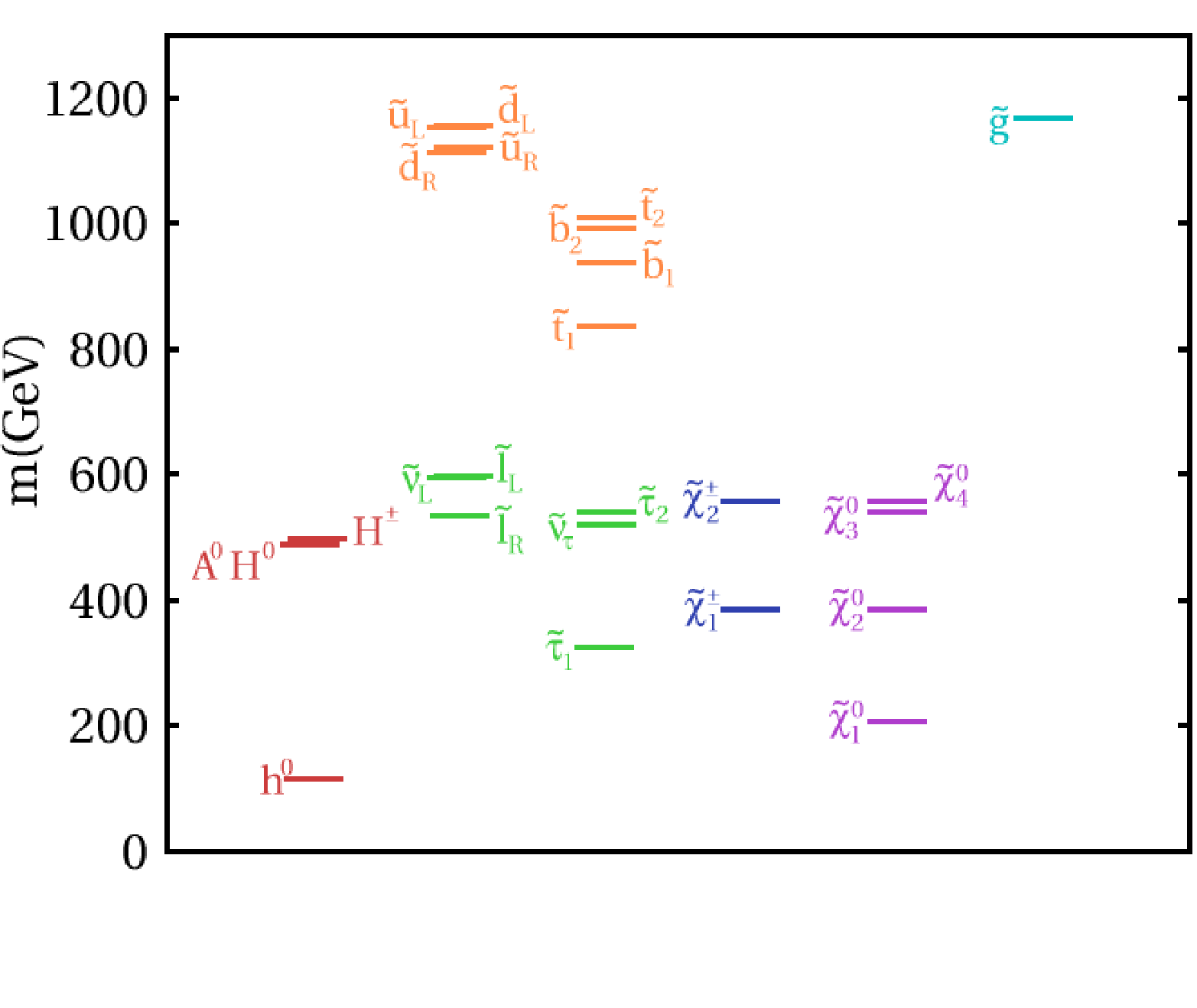}(c)
\end{center}
\caption{{\em This plot shows the running of $m_{H_1}^2$ and $m_{H_2}^2$ and a selection of sparticle/Higgs masses for $m_{1/2} = m_{16} = 500$ GeV, $A_0 = 0$, sign($\mu > 0$), $\tan\beta = 50$ and $\mathcal{D} = 0$ for {\rm (a)} $m_{10}/m_{16} = 1$ (the CMSSM case), {\rm (b)}  $m_{10}/m_{16} = 0.75$ and {\rm (c)} $m_{10}/m_{16} = 1.25$. They correspond to points A, B and C respectively in Table~\ref{benchtbl}. }\label{fig1}}
\end{figure}

To illustrate the above result, Fig.~\ref{fig1} shows the running of the Higgs soft masses from $M_G$ to $M_Z$ for the point in the parameter space $m_{1/2} = m_{16} = 500$ GeV, for different values of $m_{10}/m_{16}$. We shall use this ``typical'' point in the $(m_{1/2},m_{16})$-plane to compare the results of varying $m_{10}/m_{16}$ and $\mathcal{D}$. It satisfies most of, if not all, the experimental constraints (depending on the exact values of $\mathcal{D}$ and $m_{10}/m_{16}$) and avoids the particularly sensitive regions close to where EWSB fails. Also, all the approximations used in the analysis in this section are more or less valid in this region. Note that there is nothing ``special'' about setting $m_{16} = m_{1/2}$ since all the soft masses are renormalized so differently. The precise values of the sparticle and Higgs masses can be gleaned from Table~\ref{benchtbl}, points A, B and C, and the values of the soft mass parameters are approximately given by the square of the light family sfermion masses. The sparticle and Higgs masses are also displayed diagrammatically in Fig.~\ref{fig1}. These numerical results were produced using the method of Section~\ref{method}.

\begin{figure}[tbp]
\begin{center}
\includegraphics[width=0.47\textwidth]{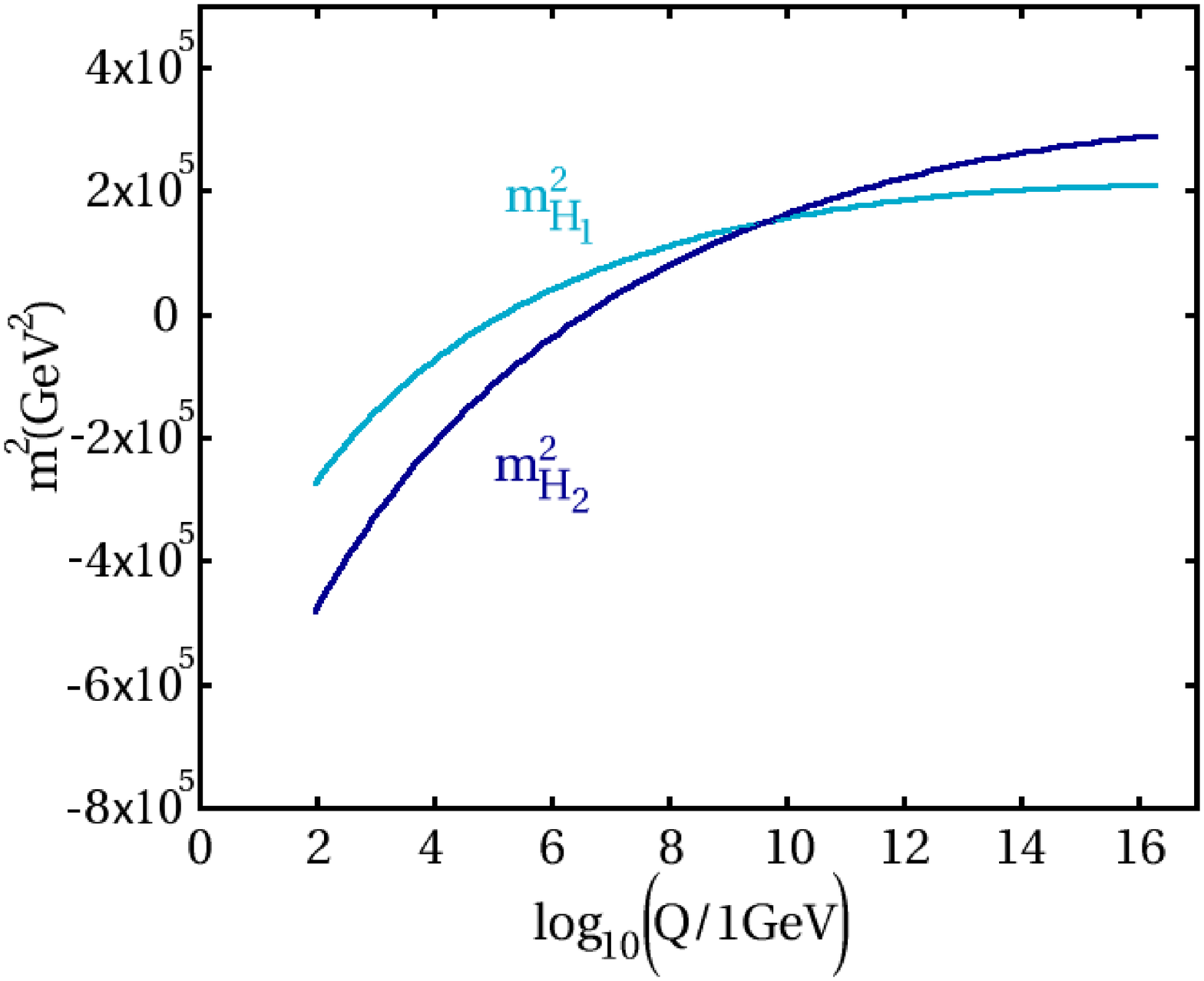}  
\includegraphics[width=0.47\textwidth]{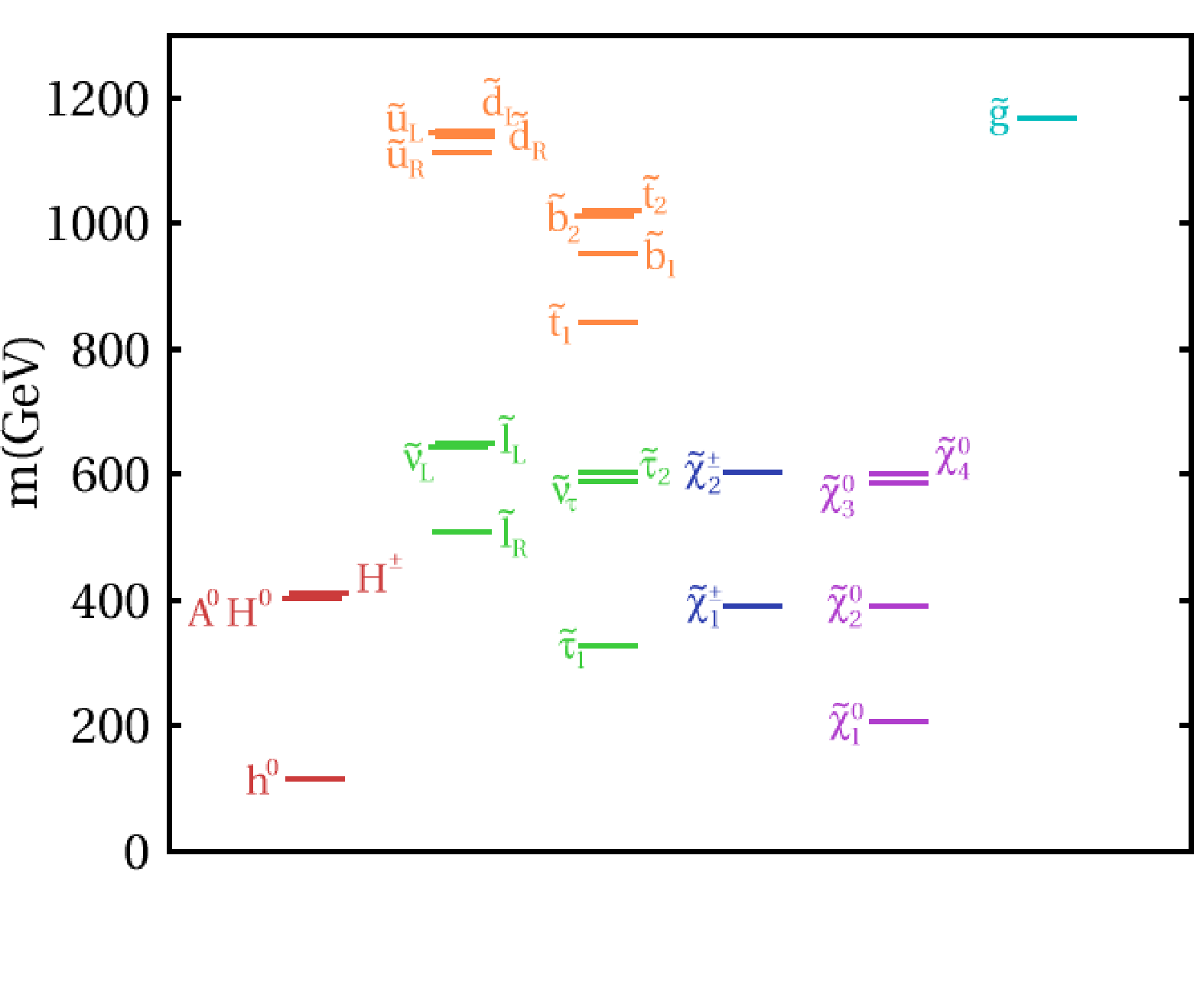}(a)  
\includegraphics[width=0.47\textwidth]{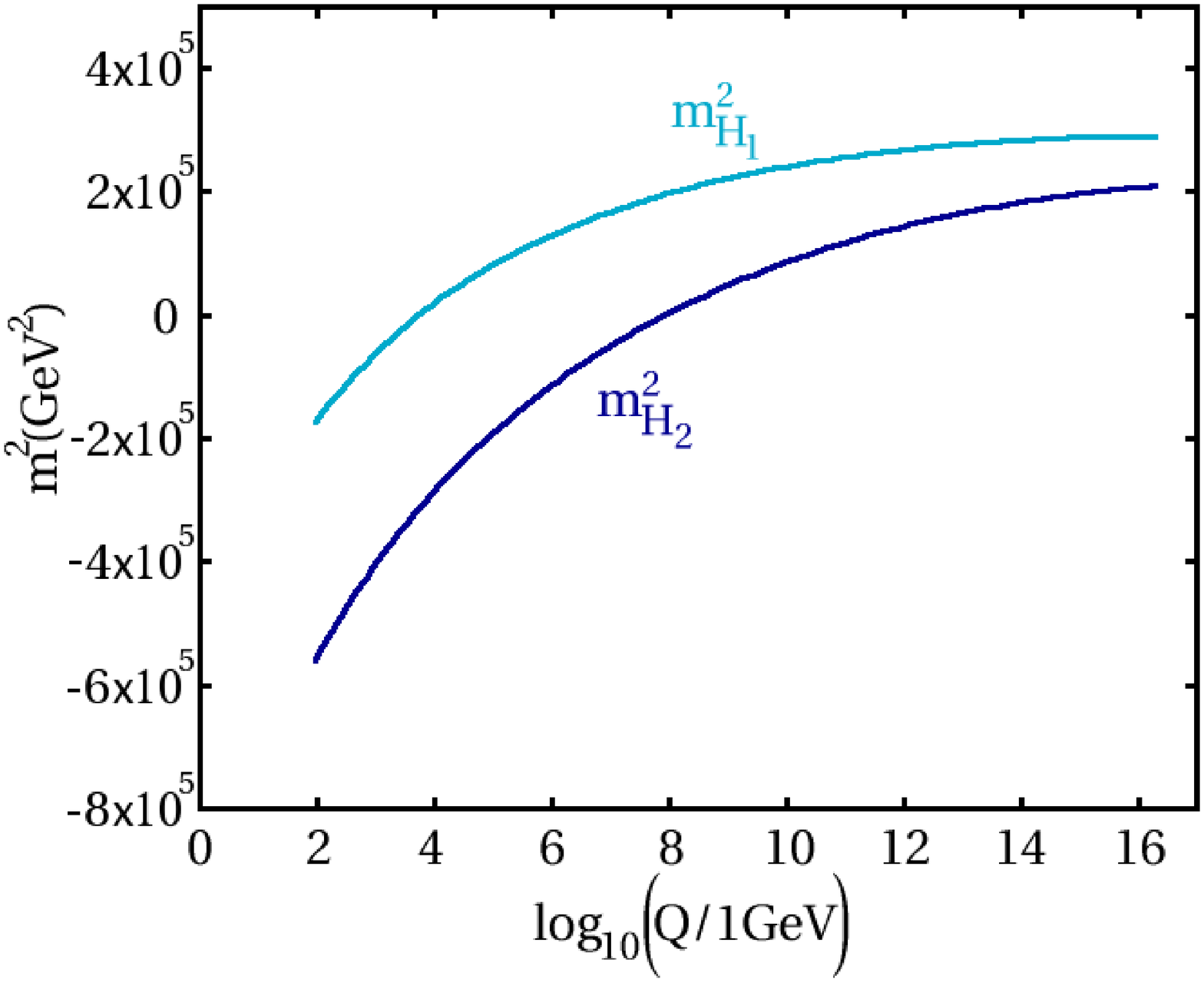} 
\includegraphics[width=0.47\textwidth]{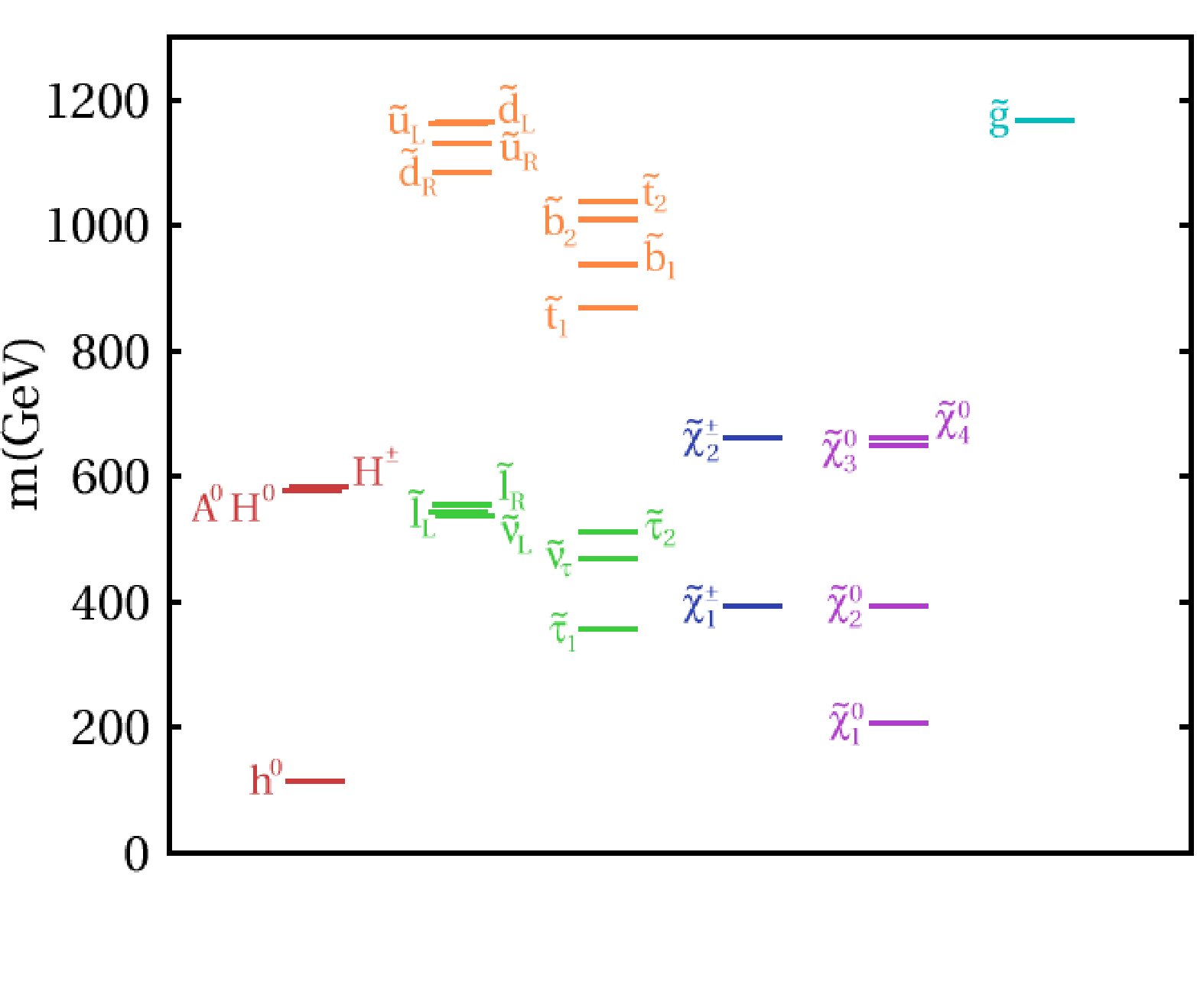}(b)
\end{center}
\caption{{\em Same as Fig.~\ref{fig1}(a), but in {\rm (a)} $\mathcal{D} = -0.4$ and in {\rm (b)} $\mathcal{D} = 0.4$. They correspond to points D and E in Table~\ref{benchtbl}.}\label{fig2}}
\end{figure}

The $D$-terms have a much larger effect on the heavy Higgs masses than $m_{10}/m_{16}$ since they split the Higgs soft masses at tree level rather than through subtleties involving one-loop corrections. At $M_G$, the splitting is
\begin{displaymath}
m_{H_1}^2 - m_{H_2}^2 = 4g_{10}^2 D^2.
\end{displaymath}
This has important implications for the breaking of electroweak symmetry; for large negative values of $D^2$, $m_{A^0}^2$ can be forced negative indicating that the electroweak symmetry has not been broken correctly; in other words, a solution to the potential minimisation conditions consistent with the measured value of $M_Z$ cannot be obtained without also having $m_{A^0}^2 < 0$. Of course, negative $D^2$ does not automatically mean that this will occur since $m_{H_2}^2$ is renormalized differently from $m_{H_1}^2$ in a way that increases the difference $m_{H_1}^2 - m_{H_2}^2$. However, it does mean that $m_{A^0}$ can be significantly smaller in this case than in the CMSSM. This is interesting because the region of parameter space where $2m_{\tilde{\chi}_1^0} \simeq m_{A^0}$, permitting resonant annihilation of LSP neutralinos via an S-channel $A^0$, can occur in different parts of the parameter space as we shall see. Fig.~\ref{fig2} shows the running of the soft Higgs masses and part of the spectrum when the $D$-terms are non-zero. Again, the full spectrum can be found in Table~\ref{benchtbl} points D and E.  

\begin{figure}[tbp]
\begin{center}
\includegraphics[width=0.47\textwidth]{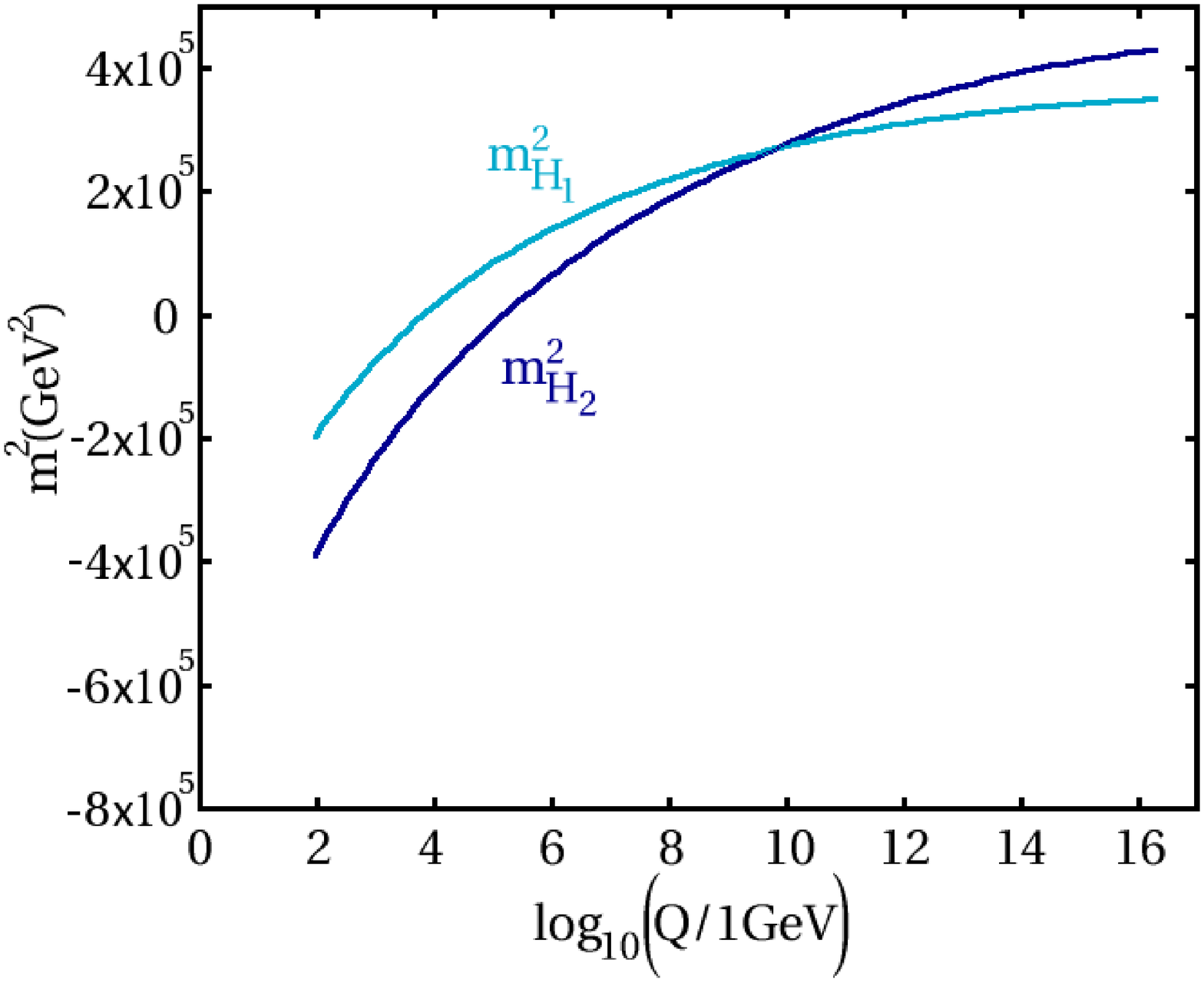}  
\includegraphics[width=0.47\textwidth]{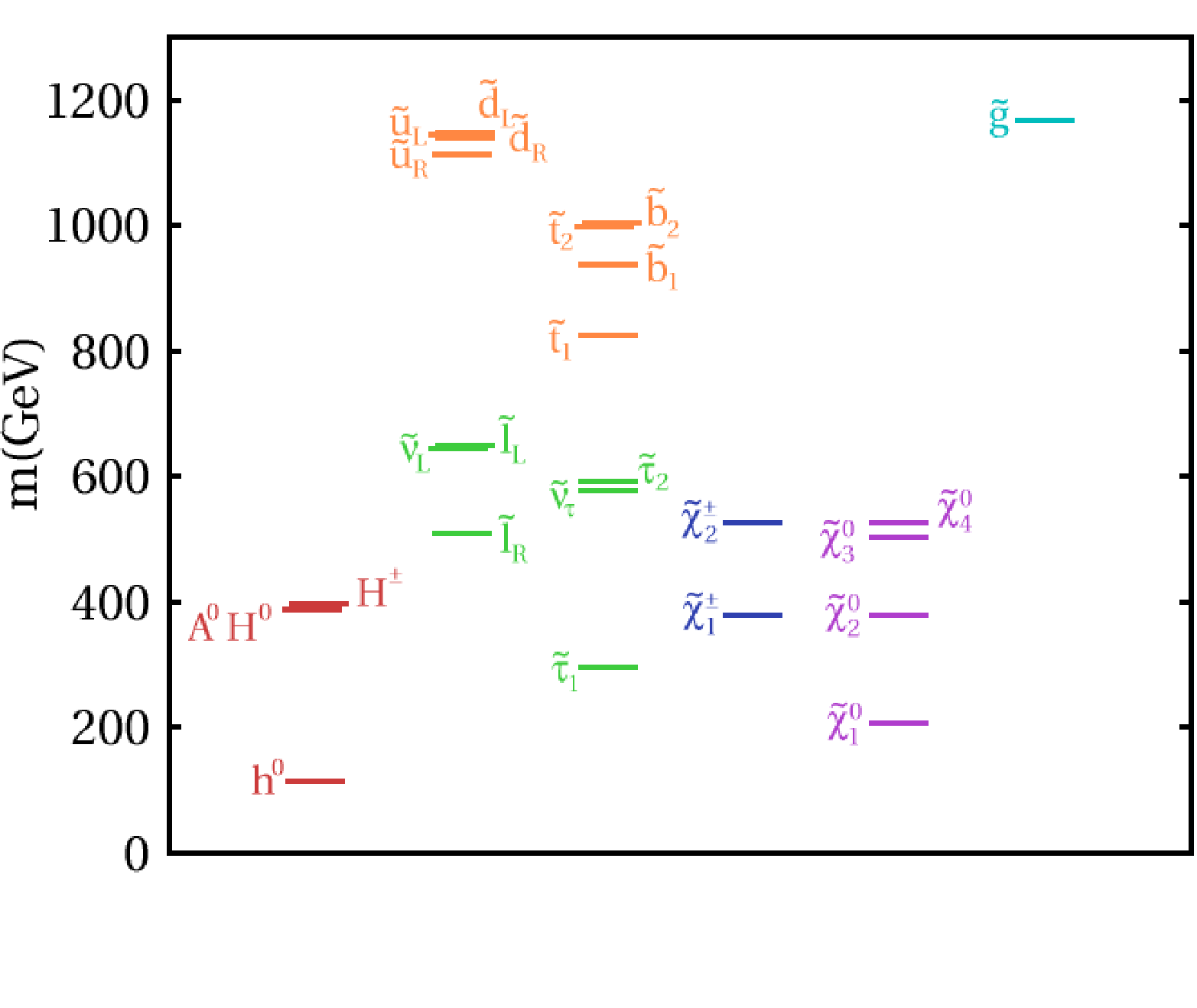}(a)  
\includegraphics[width=0.47\textwidth]{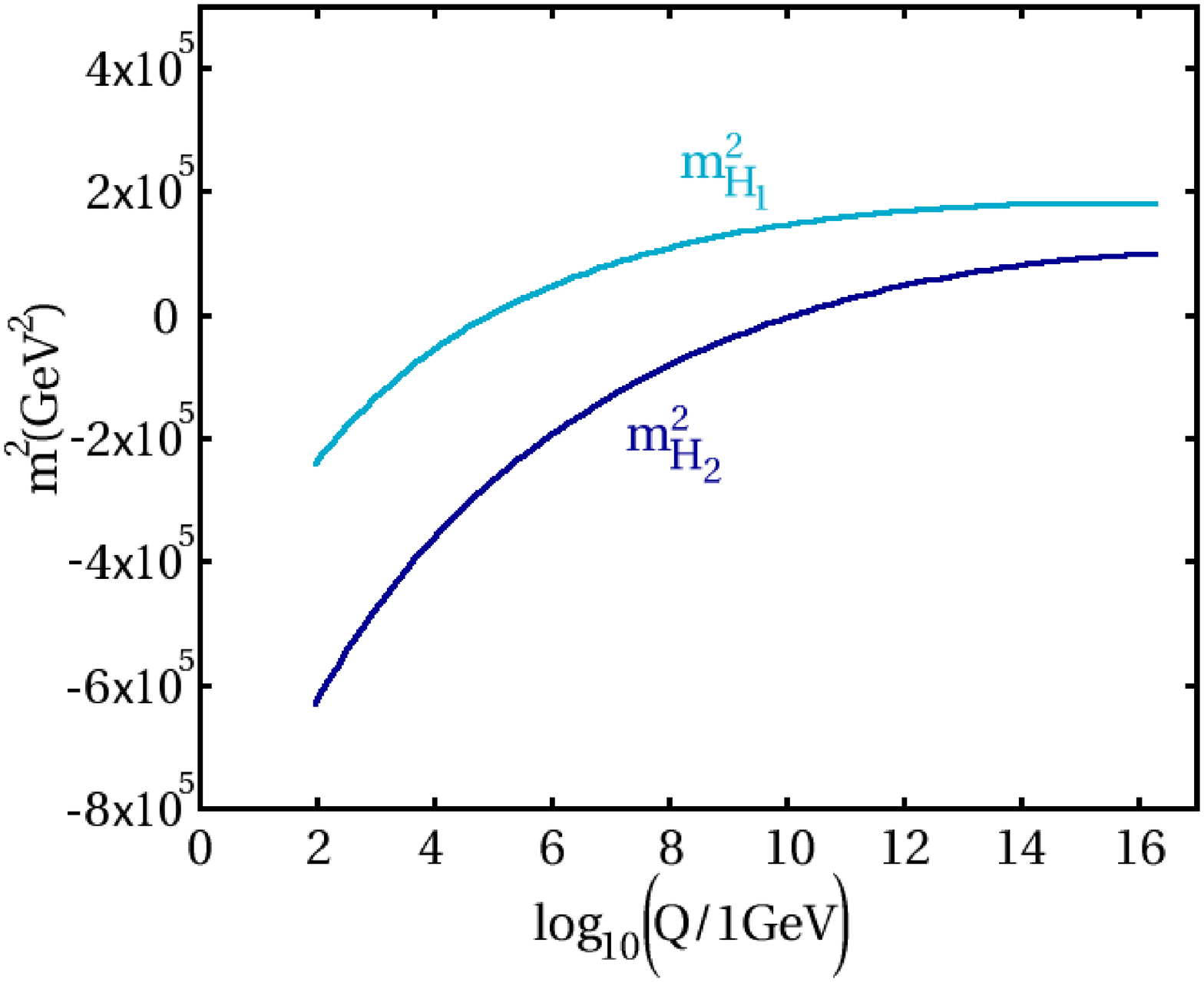} 
\includegraphics[width=0.47\textwidth]{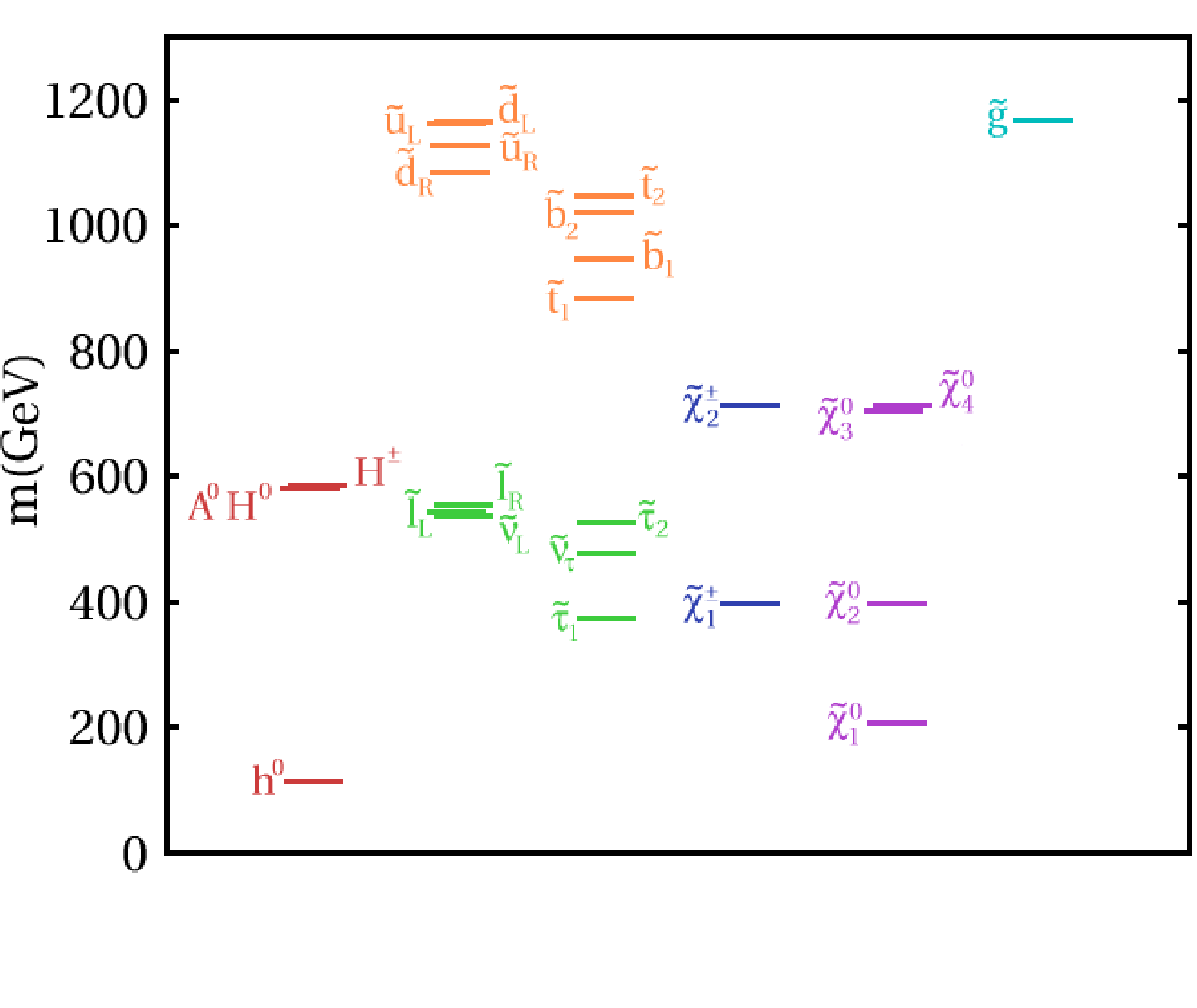}(b)
\end{center}
\caption{{\em Similar to Fig.~\ref{fig1}(a), but in {\rm (a)} $\mathcal{D} = -0.4$ and $m_{10}/m_{16} = 1.25$, while in {\rm (b)} $\mathcal{D} = 0.4$ and $m_{10}/m_{16} = 0.75$. They correspond to points F and G in Table~\ref{benchtbl}.}\label{fig3}}
\end{figure}

Finally, in order to show the full contrast of all the effects of combining the $D$-terms with $m_{10}/m_{16}$, we show the plots in Fig.~\ref{fig3} (corresponding to Table~\ref{benchtbl} points F and G).

\subsection{Electroweak Symmetry Breaking and $\mu$}\label{EWSB}
The value of $\mu$ is Highly dependent on the Higgs soft masses. The minimisation of the Higgs potential gives the following equation~\cite{Pierce:1997zz}:
\begin{equation}\label{mueq}
|\mu|^2 = \frac{\overline{m}_{H_1}^2 - \overline{m}_{H_2}^2 \tan^2\beta}{\tan^2\beta - 1} - \frac{1}{2} M_Z^2 - \frac{1}{2} \mathcal{R}e \Pi_{ZZ}^T
\end{equation}
in the same notation as previously. In the large $\tan\beta$ limit, the term involving $m_{H_1}^2$ becomes irrelevant and $m_{H_2}^2$ must be of $\mathcal{O}(M_Z^2)$ or negative at the electroweak breaking scale for an acceptable, i.e. $|\mu|^2 > 0$, solution to be found. In general, large radiative corrections due to the top Yukawa coupling achieve this. However, in the high $m_0$, low $m_{1/2}$ region of the CMSSM, sometimes referred to as the ``focus-point'' region~\cite{Feng:1999mn,Feng:1999zg,Feng:2000gh,Feng:2000bp}\footnote{``Focus-point'' refers to the fact that the RGEs exhibit an approximate focusing effect on $m_{H_2}^2$ near the weak scale as $m_0$ is varied, keeping other parameters fixed. In other words the value of $m_{H_2}^2$ is insensitive to $m_0$ and tends to be naturally of $\mathcal{O}(M_Z^2)$. Choosing $m_{1/2}$ to be very small and $m_0$ large, the soft masses and $\mu$ are naturally of order the EW scale in Eq.~\ref{mueq} in order to produce the correct result for $M_Z$, resulting in very small overall fine tuning even though the sfermion masses can be $\mathcal{O}$(several $\times$ TeV)}, the potency of these radiative corrections is reduced. This is because at small $m_{1/2}$ the relative size of the terms chiefly responsible for driving $m_{H_2}^2$ negative, i.e. those involving $|y_t|^2$ multiplied by the soft squark masses $m^2_{\tilde{Q}}$ and $m^2_{\tilde{u}_R}$ in the RGE for $m_{H_2}^2$, are smaller. One can understand this by realizing that a small value for $m_{1/2}$ means a small value for the gaugino mass $M_3$ and it is terms proportional to $g_3^2|M_3|^2$ that are the main factors involved in making the squark masses (and therefore $X_t$) large as they are RG evolved to low scales. Moreover, a large $|a_t|^2$, the square of the trilinear stop-stop-Higgs coupling, also helps to drive down $m_{H_2}^2$. Starting from $A_0 = 0$ at $M_G$, $a_t$ is driven negative below this scale by a term proportional to $M_3$. The larger $M_3$ is, the larger $|a_t|$ becomes and the stronger the effect on $m_{H_2}^2$ as the parameters are run towards $M_Z$. If $m_{1/2}$ is small enough the radiative corrections may not be large enough to drive $m_{H_2}^2$ sufficiently low to break the electroweak symmetry. For a given value of $m_{1/2}$, it is generally possible to choose a large initial value of $m_{H_2}^2$, i.e. $m_0^2$, for which $\mu \simeq 0$ at low scales. The boundary along which $\mu$ vanishes marks the border of correct EWSB, although before this line is reached, the LEP limits on the chargino mass are violated since as $\mu$ drops below $M_2$, the lightest chargino becomes $\sim \mu$ GeV. As is well known~\cite{Polonsky:1995rz,Matalliotakis:1995ft,Olechowski:1995gm,Murayama:1995fn,Rattazzi:1995gk}, by admitting non-universality in the Higgs sector, one can find successful EWSB in regions excluded in the case of universal scalar masses. 

$D$-terms, therefore, at the level of the input scale, can either help or hinder the EWSB process by increasing or decreasing $m_{H_2}^2$ at $M_G$ since $m^2_{H_2} = m_{10}^2 - 2 g_{10}^2 D^2$. Thus we may expect the $\mu = 0$ boundary to move to lower values of $m_{16}$ for $D^2 < 0$ and to higher values for $D^2 > 0$ for a given $m_{1/2}$. In practice, however, by varying the $D$-terms alone (with $m_{10}/m_{16}$ set to 1), the point $m_{A^0}^2 = 0$ is often reached before $\mu = 0$, depending on the exact location in the parameter space. On the other hand, setting the $D$-terms to 0 and increasing $m_{10}/m_{16}$, we push up the value of $m_{H_2}^2$ and therefore $|\mu|^2$ towards zero at the EWSB scale towards without decreasing the value of $m_{A^0}^2$ enormously. Therefore, at much lower values of $m_{16}$ than in the CMSSM, we find the boundary $\mu = 0$ where EWSB fails. 

We now look at a set of rather more interesting points in parameter space that illustrate some of the observations of this section --- points close to where the breakdown of the radiative EWSB mechanism occurs. We also draw attention to another important fact, namely that, even relatively far from where EWSB fails, at low $m_{1/2}$, high $m_{16}$, $\mu$ becomes increasingly sensitive to changes in the $D$-terms. This can be traced back to the fact that the $D$-term contribution essentially provides an additive constant to $m_{H_2}^2$ imposed at $M_G$, and, in our parameterization, it is proportional to $m_{16}$. In the small $m_{1/2}$, large $m_{16}$ region of parameter space, in the CMSSM case, $\mu$ is quite small and a large positive value for $D^2$ for example, will result in a relatively large correction to $m_{H_2}^2$ and therefore a relatively large increase in $\mu$. 

\begin{figure}[tbp]
\begin{center}
\includegraphics[width=0.47\textwidth]{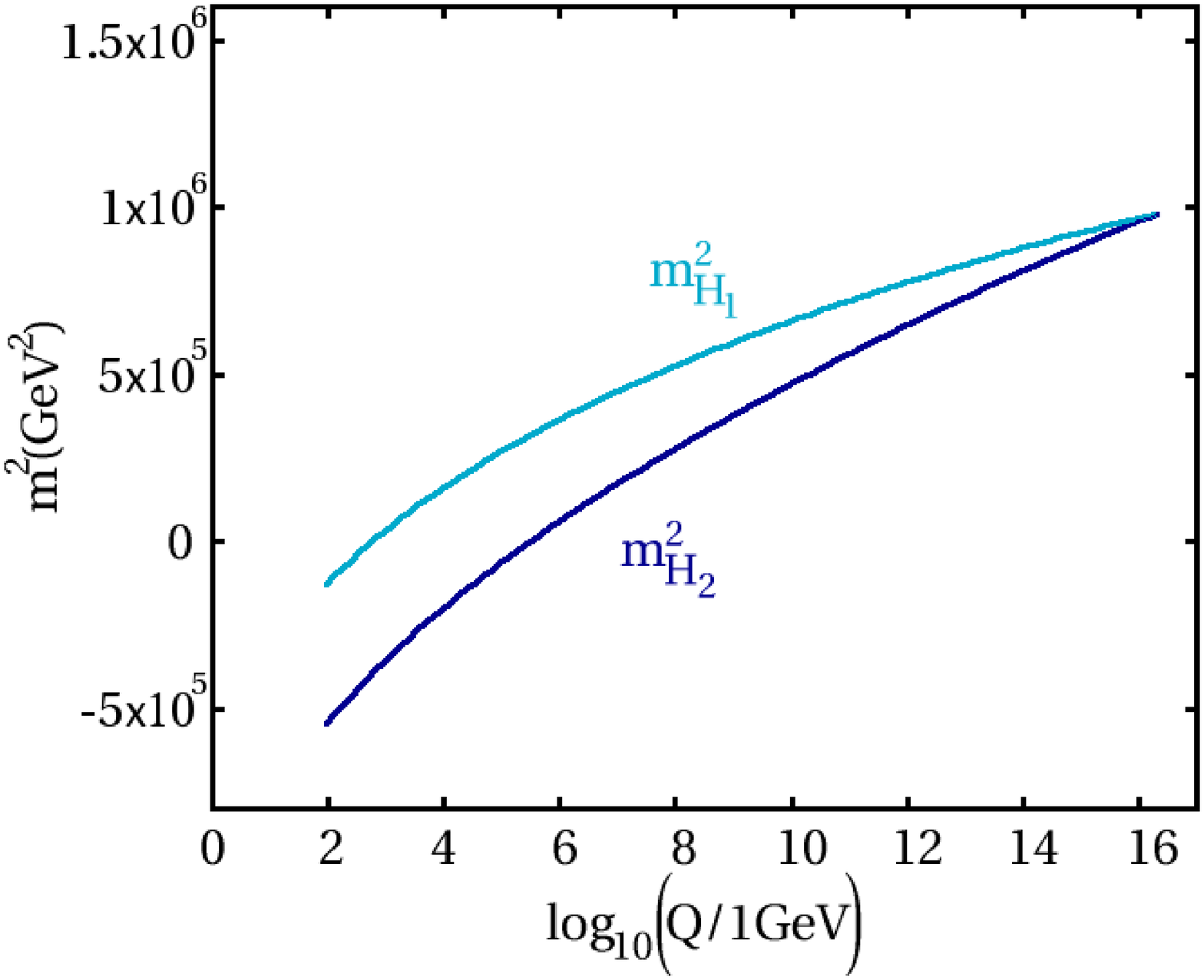}
\includegraphics[width=0.47\textwidth]{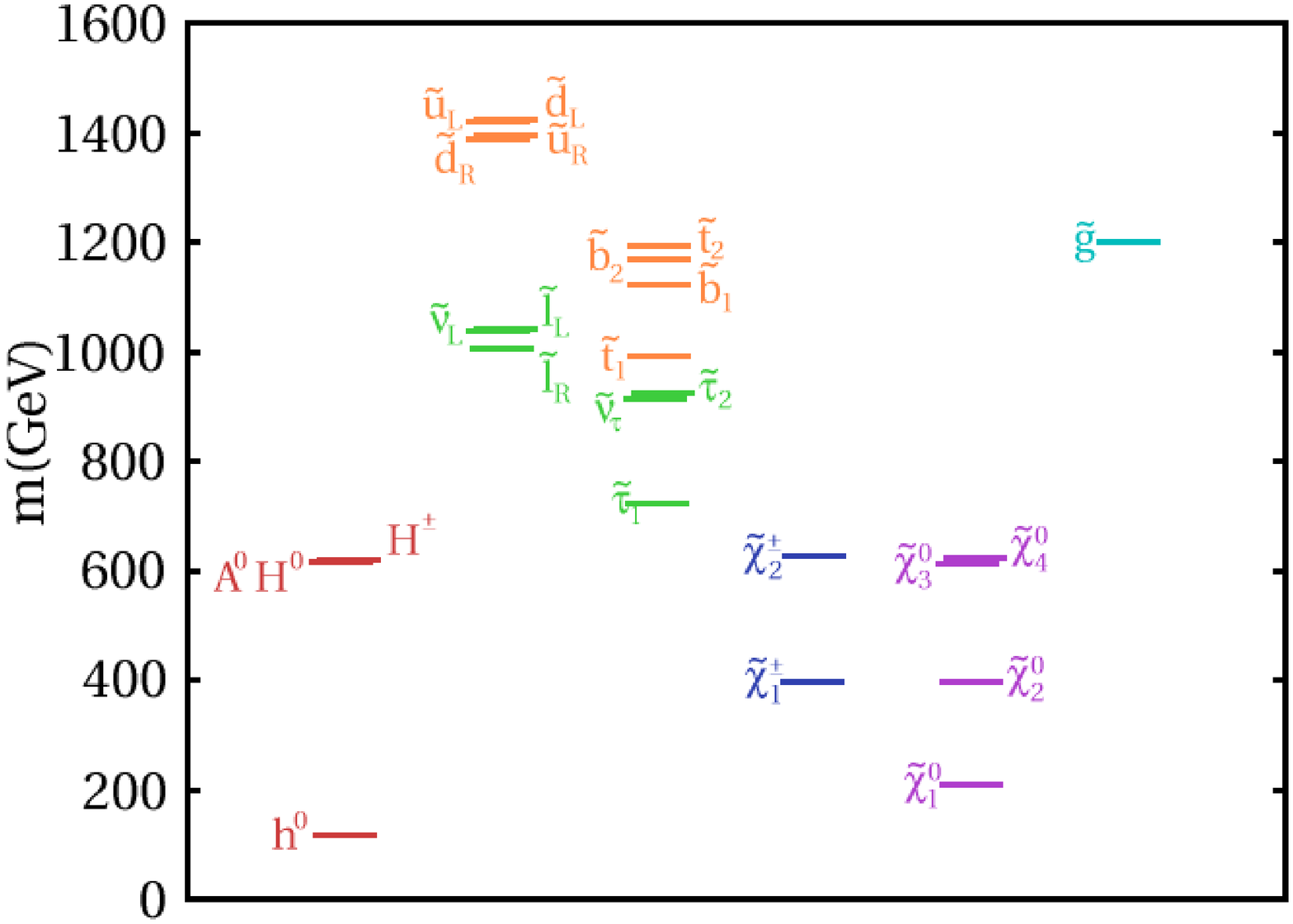}(a)
\includegraphics[width=0.47\textwidth]{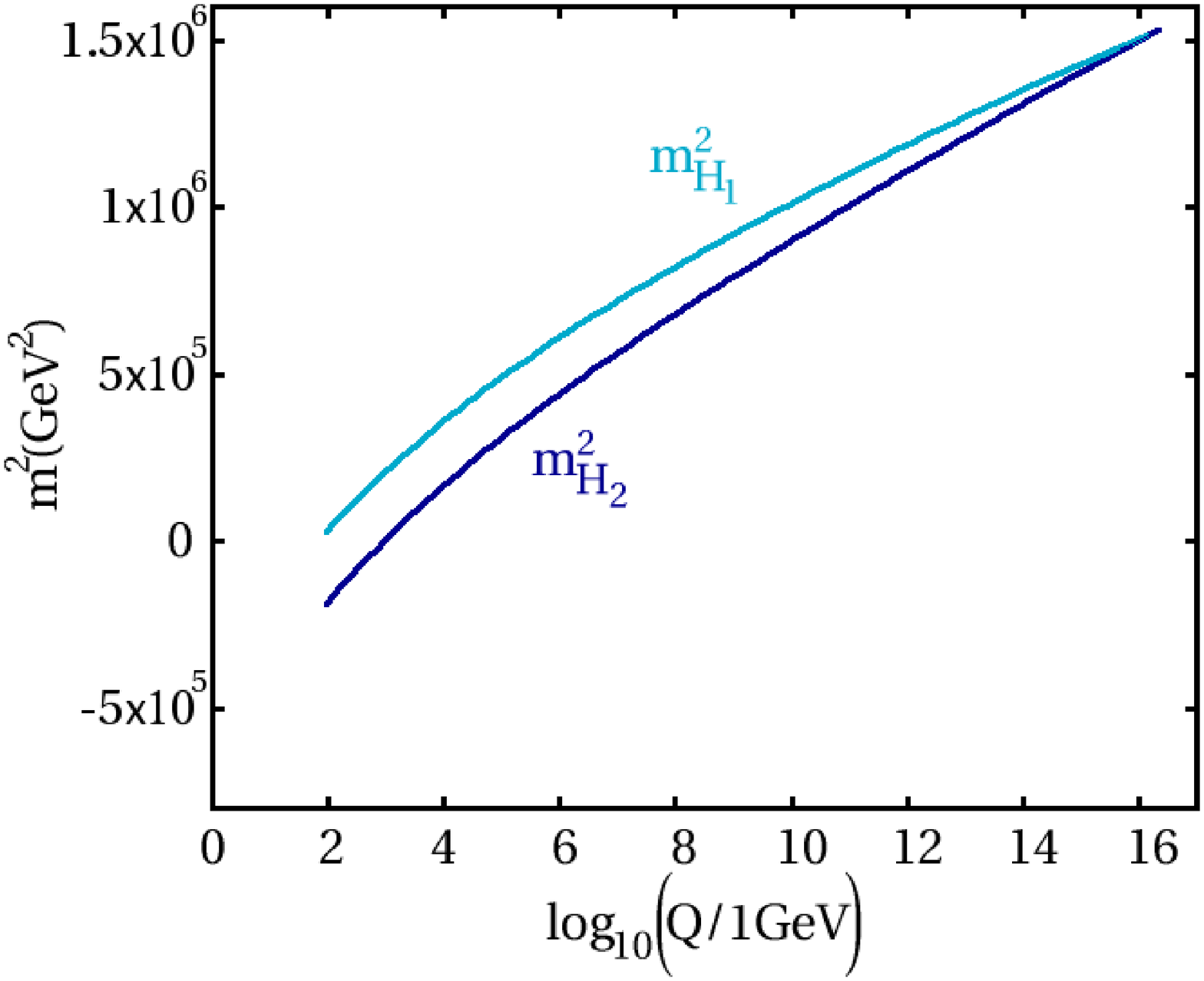} 
\includegraphics[width=0.47\textwidth]{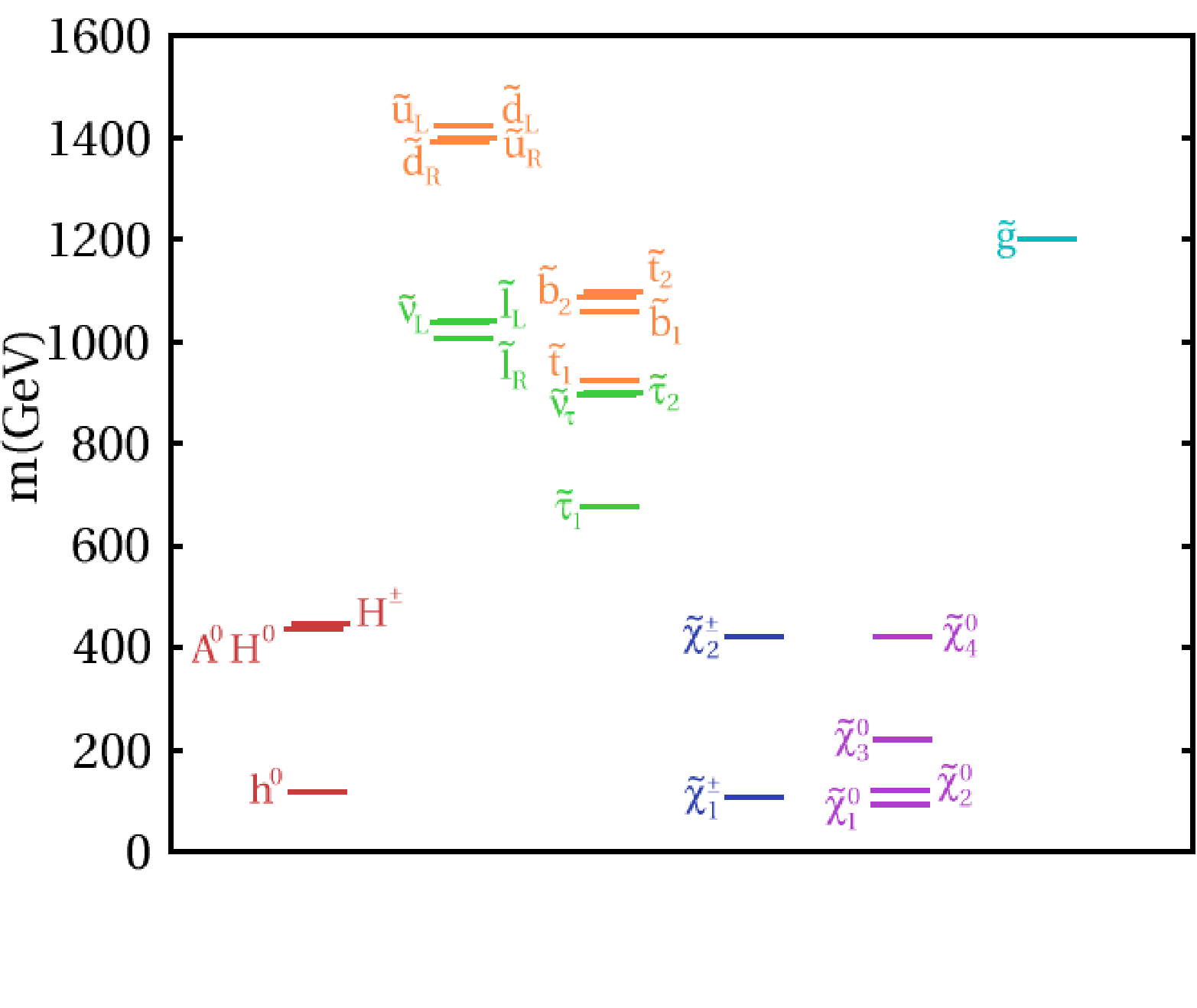}(b)
\end{center}
\caption{{\em Similar to Fig.~\ref{fig1} but in {\rm (a)} $m_{1/2} = 500$ GeV, $m_{16} = 990$ GeV, $m_{10}/m_{16} = 1$ and $\mathcal{D} = 0$, in {\rm (b)} $m_{1/2} = 500$ GeV, $m_{16} = 990$ GeV, $m_{10}/m_{16} = 1.25$ and $\mathcal{D} = 0$. These represent points {\rm U} and {\rm V} in Table~\ref{Inttbl}. Note how light the chargino and neutralino spectra are in {\rm (b)} compared to {\rm (a)}.}\label{lowmu}}
\end{figure}

To begin with, in Fig.~\ref{lowmu}(a) and (b) 
and Table~\ref{Inttbl} points U and V we compare a point in the CMSSM parameter space, $m_{1/2} = 500$ GeV, $m_{16} = 990$ GeV, where $\mu$ is still fairly sizeable and the corresponding point with $\mathcal{D} = 0$ and $m_{10}/m_{16} = 1.25$ where $\mu$ is very small. Note that this occurs for a relatively low value of $m_{16}$, much lower than the ``focus point'' region in the CMSSM, and in this region, $\mu$ is much more sensitive to the change in $m_{10}/m_{16}$ than for our points in Figs.~\ref{fig1}-\ref{fig3} and Table~\ref{benchtbl}. This will be important later on when we come to discuss observables such as $(g-2)_\mu$ and $\mathcal{B}(b\rightarrow X_s \gamma)$. Of particular importance is the lightness of the charginos and neutralinos, the lightest of which have important consequences for the cold dark matter relic density.

\begin{figure}[tbp]
\begin{center}
\includegraphics[width=0.47\textwidth]{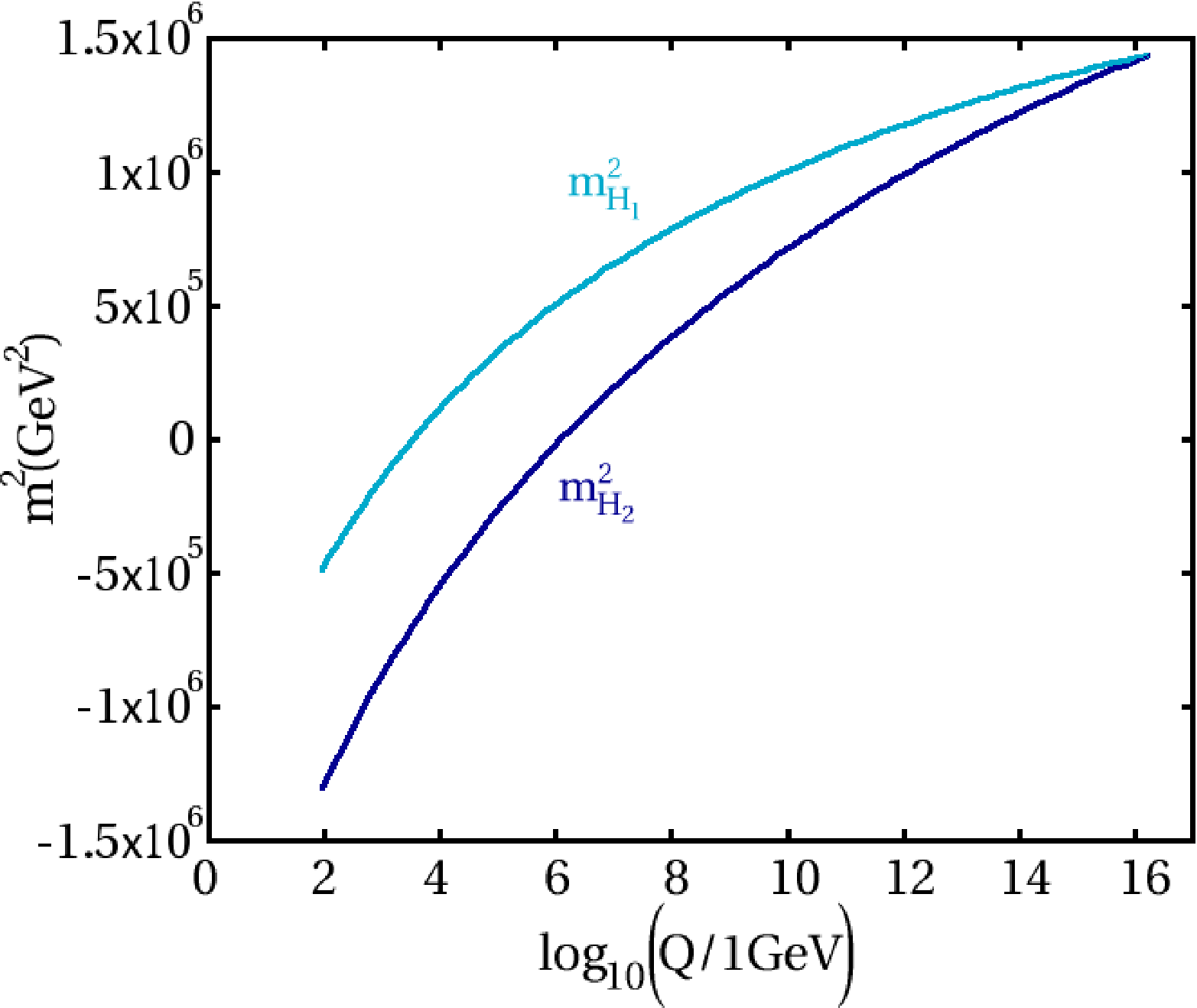}
\includegraphics[width=0.47\textwidth]{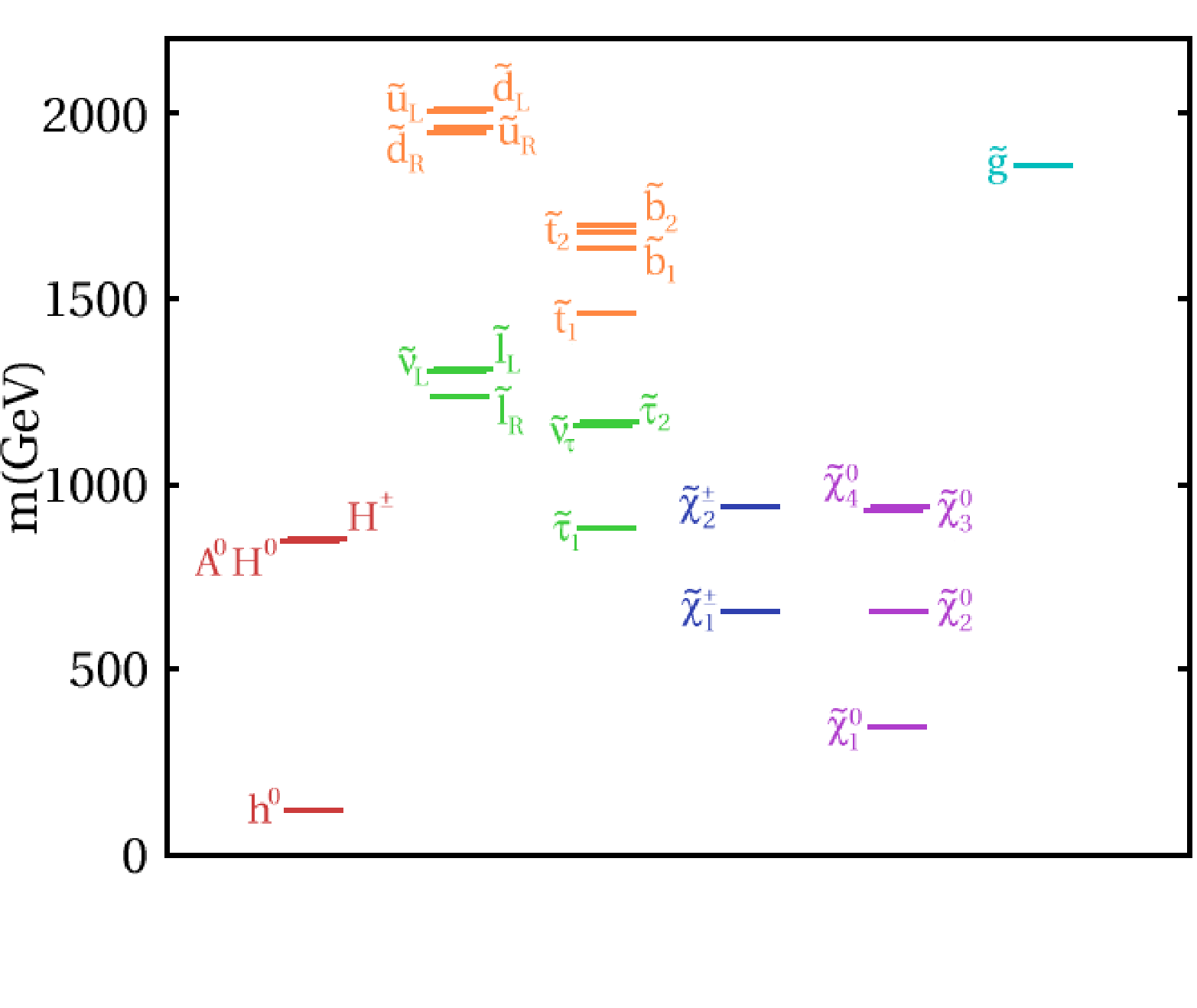}(a)
\includegraphics[width=0.47\textwidth]{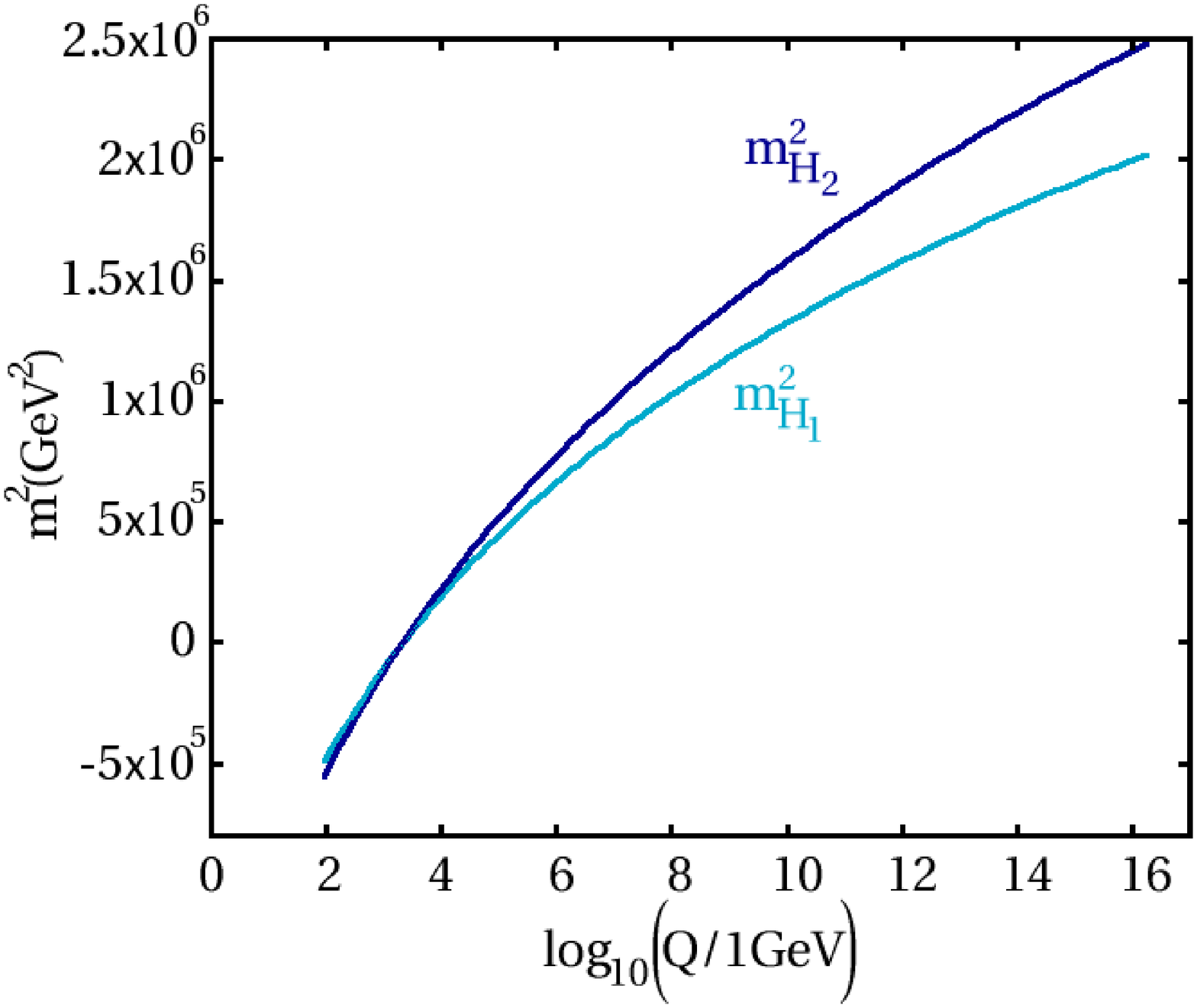} 
\includegraphics[width=0.47\textwidth]{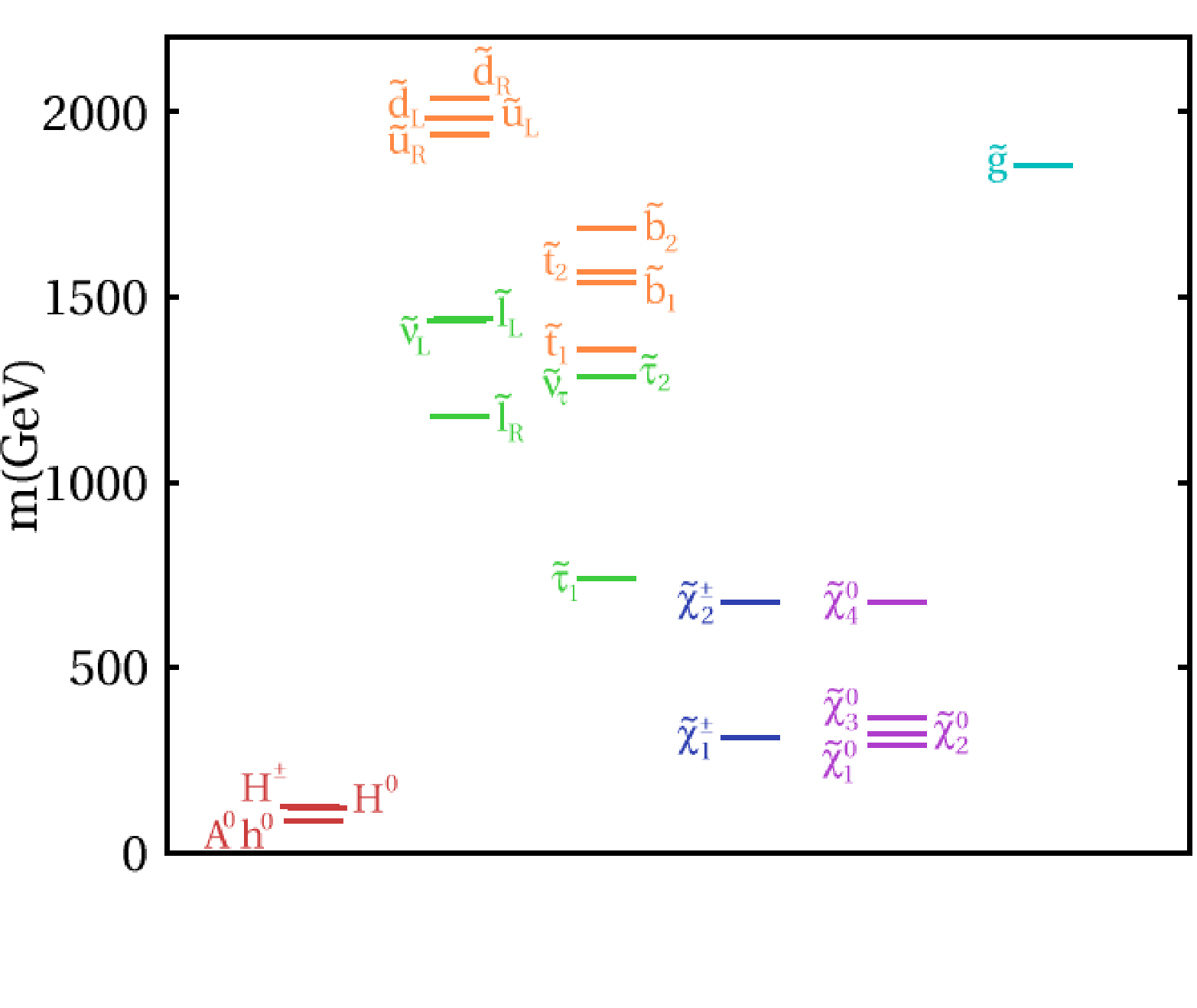}(b)
\end{center}
\caption{{\em Similar to Fig.~\ref{fig1} but in {\rm (a)} $m_{1/2} = 813$ GeV, $m_{16} = 1200$ GeV, $m_{10}/m_{16} = 1$ and $\mathcal{D} = 0$, in {\rm (b)} $m_{1/2} = 813$ GeV, $m_{16} = 1200$ GeV, $m_{10}/m_{16} = 1.25$ and $\mathcal{D} = -0.4$. These represent points {\rm W} and {\rm X} in Table~\ref{Inttbl}. Notice how light the heavy Higgses are in {\rm (b)}.}\label{lowma}}
\end{figure}

In Fig.~\ref{lowma}(a) and (b) 
and Table~\ref{Inttbl} we have points W and X. W is the CMSSM point at $m_{1/2} = 813$ GeV and $m_{16} = 1200$ GeV. X shows the same point, but with $\mathcal{D} = -0.4$ and $m_{10}/m_{16} = 1.25$. For X, due to the large effect of the negative $D$-terms on $m_{A^0}$, the heavy Higgs masses are tiny. Note that, even though $m_{10}/m_{16} = 1.25$, $\mu$ is still far from zero, around $300$ GeV. 

\begin{figure}[tbp]
\begin{center}
\includegraphics[width=0.47\textwidth]{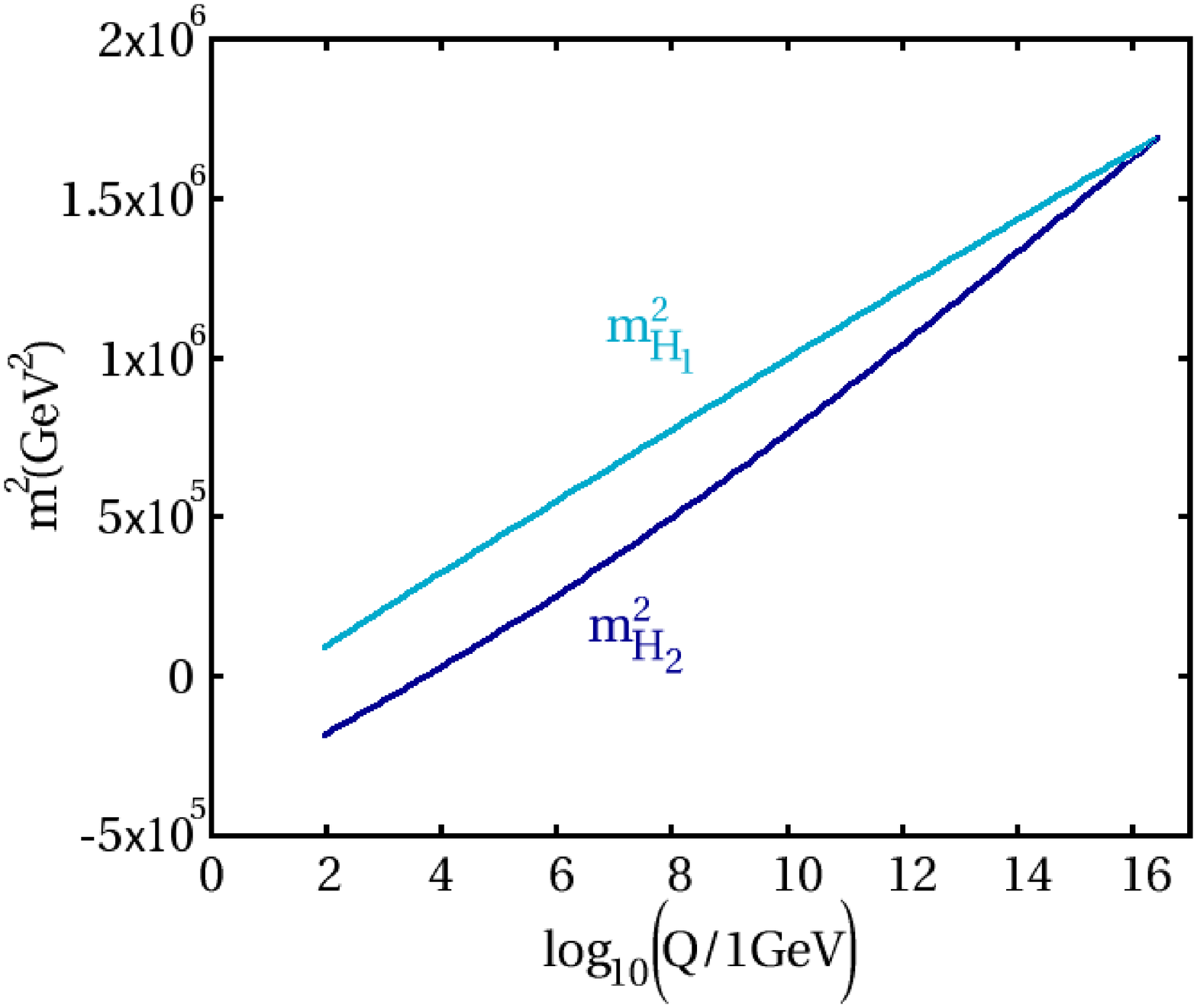}
\includegraphics[width=0.47\textwidth]{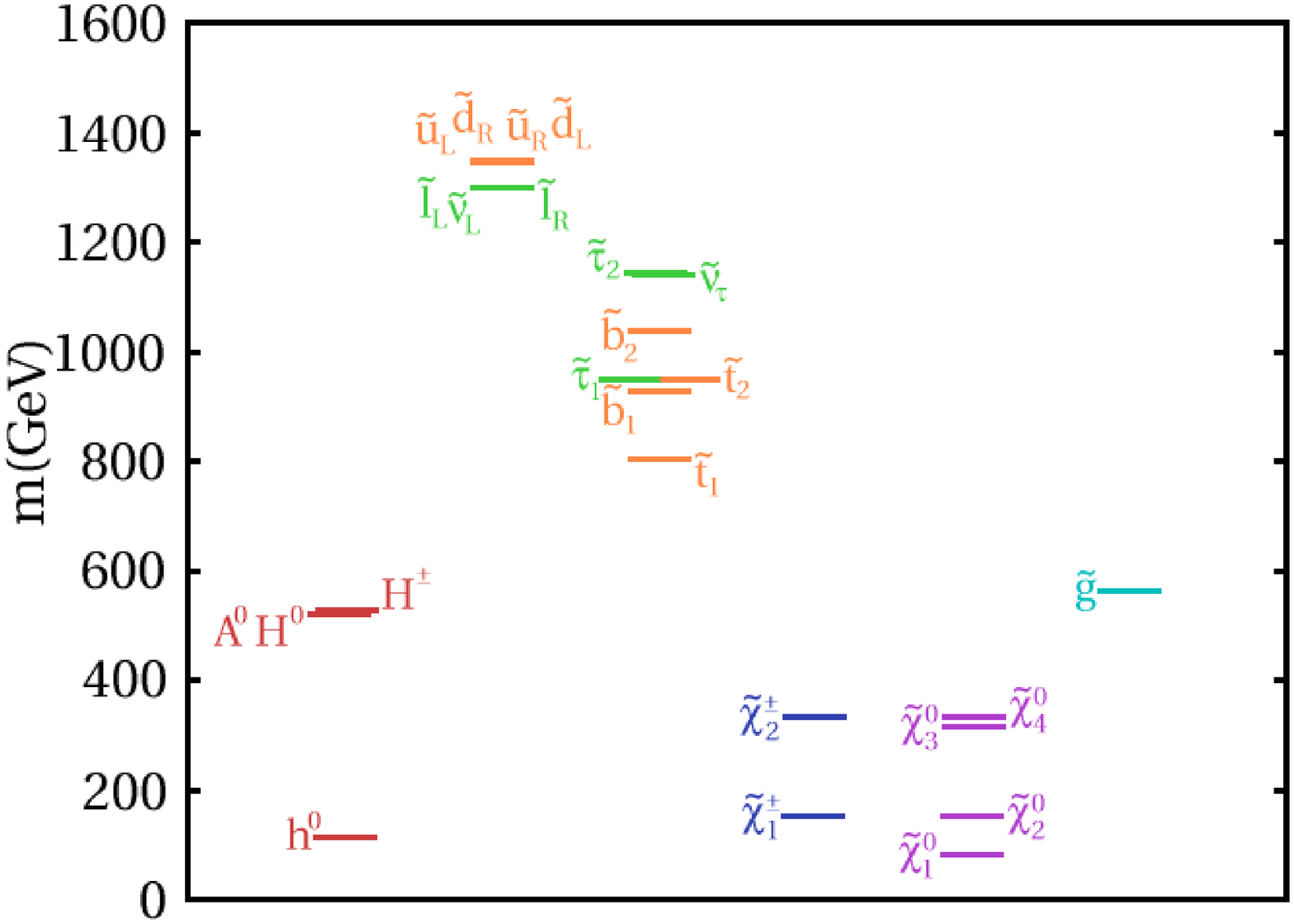}(a)
\includegraphics[width=0.47\textwidth]{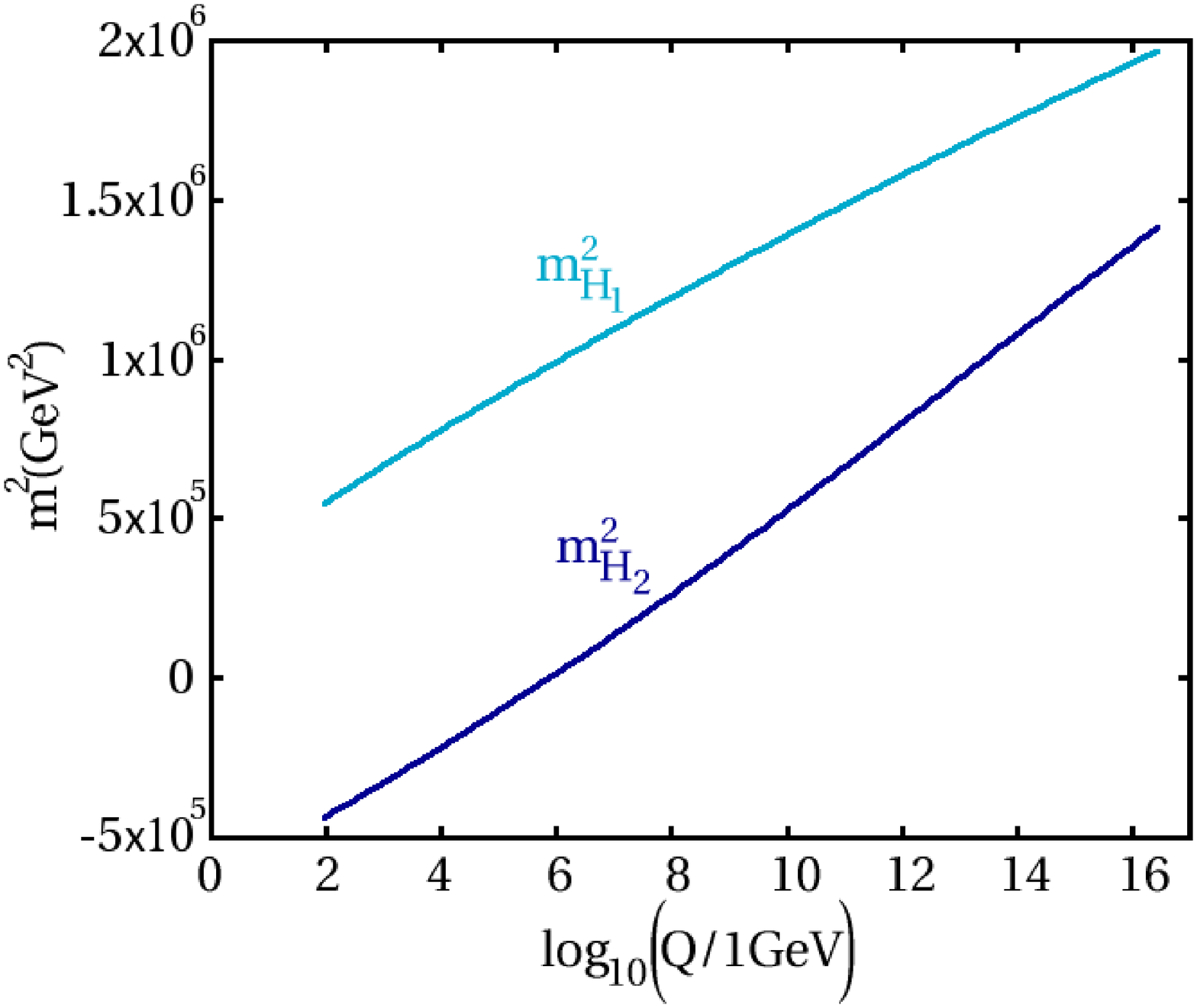} 
\includegraphics[width=0.47\textwidth]{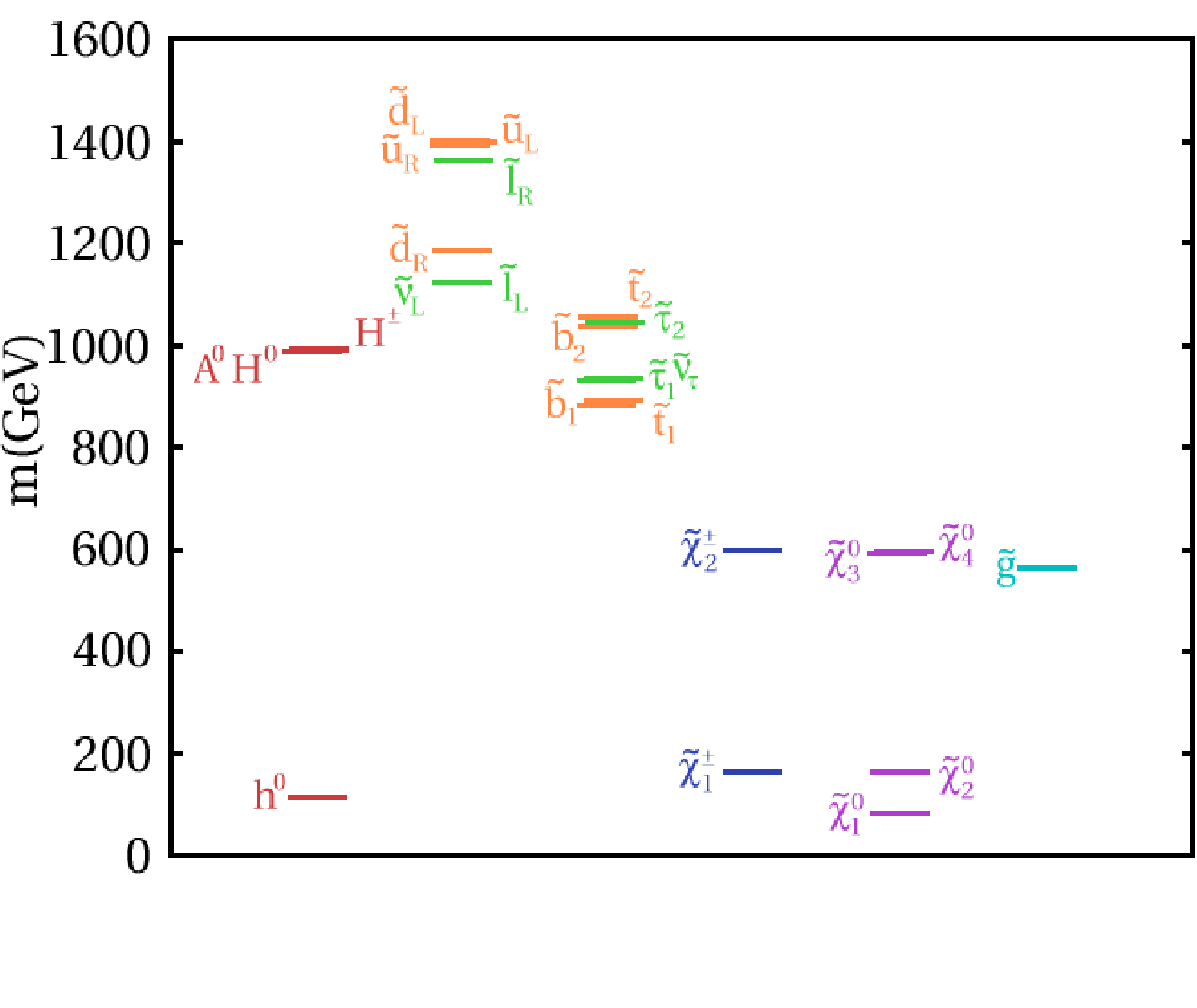}(b)
\end{center}
\caption{{\em Similar to Fig.~\ref{fig1} but in {\rm (a)} $m_{1/2} = 200$ GeV, $m_{16} = 1300$ GeV, $m_{10}/m_{16} = 1$ and $\mathcal{D} = 0$, in {\rm (b)} $m_{1/2} = 200$ GeV, $m_{16} = 1300$ GeV, $m_{10}/m_{16} = 1$ and $\mathcal{D} = 0.4$. These represent points {\rm Y} and {\rm Z} in Table~\ref{Inttbl}. The $D$-terms affect $\mu$ about as much as they do $m_{A^0}$ in this region of parameter space.}\label{lowhi}}
\end{figure}

The final pair of points we will consider are at $m_{1/2} = 200$ GeV, $m_{16} = 1300$ GeV, firstly for zero $D$-terms and $m_{10}/m_{16} = 1$ and secondly for $\mathcal{D} = 0.4$ and $m_{10}/m_{16} = 1$. The results are shown in Fig.~\ref{lowhi}(a) and (b) 
and Table~\ref{Inttbl}, points Y and Z, demonstrating the point made earlier: that $\mu$ is more sensitive to changes in $D^2$ (and $m_{10}/m_{16}$) at small $m_{1/2}$ and large $m_{16}$ than for regions where $m_{1/2} \sim m_{16}$, and is especially so close to the no EWSB boundary. Again, this fact will be of importance in our discussion of $\tan\beta$-enhanced amplitudes. 

\subsection{The Sfermions}
We now consider the effects of the $D$-terms and $m_{10}/m_{16}$ on the sfermion spectrum. For a graphical representation of what follows the reader is referred back to Figs.~\ref{fig1} - \ref{lowhi}. Due to the fact that they almost completely cancel out of the RGEs, the $D$-term contributions to the masses squared of the sfermions evaluated at $M_{{\textrm{SUSY}}}$ are numerically equivalent to their contributions to the soft mass squared parameters at the GUT scale, Eqs.~\ref{Deqs}, to a good approximation. As we have discussed already, increasing $m_{10}/m_{16}$ will in general lead to a decrease in the third family sparticle masses as a result of larger values of the $X_i$ in the RGEs. However, as we shall see, by far the most significant effect from the point of view of phenomenology is the effect this has on $\mu$. We now turn to examine the sparticle masses in detail. In all of what follows, we will treat the first two families of squarks and sleptons as being degenerate between the generations, i.e. $m_{\tilde{d}_R} = m_{\tilde{s}_R}$, $m_{\tilde{u}_L} = m_{\tilde{c}_L}$, $m_{\tilde{\nu}_e} = m_{\tilde{\nu}_\mu}$, etc., which is accurate to a very good approximation.

\subsubsection*{The Squarks}\label{squarks}
 In the case of the squarks the $D$-term effects are simple for the two light generations and the stops; we can expect the right-handed down type squarks to be lighter/heavier than in the CMSSM as follows: 
\begin{displaymath}
m_{\tilde{d}_R}^{(D)} \simeq m_{\tilde{d}_R}\left(1 - \frac{3}{2}g_{10}^2 \frac{D^2}{{m_{\tilde{d}_R}}^2}\right),
\end{displaymath}
where the soft masses on the RHS are the mSugra soft masses at $M_Z$ and $g_{10}$ is the unified gauge coupling evaluated at $M_G$. For the up-type squarks and the left-handed down squarks: 
\begin{displaymath}
m_{\tilde{q}}^{(D)} \simeq m_{\tilde{q}}\left(1 + \frac{1}{2}g_{10}^2 \frac{D^2}{m_{\tilde{q}}^2}\right),
\end{displaymath}
 where $m_{\tilde{q}}^2$ is one of $m_{\tilde{u}_R}^2$ or $m_{\tilde{Q}}^2$. Since for the two light families there are strong radiative corrections coming from the gluino ($M_3$) term in the RGEs untempered by large Yukawa terms, the squark masses are always significantly larger than $m_{16}$ and so the corrections coming from the $D$-terms are relatively small. From our points in Figs.~\ref{fig1}-\ref{lowhi} one can see that this is a barely perceptible effect, in fact a few percent at most (in the case of the right-handed down squarks).

The stops receive a similar correction to the up-type squarks above:
\begin{displaymath}
\delta m_{\tilde{t}_i}^{(D)} \simeq m_{\tilde{t}_i}\left(1 + \frac{1}{2}g_{10}^2 \frac{D^2}{{m_{\tilde{t}_i}^2}}\right),
\end{displaymath}
where $i=1,2$, though it is a relatively larger effect, e.g. $\sim 2-3\%$ instead of $\lesssim 1\%$ for the first two family squarks (except for $\tilde{d}_R$) in Table~\ref{benchtbl}, on account of the fact that the stop soft masses are substantially suppressed by the factors $X_i$ in the relevant RGEs while the $D$-terms are unsuppressed. For $m_{1/2} \ll m_{16}$ these percentages are much larger since, in this regime, $M_3$ is small and the sfermion soft masses are no longer strongly enhanced at the weak scale by the corresponding terms in the RGEs; meanwhile, the $D$-terms are relatively large, having been chosen proportional to $m_{16}$. As a result, the $D$-term correction to the squark masses is relatively large. The $D$-terms do not significantly affect the stop mass mixing since they cancel out of the mixing term in the quadratic formula derived from the diagonalization of the mass matrix; the mixing remains almost entirely dependent on the trilinear coupling $A_t$ which is insensitive to the sfermion soft masses. 

Changing $m_{10}/m_{16}$, which mainly affects $\mu$, will not have a strong tree-level effect on the stop masses since we are in the large $\tan\beta$ regime where the $\mu$-dependent terms in the off-diagonal elements of the stop mass matrix are heavily suppressed. However, there will be a small decrease in the overall masses as noted above (around $\sim 1-2\%$ for the points in Table~\ref{benchtbl}), due to large values of the $X_i$ in the RGEs. Again, the mixing angle is almost completely unaffected, being largely dependent on $A_t$ which remains relatively constant.

For the sbottoms things are considerably more complicated. First of all, we consider the {\em average} sbottom mass squared as compared to the CMSSM. From diagonalizing the tree-level sbottom mass matrix one obtains (ignoring EWSB $D$-terms and the bottom mass except where it is enhanced by $\tan\beta$)
\begin{equation}
\overline{m}_{\tilde{b}}^{2(D)} \simeq \frac{1}{2}\left(m_{\tilde{b}_L}^2 + m_ {\tilde{b}_R}^2 - 2g_{10}^2 D^2\right). \label{avsb}
\end{equation}
Here we have used a notation where $m_{\tilde{b}_L}^2 \equiv m_{\tilde{Q}_{33}}^{2(\textrm{CMSSM})}(M_Z)$, etc. to avoid notational clutter. Therefore the average sbottom mass is lowered in the case of positive $\mathcal{D}$ and raised in the event of negative $\mathcal{D}$. Due to the distribution of the $D$-terms in the sbottom mass matrix, the sbottom mass eigenstate that is mainly $\tilde{b}_R$ will vary a lot more than the one that is mainly $\tilde{b}_L$ if the L-R mixing is not too large. As far as the sbottom mass {\em splitting} is concerned one finds 
\begin{equation}
\Delta m_{\tilde{b}}^{2(D)} \simeq \sqrt{\left(m_{\tilde{b}_L}^2 - m_ {\tilde{b}_R}^2 + 4g_{10}^2 D^2 \right)^2 + 4|\mu|^2 m_b^2 \tan^2\beta}.\label{splsb}
\end{equation} 
First off, we cannot say in general which of the two terms under the square root is the larger and cannot simply approximate the square root by the binomial expansion. Any effect on the splitting induced by the $D$-term depends on whether $m_{\tilde{b}_L}^2 > m_{\tilde{b}_R}^2$ or vice-versa, and also on the magnitude of $m_{\tilde{b}_L}^2 - m_{\tilde{b}_R}^2$ relative to the magnitude of the $D$-term contribution. We identify two contrasting regimes:
\begin{itemize}
\item{Large $m_{1/2}$ and small $m_{16}$: we find $m_{\tilde{b}_L}^2$ to be larger than $m_{\tilde{b}_R}^2$ due to the effect of a large value of the gaugino mass $(M_2)$ term in the RGE for $m_{\tilde{b}_L}^2$ which is absent from the RGE for the electroweak singlet mass squared $m_{\tilde{b}_R}^2$. The fact that the difference between the left- and right-handed masses is due to, and grows with, $m_{1/2}$ in this regime, and that we assume that the $D$-term grows with $m_{16}$, means that the $D$-term is therefore relatively small compared to the mass splitting, even for $|\mathcal{D}| = 0.4$, and has a much reduced effect. Nevertheless, in this region of parameter space, a positive $D^2$ will increase the splitting between the masses and a negative $D^2$ will reduce the splitting.}
\item{Small $m_{1/2}$ and large $m_{16}$: we find the opposite is true --- $m_{\tilde{b}_L}^2$ is less than $m_{\tilde{b}_R}^2$ due to the large $X_t$ term in the $m_{\tilde{b}_L}^2$ RGE, a term which is again absent from the corresponding equation for $m_{\tilde{b}_R}^2$. Here, the $D$-term can be relatively large compared to the difference between the soft masses and can easily be the dominant part of the first term under the square root. In this case, at least for large $\mathcal{D}$, both signs of $D^2$ increase the mass splitting though negative $D^2$ will clearly have a stronger effect.} 
\end{itemize}
In certain regions of parameter space $D^2$ is such that it exactly cancels out the difference between the soft masses squared, leaving the minimum possible splitting
\begin{displaymath}
\Delta m_{\tilde{b}}^{(D)} \simeq 2|\mu| m_b \tan\beta. 
\end{displaymath}
In intermediate cases things depend on the exact numbers. As it turns out, around the point $m_{1/2} \simeq m_{16} \simeq 500$, $m_{\tilde{b}_L}^2 \simeq m_{\tilde{b}_R}^2$ and in this vicinity we cannot make a general connection between the sign of $\mathcal{D}$ and the magnitude of the first term in the square root. 
There is an additional effect that comes into play. As we mentioned earlier, decreasing $\mathcal{D}$ results in a decrease in the size of $\mu$, tending to decrease the contribution of the $\mu$-dependent term to $\Delta m_{\tilde{b}}^{(D)}$ though this a sub-dominant effect throughout most of the parameter space. Although the $D$-terms cause changes in the weak scale value of $\mu$ a stronger effect can be achieved by varying $m_{10}/m_{16}$. Indeed we expect significantly smaller splitting of the sbottom masses for large values of this parameter where $\mu$ is suppressed. One can see these effects in the Figs.~\ref{fig1} - \ref{lowhi} and the corresponding tables.

The sbottom mixing angle is likewise difficult to predict in the general case, also being sensitive to the difference between the two soft sbottom masses and to the off-diagonal $\mu$ dependent term.  

\subsubsection*{The Sleptons}
We begin by considering the sneutrinos and sleptons of the first two families. The left-handed slepton masses are given approximately by
\begin{equation}
m_{\tilde{\nu}_i}^{(D)} \simeq m_{\tilde{e}_{Li}}^{(D)} \simeq m_{\tilde{L}_i}\left(1 - \frac{3}{2}g_{10}^2 \frac{D^2}{m_{\tilde{L}_i}^2}\right),\label{sntaumass}
\end{equation}
whereas the right-handed slepton masses are given by
\begin{displaymath}
m_{\tilde{e}_{Ri}}^{(D)} \simeq m_{\tilde{e}_{Ri}}\left(1 + \frac{1}{2}g_{10}^2 \frac{D^2}{m_{\tilde{e}_{Ri}}^2}\right),
\end{displaymath}
where $i=1,2$ and again the soft masses on the RHS are the mSugra masses at the EWSB scale and $g_{10}$ is the unified gauge coupling evaluated at $M_G$. If $m_{1/2}$ is very small then $m_{\tilde{L}_i} \simeq m_{\tilde{e}_{Ri}}$ is of roughly equal magnitude to $m_{16}$ at the EWSB scale, at least for moderate to large values of $m_{16}$. In this case the right- and left-handed sleptons are nearly degenerate. Otherwise $m_{\tilde{L}_i}$ is substantially larger than both $m_{\tilde{e}_{Ri}}$ and $m_{16}$, radiatively driven by the wino mass $M_2$. The right-handed masses do not have a term proportional to $M_2$ in the RGE and are always smaller than their left-handed counterparts in the CMSSM although they may still be significantly bigger than $m_{16}$ if $m_{1/2}$ is large. Since the slepton masses are smaller than the squark masses as a result of not having an $M_3$ term in their RGEs, the relative effect of the $D$-terms is much larger and can be as much as $\sim 12\%$ for $\mathcal{D} = 0.4$ for the left-handed sleptons and $\sim 4\%$ for the right-handed sleptons in Table~\ref{benchtbl}. Consequently, the right-handed charged sleptons can be heavier than the left-handed charged sleptons even at moderate values of $m_{1/2}$. Again, these effects can be seen graphically in Figs.~\ref{fig1} - \ref{lowhi} and numerically in the tables.

Again, the third family is more complicated, but far more clear-cut than for the sbottoms. The $\tau$ Yukawa coupling plays an important r\^ole by balancing the $M_2$ term in the RGE for $m_{\tilde{\tau}_L}^2$ and making sure that $m_{\tilde{\tau}_R}$ is always pushed lower than $m_{16}$ at $M_Z$ and for low values of $m_{16}$ it is usually found that the stau is the LSP or is even tachyonic. The $\tau$ sneutrino mass is given by Eq.~\ref{sntaumass} with $m_{\tilde{L}_i}$ replaced by $m_{\tilde{\tau}_L}$. The average stau mass is given by Eq.~\ref{avsb} with the squark soft masses replaced by the slepton soft masses. Again, increasing the value of $D^2$ will lower the average mass, with the predominantly $\tilde{\tau}_L$ mass eigenvalue (assuming such a distinction exists) varying more than the mainly right-handed one. For the mass splitting, the situation is clearer:
\begin{equation}
\Delta m_{\tilde{\tau}}^{2(D)} \simeq \sqrt{\left(m_{\tilde{\tau}_L}^2 - m_ {\tilde{\tau}_R}^2 - 4g_{10}^2 D^2 \right)^2 + 4|\mu|^2 m_\tau^2 \tan^2\beta}\label{splstau}
\end{equation}
in the same notation as before. Note the change of sign of the $D$-term relative to Eq.~\ref{splsb}. Now, however, $m_{\tilde{\tau}_L}^2$ is {\em always} bigger than $m_{\tilde{\tau}_R}^2$ and we find that throughout the parameter space we explore that a positive value of $D^2$ always reduces the mass splitting (although it may swap which of the mass eigenstates is dominantly right- or left-handed) and a negative value will always increase the difference. 

For the 3rd family sleptons, although the $\mu$-dependent term in Eq.~\ref{splstau} grows as one decreases $m_{10}/m_{16}$, it is still very small relative to the difference in the soft masses squared due to the $\tau$ mass squared suppression (even taking into account its $\tan^2\beta$ enhancement). In fact, since $m_{\tilde{\tau}_R}^2$ is twice as dependent on the factor $X_\tau$ through the RGEs as $m_{\tilde{\tau}_L}^2$, and $X_\tau$ is substantially smaller for $m_{10}/m_{16} = 0.75$ than in the CMSSM case, the difference $m_{\tilde{\tau}_L}^2 - m_{\tilde{\tau}_R}^2$ decreases. As a result the stau mass difference decreases as $m_{10}/m_{16}$ decreases. The average masses, on the other hand, grow with decreasing $m_{10}/m_{16}$, also due to the smaller $X_\tau$ in the RGEs. Again the reader is referred to Figs.~\ref{fig1} - \ref{lowhi} and the Tables~\ref{benchtbl} and~\ref{Inttbl}. 

\subsection{The Charginos and Neutralinos} 
$D$-terms have a relatively small influence on the charginos and neutralinos. The soft gaugino masses $M_i$ are almost entirely unaffected, but there is some effect on $\mu$ which governs the masses of the Higgsino-like charginos and neutralinos. We note that we can make a distinction between Higgsino-like and gaugino-like charginos and neutralinos when we have the hierarchy $\mu^2 \gg M_2^2 \gg M_W^2$, where $M_W$ is the $W^{\pm}$ boson mass. This occurs in most of the parameter space that we examine and the mixing between gauginos and Higgsinos is relatively small. Very roughly, in the situation where this hierarchy arises, we have 
\begin{eqnarray}
m_{\tilde{\chi}_1^0} &\simeq& M_1 \nonumber \\
m_{\tilde{\chi}_2^0} \simeq m_{\tilde{\chi}_1^{\pm}} &\simeq& M_2 \nonumber \\
m_{\tilde{\chi}_3^0} \simeq m_{\tilde{\chi}_4^0} \simeq m_{\tilde{\chi}_2^{\pm}} &\simeq& \mu
\end{eqnarray}
An increase in $D^2$ will cause a corresponding increase in $\mu$ relative to $M_2$ via the mechanism detailed earlier in Section~\ref{EWSB}. As we mentioned before, this is a relatively small effect. Varying $m_{10}/m_{16}$, on the other hand, creates substantial changes in $\mu$ and, as a result, the Higgsino-like neutralinos and charginos can be significantly lighter or heavier than in the mSugra case. This is of vital importance for the calculation of the neutralino cosmological relic density where, as $\mu$ approaches $M_1$, the lightest neutralino may have a significant Higgsino component, enhancing the highly efficient Higgs-exchange annihilation channels. It is also important for any process with chargino and/or neutralino loops such as $(g-2)_\mu$, $\mathcal{B}(b \rightarrow X_s \gamma )$ and the SUSY correction to the bottom mass, $\delta m_b$, which we have already touched on briefly. We will discuss these constraints in a little more detail in later sections. 

Another matter that needs to be taken into account in order to discuss low energy observables is the dependence of the mixing angles in the gaugino sector on $\mu$ and the $M_i$. In the chargino sector, in the basis $(\tilde{W}^{\pm},\tilde{H}^{\pm})$, at tree level, the chargino mass matrix 
\begin{equation}
\mathcal{M}_{\tilde{\chi}^{\pm}} = \left( \begin{array}{cc}
M_2 & \sqrt{2}M_W\sin\beta\\
\sqrt{2}M_W\cos\beta & \mu \end{array} \right)
\end{equation}
is diagonalized by a bi-unitary transformation with diagonalization matrices $U$ and $V$, i.e. $\mathcal{M}_{\tilde{\chi}^{\pm}}^{\mathrm{diag}} = U^* \mathcal{M}_{\tilde{\chi}^{\pm}} V^{-1}$. In the case where all the parameters are real, which is an assumption we make here, $U$ and $V$ are both orthogonal and parameterized by angles $\theta_R$ and $\theta_L$:
\begin{equation}
\mathcal{M}_{\tilde{\chi}^{\pm}}^{\mathrm{diag}} = \left( \begin{array}{cc} \cos\theta_L & \sin\theta_L\\ -\sin\theta_L & \cos\theta_L \end{array} \right) \left( \begin{array}{cc} M_2 & \sqrt{2}M_W\sin\beta\\ \sqrt{2}M_W\cos\beta & \mu \end{array} \right) \left( \begin{array}{cc} \cos\theta_R & -\sin\theta_R\\ \sin\theta_R & \cos\theta_R \end{array} \right)\label{chgmm}
\end{equation}
 To order $M_W^2$ divided by powers of soft masses (assuming that the difference $\left(\mu^2 - M_2^2\right) \gg M_W^2$ which is the case in most of the parameter space we analyse), and in the large $\tan\beta$ regime where $\sin\beta \simeq 1$ and $\cos\beta \simeq 0$, it can easily be shown that 
\begin{eqnarray}
\cos\theta_L &\simeq& 1 - \frac{\mu^2 M_W^2}{\left(\mu^2 - M_2^2\right)^2}\nonumber\\
\sin\theta_L &\simeq&  - \frac{\sqrt{2}\mu M_W}{\mu^2 - M_2^2}\nonumber\\
\cos\theta_R &\simeq& 1 - \frac{M_2^2 M_W^2}{\left(\mu^2 - M_2^2\right)^2}\nonumber\\
\sin\theta_R &\simeq& - \frac{\sqrt{2}M_2 M_W}{\mu^2 - M_2^2}.
\end{eqnarray}  
 From a quick glance, we can deduce that in this regime the chargino mixing angles will grow as $\mu$ decreases as a result of increasing $m_{10}/m_{16}$ or decreasing $D^2$. When $\mu$ approaches $M_2$, as in the region close to where $\mu$ vanishes at the boundary of EWSB for large $m_{10}/m_{16}$ these approximations break down and this simple power series expansion is no longer applicable for the mixing angles.   

In the neutralino sector things are much more complicated since the mass matrix is $4\times 4$. In the basis $(\tilde{B}^0, \tilde{W}^0, \tilde{H}^0_1, \tilde{H}^0_2)$, at tree-level, 
\begin{displaymath}
\mathcal{M}_{\tilde{\chi}^{0}} = \left( \begin{array}{cccc}
M_1 & 0 & -M_Z\cos\beta \sin\theta_W & M_Z\sin\beta \sin\theta_W \\
0 & M_2 & M_Z\cos\beta \cos\theta_W & -M_Z\sin\beta \cos\theta_W \\
-M_Z\cos\beta \sin\theta_W & M_Z\cos\beta \cos\theta_W & 0 & -\mu \\
M_Z\sin\beta \sin\theta_W & -M_Z\sin\beta \cos\theta_W & -\mu & 0
 \end{array} \right),
\end{displaymath}
where $\tan\theta_W = \sqrt{\frac{3}{5}}\frac{g_1}{g_2}$ is the tangent of the Weinberg angle. In the approximation where all of the soft breaking parameters are real, this matrix can be diagonalized by an orthogonal matrix $O$, such that
\begin{displaymath}
\mathcal{M}_{\tilde{\chi}^{0}}^{\mathrm{diag}} = O^T \mathcal{M}_{\tilde{\chi}^{0}} O.
\end{displaymath}

In what follows, we will be interested only in the lightest neutralino, $\tilde{\chi}_1^0$. In the limit of large $\tan\beta$, with $\mu^2 > M_1^2 \gg M_Z^2$ and $\mu^2 - M_1^2 \gg M_Z^2$ (these relations again hold throughout the majority of the parameter space probed) an approximate solution can be found for the lightest mass eigenvalue and we can express roughly the components of the lightest neutralino, $m_{\tilde{\chi}^0_1} \simeq M_1$, in the basis $(\tilde{B}^0, \tilde{W}^0, \tilde{H}^0_1, \tilde{H}^0_2)$ as
\begin{equation}\label{neutevec}
\psi_{\tilde{\chi}^0_1} \propto \left(\begin{array}{c}
1\\
0\\
\frac{\mu M_Z \sin\theta_W}{\mu^2 - M_1^2}\\
-\frac{M_1 M_Z \sin\theta_W}{\mu^2 - M_1^2}.
\end{array}\right)
\end{equation}
Thus, by decreasing $\mu$ relative to $M_1$, from an increase in $m_{10}/m_{16}$ or a decrease in $D^2$, we obtain an increase in the Higgsino fraction of the lightest neutralino. As we shall see, this can have an enormous effect on the neutralino relic density.

The effects on the charginos and neutralinos of varying $D^2$ and $m_{10}/m_{16}$ can be seen in Figs.~\ref{fig1}-\ref{lowhi} and in the tables. 

%\subsection{$(g-2)_\mu$, $\mathcal{B}(b \rightarrow X_s \gamma)$ and $\delta m_b$}
\section{$\tan\beta$-enhanced Amplitudes}\label{tanb}
We now try to draw some general conclusions about how the above changes in the masses and mixing angles should affect important observables such as the muon anomalous magnetic moment and the branching ratio $\mathcal{B}(b \rightarrow X_s \gamma)$.

It is well known that the Feynman diagrams for the quantities $(g-2)_\mu$~\cite{Chattopadhyay:1996ae,Moroi:1996yh,Lopez:1994vi}, $\mathcal{B}(b\rightarrow X_s \gamma)$~\cite{Diaz:1993ww,Lopez:1994uz,Borzumati:1994zg,Oshimo:1993zd}, and the SUSY threshold correction to the bottom mass~\cite{Hall:1994gn,Hempfling:1994kv,Carena:1994bv} all contain, and in this case are dominated by, $\tan\beta$-enhanced contributions.  This comes about when the required chirality flip for these processes arises from mass insertions in the gaugino or sfermion line as opposed to a mass insertion on the external fermion line.  

\subsection{The Muon Anomalous Magnetic Moment}\label{g2m}
\begin{figure}[tbp]
\begin{center}
\includegraphics[width=0.6\textwidth]{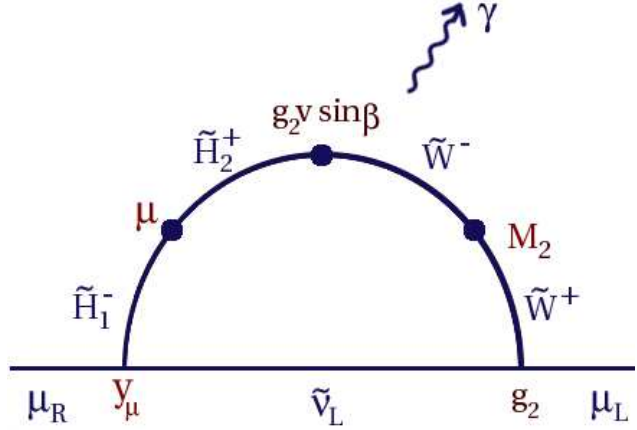}  
\end{center}
\caption{{\em The $\tan\beta$-enhanced mass insertion diagram for the chargino contribution to $(g-2)_\mu$. The various factors show the couplings at each vertex/insertion. From the factor $y_\mu v \sin\beta$ we deduce the diagram is proportional to $m_\mu \tan\beta$. Contributions for which the helicity flip occurs on the external muon line carry a relative suppression of $\tan\beta$.}\label{g2chg}}
\end{figure}

We will consider in detail $a_\mu \equiv \frac{(g-2)_\mu}{2}$. Throughout our parameter space, the dominant diagram is a chargino-sneutrino loop. In the absence of right-handed sneutrinos, the mass insertions can only occur in the gaugino line, with a $\mu_R$-$\tilde{\nu}_L$-$\tilde{H}$ vertex at one end and a $\tilde{\nu}_L$-$\mu_L$-$\tilde{W}$ vertex at the other --- see Fig.~\ref{g2chg} --- and the diagram is proportional to $\mu M_2 \tan\beta$.

Switching now to the mass eigenstate basis for the exact 1-loop calculation of the chargino part of the amplitude, the $\tan\beta$ enhanced diagrams come from two loops, one containing $\tilde{\chi}_1^\pm $ and one containing $\tilde{\chi}_2^\pm$, in each case connecting left- to right-handed muons so that the helicity flip occurs in the chargino line. The part of the amplitude containing the $\tilde{\chi}_1^\pm$ loop is proportional to a factor $\cos\theta_R \sin\theta_L$ and the part containing a $\tilde{\chi}_2^\pm$ loop is proportional to a factor $-\sin\theta_R \cos\theta_L$. Throughout our parameter space we find these factors to be opposite in sign and so there is some cancellation between the loop diagrams involving the different chargino mass eigenstates. The charginos contribute factors to the amplitude that can be written in the form (compare with the formulae in, for example,~\cite{Hisano:1996cp}):

\begin{eqnarray}\label{g2exact}
\delta a_{\mu}^{\tilde{\chi}_1^{\pm}} &=& -C \frac{m_{\tilde{\chi}_1^{\pm}}}{m_{\tilde{\nu}_\mu}^2}\cos\theta_R\sin\theta_L F\left(\frac{m_{\tilde{\chi}_1^{\pm}}^2}{m_{\tilde{\nu}_\mu}^2}\right)\\ 
\delta a_{\mu}^{\tilde{\chi}_2^{\pm}} &=& C \frac{m_{\tilde{\chi}_2^{\pm}}}{m_{\tilde{\nu}_\mu}^2}\sin\theta_R\cos\theta_L F\left(\frac{m_{\tilde{\chi}_2^{\pm}}^2}{m_{\tilde{\nu}_\mu}^2}\right).\label{g2exact2} 
\end{eqnarray}

In this equation, C is a positive constant with respect to varying $D$-terms or $m_{10}/m_{16}$ (at least up to negligible corrections), and involves $m_\mu$, $\beta$, the Higgs VEV $v$ and the gauge coupling $g_2$. The function $F$ is a phase space integral dependent on the relevant mass-squared ratio:
\begin{equation}\label{Feq}
F(x) =   \frac{-3 + 4x - x^2 - 2\ln{x}}{(1-x)^3}.
\end{equation}

We can project out the $\tan\beta$-enhanced part of the amplitude by setting $\sin\beta \simeq 1$ and $\cos\beta \simeq 0$.  The leading-order terms in the soft SUSY-breaking mass parameters are given by:

\begin{eqnarray}
\delta a_{\mu}^{\tilde{\chi}_1^{\pm}} &\simeq& C \frac{M_2}{m_{\tilde{L}_\mu}^2}\frac{\sqrt{2}\mu M_W}{\mu^2 - M_2^2} F\left(\frac{M_2^2}{m_{\tilde{L}_\mu}^2}\right)\\ \nonumber
\delta a_{\mu}^{\tilde{\chi}_2^{\pm}} &\simeq& -C \frac{\mu}{m_{\tilde{L}_\mu}^2} \frac{\sqrt{2}M_2 M_W}{\mu^2 - M_2^2} F\left(\frac{\mu^2}{m_{\tilde{L}_\mu}^2}\right),
\end{eqnarray}

where $m_{\tilde{L}_\mu}^2 \equiv m_{\tilde{L}_{22}}^2$. And so

\begin{equation}\label{amu}
\delta a_\mu \simeq \delta a_{\mu}^{\tilde{\chi}^{\pm}} = C^\prime \frac{M_2\mu}{m_{\tilde{L}_\mu}^2\left(\mu^2-M_2^2\right)}\left(F\left(\frac{M_2^2}{m_{\tilde{L}_\mu}^2}\right) - F\left(\frac{\mu^2}{m_{\tilde{L}_\mu}^2}\right)\right).
\end{equation}

Here, $C^\prime$ is a another positive constant:
\begin{displaymath}
 C^\prime \simeq \frac{g_2^2 m_\mu^2 \tan\beta}{16 \pi^2} \simeq 1.44 \times 10^{-3}.
\end{displaymath}

\begin{figure}[tbp]
\begin{center}
\includegraphics[width=0.7\textwidth]{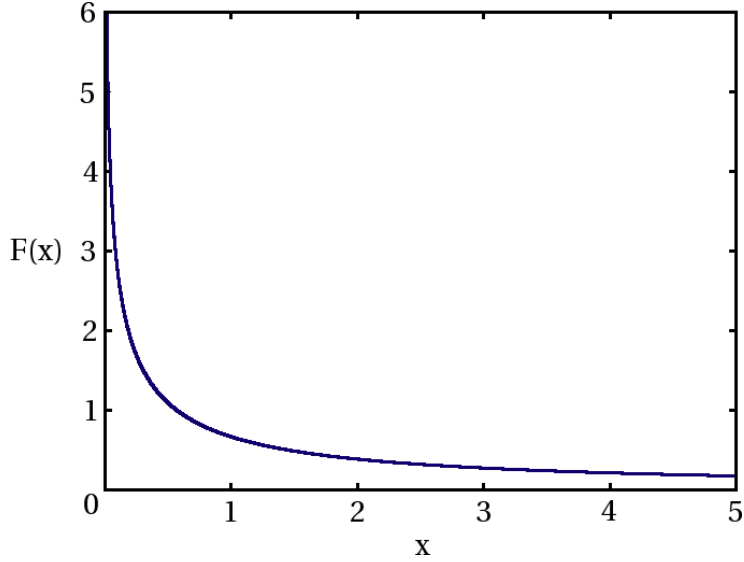}  
\end{center}
\caption{{\em The loop function F(x), Eq.~\ref{Feq}.}\label{Fx}}
\end{figure}

Unfortunately, from the point of view of getting an intuitive idea of how the $D$-terms and sfermion-Higgs splitting will affect $a_\mu$, it doesn't make much sense to series expand the function $F$ in terms of the mass ratios since they are frequently close to the critical value 1. However, $F(x)$ is a monotonically decreasing function of $x$, and is shown in Fig.~\ref{Fx}.  Now we consider what happens when we increase the value of $\mathcal{D}$ from 0 to 0.4 for the point $m_{16} = m_{1/2} = 500$  GeV. Two main effects occur. The first is that value of $m_{\tilde{L}_\mu}$ decreases (e.g. by $\sim 55$GeV (around 10\%) for point E in table~\ref{benchtbl} as compared with point A). The second is that $\mu$ increases. The change is less significant for $\mu$ than for $m_{\tilde{L}_\mu}$ since $\mu$ is only affected at loop order through RGE effects (e.g. by around $30$ GeV (around 5\%) for point E compared with A). $M_2$ remains basically unchanged. Consequently, the factor outside the large parentheses of Eq.~\ref{amu}, $M_2\mu/(m_{\tilde{L}_\mu}^2(\mu^2 - M_2^2))$, grows as $\mathcal{D}$ increases. However, this is countered somewhat by what happens inside the parentheses. By increasing $D^2$ the mass ratios increase and one moves along the x-axis of Fig.~\ref{Fx} to the right causing a general decrease in the magnitude of the function in Eq.~\ref{Feq}. One must be careful though because $\mu$ will increase as a result of increasing $D^2$, resulting in an increase in $\mu^2/m_{\tilde{L}_\mu}^2$ relative to $M_2^2/m_{\tilde{L}_\mu}^2$ therefore tending to create a larger difference between the two $F(x)$ terms and cancelling out the effect of the smaller overall magnitude. Comparing the points E and A, the overall decrease narrowly wins and the difference between the loop functions $F(x)$ is larger in A than in E. The external factor $M_2\mu/(m_{\tilde{L}_\mu}^2(\mu^2 - M_2^2))$, on the other hand, is larger in E than in A and overall the contribution to $a_\mu$ is greater for $\mathcal{D} = 0.4$ than for $\mathcal{D} = 0$. However, when one looks at point D for which $\mathcal{D} = -0.4$, one can see that the cancellation between the loop function terms dominates the larger overall magnitude of each contribution and the difference between the $F(x)$ for point D is also smaller than for point A (note, though, that our approximation is somewhat worse for point D than for points A and E due to the fact that $\mu$ is much closer in magnitude to $M_2$). Table~\ref{Ftbl} shows the corresponding values for our approximation as compared to the calculated figures.

We note here than in different parts of parameter space things change somewhat. For example, for $m_{16} \gg m_{1/2}$, at least for $m_{10}/m_{16} = 1$, increasing $D^2$ has a stronger effect on $\mu$ than it does on $m_{\tilde{L}_\mu}^2$ and the above situation is reversed; although this time the loop function for $\mathcal{D} = 0.4$ is much larger than that for $\mathcal{D} = 0$, the external factor $M_2\mu/m_{\tilde{L}_\mu}^2(\mu^2-M_2^2)$ is much smaller and the amplitude for $\mathcal{D} = 0.4$ is smaller than that for $\mathcal{D} = 0$. Note that in this region of parameter space, $\mathcal{D} = -0.4$ is excluded by the EWSB requirements and in any case, the above approximation would break down. 

To summarize: in general, it is somewhat complicated to understand intuitively how the $D$-terms affect $a_\mu$, even though it is a relatively simple calculation as compared with, for example, $\mathcal{B}(b\rightarrow X_s \gamma)$. It all depends on what part of the parameter space you are in, where on the curve in Fig.~\ref{Fx} you are and also on the relative effects of the $D$-terms on $\mu$ with respect to $m_{\tilde{L}_\mu}^2$. 

Increasing $m_{10}/m_{16}$ has a more straightforward effect although it is still not immediately obvious what will happen from our approximation Eq.~\ref{amu}. Since an increase in this variable will tend to decrease $\mu$ while leaving the sneutrino mass and $M_2$ relatively unchanged, from the above discussion it is clear that the difference between the loop functions will become smaller whereas the factor $M_2\mu/m_{\tilde{L}_\mu}^2(\mu^2 - M_2^2)$ will grow. We find that throughout the parameter space, the second effect is greater and that increasing $m_{10}/m_{16}$ causes $a_\mu$ to grow. 

\subsection{$\mathcal{B}(b \rightarrow X_s \gamma)$}\label{bsg}
A similar discussion on $\mathcal{B}(b \rightarrow X_s \gamma)$ to that above would be rather more involved and not any more illuminating. We will limit ourselves to making a few general remarks.

In the Standard Model, the calculated branching ratio $\mathcal{B}(b\rightarrow X_s \gamma)$ accounts very well for the experimentally observed value. In the MSSM one must also include contributions from the charged Higgs diagram, which always has the same sign as the Standard Model $W^{\pm}$ loop, and the chargino diagram. Either these additional contributions cannot be too large, or must cancel. In the positive $\mu$ regime, the chargino amplitude has the opposite sign to the charged Higgs, allowing this cancellation and helping to keep the amplitude within the experimental bounds. However, in the CMSSM often the charged Higgs mass is large compared to the chargino masses at low values of $m_0$ and $m_{1/2}$ and the charged Higgs contribution is not sufficient to compete with the chargino contribution. As a consequence the calculated value for $\mathcal{B}(b\rightarrow X_s \gamma)$ undershoots the experimental result and excludes a significant region of the parameter space. 

The $\tan\beta$-enhanced part of the chargino contribution to $\mathcal{B}(b \rightarrow X_s \gamma)$ is a more complicated expression than the one for the $(g-2)_\mu$ chargino contribution for two reasons. Firstly, the approach to the calculation of $\mathcal{B}(b \rightarrow X_s \gamma)$ is different --- the strength of the QCD interaction necessitates operator product methods instead of straightforward one-loop renormalized perturbation theory and one has to calculate the SUSY and charged Higgs contributions to two different Wilson coefficients. Secondly, one also has to consider the fact that there are two different stop mass eigenstates propagating in the loop as opposed to the solitary sneutrino state, both of which contribute to $\tan\beta$-enhanced diagrams.  

When comparing the $D$-term- and $m_{10}/m_{16}$-corrected case to the standard mSugra scenario, an important factor in the discussion is the comparative sizes of the charged Higgs and Higgsino-like chargino masses, i.e. $m_{A^0}$ and $\mu$ respectively. It turns out that in the $b\rightarrow X_s \gamma$ case, unlike in the case for $(g-2)_\mu$, we find that the chargino amplitude always grows as $\mu$ becomes smaller, whether due to $D$-terms or $m_{10}/m_{16}$. Likewise, the charged Higgs contribution always increases with decreasing $m_{H^{\pm}}$ (or equivalently $m_{A^0}$). However, the question is not only to do with which of $m_{A^0}$ or $\mu$ changes most. Another factor is the overall size of the amplitudes. Even if the charged Higgs contribution becomes smaller relative to the chargino contribution and is less effective in cancelling it, it could be that at the same time the chargino contribution becomes too small to ruin the Standard Model prediction. In any case, it is the sum of the two amplitudes compared to the $W^{\pm}$ loop that is important rather than their relative changes in magnitude.

If, for example, we decrease $\mathcal{D}$ we cause a decrease in $\mu$, but also a decrease in $m_{H^{\pm}}$. This effect is relatively larger for $m_{H^{\pm}}$ than for $\mu$ over much of the parameter space (except for $m_{1/2} \ll m_{16}$) and so one may expect the Higgs term to cancel the excessively large chargino contribution more successfully. However, as mentioned above, the relative size of the changes is not the only important factor. The absolute size of the amplitudes matters too. It turns out that in the region $m_{1/2} \simeq m_{16} \simeq 500$ GeV, although the Higgs contribution grows {\em relative} to the chargino contribution with decreasing $D^2$, they both increase by a roughly equal amount in {\em absolute} magnitude and their sum remains about the same. One can see that the $\mathcal{B}(b\rightarrow X_s \gamma)$ values quoted for points A, D and E in Table~\ref{benchtbl} are almost equal. 

For the region $m_{1/2} \ll m_{16}$, increasing $\mathcal{D}$ from 0 to 0.4 ($\mathcal{D} = -0.4$ is ruled out by EWSB restrictions) affects $\mu$ just about as much as $m_{H^{\pm}}$ and by quite a substantial amount. Here, the overall size of the SUSY contribution decreases; in addition, the charged Higgs contribution, although it decreases, cancels out the chargino contribution more successfully. As we shall see, this results in the excluded region becoming smaller as $D^2$ increases.  
 
A particularly interesting case is when $D^2$ is large and negative, close to the boundary where $m_{A^0}^2$ vanishes. Here, $m_{H^{\pm}}$ is also very small although $\mu$ can remain several hundred GeV. In this case, one finds a narrow sliver of acceptable parameter space where the charged Higgs contribution is large enough to counter the chargino amplitude even though the chargino amplitude is very large. We also found regions where the Higgs contribution is so large that the upper bound on the branching fraction is violated.

\subsection{Bottom Mass Corrections: $\delta m_b$}
Sparticle loop diagrams result in a finite threshold correction to the bottom mass. The most important contributions come from $\tan\beta$-enhanced terms which alter the relation between the (running $\overline{DR}$) bottom mass and the bottom Yukawa coupling. The corrected expression can be approximated by (see, for example~\cite{Hall:1994gn})
\begin{equation}
m_b \simeq \frac{y_b v \cos\beta}{\sqrt{2}}\left(1 + \delta m_b^{\tilde{b}\tilde{g}} + \delta{m_b^{\tilde{t}\tilde{\chi}^{\pm}}}\right),
\end{equation}
where
\begin{equation}
\delta{m_b^{\tilde{b}\tilde{g}}} \simeq \frac{2\alpha_s m_{\tilde{g}}\mu \tan\beta}{3\pi} I\left(m_{\tilde{b}_1}^2, m_{\tilde{b}_2}^2, m_{\tilde{g}}^2\right),
\end{equation}
and
\begin{equation}
\delta{m_b^{\tilde{t}\tilde{\chi}^{\pm}}} \simeq \frac{y_t^2 \mu A_t \tan\beta}{16\pi^2} I\left(m_{\tilde{t}_1}^2, m_{\tilde{t}_2}^2, \mu^2\right),
\end{equation}
with
\begin{equation}
I(x,y,z) = -\frac{xy \ln{\frac{x}{y}} + yz \ln{\frac{y}{z}} + zx\ln{\frac{z}{x}}}{(x-y)(y-z)(z-x)}.
\end{equation}
In magnitude the gluino loop tends to dominate the chargino loop due to the largeness of $\alpha_s m_{\tilde{g}}$. Since $A_t$ is always negative in the cases we consider, the chargino loop does partly cancel the gluino loop, but overall the sparticle loops contribute with a positive sign in the case of positive $\mu$ ($I(x,y,z)$ is positive). This means that, in order to account for the experimentally observed value of the bottom mass, the bottom Yukawa coupling $y_b$ must decrease. In the CMSSM case, this decrease in $y_b$ ruins the prospect of bottom-tau Yukawa unification. With variations in the $D$-terms and/or $m_{10}/m_{16}$ the squark masses change somewhat (as detailed in Section~\ref{squarks}), but $\mu$ changes more. The gluino mass $m_{\tilde{g}}$ is almost invariant. Both of these contributions grow with $\mu$ times a loop function and as $\mu$ increases the overall contribution will tend to increase. Additionally, in most of the CMSSM parameter space, there exists a hierarchy where the squark masses are heavier than $\mu$. In this case the function $I\left(m_{\tilde{b}_1}^2,m_{\tilde{b}_2}^2,\mu^2\right)$ decreases as $\mu$ increases, helping to suppress the chargino contribution relative to the gluino contribution and thus enhancing the correction to $m_b$ and resulting in an even smaller $y_b$. The best chance of obtaining bottom-tau Yukawa unification is therefore with $m_{10}/m_{16} = 1.25$ and $\mathcal{D} = -0.4$, but even in this case, throughout the $(m_{1/2},m_{16})$-plane we are still far from unification --- off by around $25\%$ at best. Other authors~\cite{Blazek:2002ta,Auto:2003ys,Auto:2004km} have found that, even allowing the $D$-terms and $m_{10}/m_{16}$ to vary much more than we have, it is impossible to reconcile Yukawa unification with the positive $\mu$ preferred by $(g-2)_\mu$ and $\mathcal{B}(b\rightarrow X_s \gamma)$. They have found, however, that if one splits the Higgs masses at the GUT scale while leaving the $D$-terms zero, such unification is possible. One group~\cite{Auto:2004km} find that even then there are problems fulfilling the neutralino relic density constraint due to the large masses involved in the favoured region of parameter space. In the same paper they discuss possible solutions to this problem. However, it may be due to the limitations of the bottom-up procedure used in their (and indeed our) analysis which often fails to converge close to the region of incorrect EWSB. Another group~\cite{Blazek:2002ta,Dermisek:2003vn}, using a top-down algorithm, find no such problems and have identified regions of this parameter space consistent with both Yukawa unification and an acceptable neutralino relic density.

\section{Calculation and Constraints\label{method}}
We use \texttt{SOFTSUSY} v.1.8.7~\cite{Allanach:2001kg}, one of several publicly available codes, to calculate the sparticle spectrum and mixings. The code has been augmented to include our ${\textrm{SO}}(10)$-inspired boundary conditions and a routine for the calculation of the SUSY contribution to the muon anomalous magnetic moment using the formulae in~\cite{Hisano:1996cp}. \texttt{SOFTSUSY} uses a bottom-up routine in which various low energy observables such as $M_Z$, fermion masses and gauge couplings are input as constraints in addition to the GUT scale boundary conditions. An iterative algorithm proceeds from an initial guess to find a set of sparticle masses and mixings consistent with the high and low scale constraints. We use full 2-loop renormalization group equations for the gauge and Yukawa couplings and the $\mu$ parameter. For the soft masses we use the full 1-loop RGEs and include the  2-loop contributions in the 3rd family approximation. Full details can be found in~\cite{Allanach:2001kg}. A comparison between \texttt{SOFTSUSY} and similar programs, for example~\cite{Porod:2003um,Djouadi:2002ze,Paige:2003mg} was made in~\cite{Allanach:2003jw} and one can directly compare the codes online at~\cite{Kramlweb:2004}. 

For the calculation of the neutralino relic density and $\mathcal{B}(b\rightarrow X_{s}\gamma)$ we use \texttt{micrOMEGAs} v.1.3.1~\cite{Belanger:2004yn}, linked to \texttt{SOFTSUSY} via an interface conforming with the Les Houches Accord~\cite{Skands:2003cj} standard that contains all the relevant parameters from \texttt{SOFTSUSY} necessary for the relic density calculation. For details of these calculations, see~\cite{Belanger:2004yn} and the papers on which they were based~\cite{Gondolo:1991dk,Edsjo:1997bg,Bertolini:1991if,Kagan:1998ym,Gambino:2001ew,Degrassi:2000qf,Chetyrkin:1997vx,Ciuchini:1998xe,Ciuchini:1998xy,Carena:1999py}. 

In our analysis we impose the following constraints:
\begin{itemize}
\item {\textbf{Direct searches}\newline
The following lower limits from LEP provide the strongest constraints on sparticle masses from direct searches~\cite{Eidelman:2004wy}: 
\begin{equation*} 
m_{\tilde{\chi}^{\pm}} \ge 103 \mathrm{GeV} \qquad m_{\tilde{e}_R} \ge 99 \mathrm{GeV}.
\end{equation*}
We include these lower bounds in our plots.}

\item {\textbf{Muon anomalous magnetic moment} \newline 
We include the $2\sigma$ bounds on the discrepancy between experiment and Standard Model theory assuming the latest results of the calculation based on $e^+ e^-$ data for the hadronic contribution~\cite{Akhmetshin:2003zn} and the most recent data from the BNL E821 experiment incorporating the results from negative muons~\cite{Bennett:2004pv}. We use the values from~\cite{Hagiwara:2003da} which include the recently recalculated $\alpha^4$ QED correction~\cite{Kinoshita:2004wi} and the most recent hadronic light-by-light contribution~\cite{Melnikov:2003xd}. Similar values were obtained by an independent calculation~\cite{Davier:2003pw}. However this second paper does not take the new theoretical results~\cite{Kinoshita:2004wi,Melnikov:2003xd} into account. From~\cite{Hagiwara:2003da}, 
\begin{equation*}
a_{\mu}^{{\textrm{exp}}} - a_{\mu}^{{\textrm{SM}}} = (24.5\pm9.0)\times 10^{-10},
\end{equation*}
where $a_{\mu} \equiv \frac{(g-2)_{\mu}}{2}$. We use the $2\sigma$ bound,
\begin{equation*}
6.5\times 10^{-10} < \delta a_{\mu} < 42.5\times 10^{-10},
\end{equation*}
as the allowed range of the SUSY contribution. Due to the inconsistency between these results and those obtained by using $\tau$ decay data, and taking into account the susceptibility to change of the measurement of the $e^+ e^-$ cross section~\cite{Akhmetshin:2003zn}, the $(g-2)_{\mu}$ constraint should perhaps be viewed more provisionally than the others. This is unfortunate since it is one of the most important, being the only one that unambiguously determines the sign of $\mu$.}

\item {{\textbf{Branching Ratio}} $\mathcal{B}(b\rightarrow X_{s}\gamma )$ \newline
The most recent world average for the branching ratio is~\cite{Jessop:2002ha}
\begin{equation*}
\mathcal{B}(b\rightarrow X_{s}\gamma )_{{\textrm{exp}}}=(3.34\pm 0.38)\times 10^{-4},
\end{equation*}
while the current Standard Model theory value is~\cite{Bieri:2003jm}\footnote{This takes into account only those results that include the improved ratio $m_c^{\overline{{\textrm{MS}}}}(m_b/2)/m_b^{{\textrm{pole}}}$ as opposed to $m_c^{{\textrm{pole}}}/m_b^{{\textrm{pole}}}$ in the $\langle X_s \gamma |(\bar{s}c)_{V-A}(\bar{c}b)_{V-A}|b\rangle$ matrix element. For details see~\cite{Gambino:2001ew}.} 
\begin{equation*}
\mathcal{B}(b\rightarrow X_{s}\gamma )_{{\textrm{SM}}}=(3.70\pm 0.30)\times 10^{-4}.
\end{equation*}
We will use this Standard Model estimate of the theoretical error in our calculation as representative of the error to be expected in our calculation which includes both Standard Model and SUSY contributions. We do this by combining the experimental and theoretical errors in quadrature to obtain the following upper and lower bounds on the branching ratio at $2\sigma$:
\begin{equation*}
2.40\times 10^{-4}<\mathcal{B}(b\rightarrow X_{s}\gamma )<4.28\times 10^{-4}.
\end{equation*}
}

\item {\textbf{Neutralino dark matter} \newline
The analysis of the data from WMAP gives a best fit value for the matter density of the universe of $\Omega_{m}h^2 = 0.135^{+0.008}_{-0.009}$ and for the baryon density, $\Omega_{b}h^2 = 0.0224\pm 0.0009$~\cite{Spergel:2003cb}. This implies that the CDM density is
\begin{equation*}
\Omega_{{\textrm{CDM}}}h^2 = 0.1126^{+0.0161}_{-0.0181}
\end{equation*}
at the $2\sigma$ level. This can be an extremely stringent bound on the MSSM parameter space, especially in the case of small $\tan\beta$. However, for large $\tan\beta$ it is less restrictive due to the presence of the $A^0$ Higgs resonance, and much less so if we allow for a source of cold dark matter other than neutralinos such as axions, or some relic density enhancement mechanism such as non-thermal production of neutralinos (see~\cite{Profumo:2004at} and references therein for more examples). In these instances, the lower bound on $\Omega_{m}h^2$ can be neglected. We plot values for which 
\begin{equation*}
0.0945 < \Omega_{{\textrm{CDM}}}h^2 < 0.1287,
\end{equation*}
and indicate the allowed regions if we choose to discard the lower bound. We also plot the locus of points for which $m_{A^0} = 2m_{\tilde{\chi}_1^0}$ marking the position of the $A^0$ resonance.}

\item {\textbf{Lightest Higgs Mass} $\mathbf{m_{h^0}}$ \newline
Where appropriate, we also display the contour 
\begin{equation*}
m_{h^0} = 114.1 \mathrm{GeV},
\end{equation*}
corresponding to the LEP bound on the lightest SM Higgs boson~\cite{Eidelman:2004wy} in the regions of parameter space where the lightest MSSM Higgs boson is Standard Model like, i.e. \mbox{$\sin(\beta - \alpha)$} is almost exactly equal to 1, where $\alpha$ is the mixing angle relating the mass eigenstates to the gauge eigenstates in the CP-even neutral Higgs sector. As we shall see in the following analysis, there are regions of parameter space where the mass of the heavy CP-even Higgs boson $m_{H^0}$ becomes comparable in mass to $m_{h^0}$ and $\sin(\beta - \alpha)$ is not close to unity. In these regions the LEP bound on the lightest Higgs does not apply and we omit the contour in our plots.}

\item {\textbf{Correct EWSB / Tachyons}\newline
The boundary on which $|\mu|^2$ vanishes, marking the border of correct radiative electroweak symmetry breaking has been plotted. In the region where $|\mu|^2 < 0$ a global minimum of the two loop effective Higgs potential cannot be found. Similarly, any regions in which $m_{A^0}^2 < 0$, also signalling that the electroweak symmetry has not been broken correctly, have been excluded. Regions with tachyonic sfermions are likewise omitted.}  
\end{itemize}

\section{The Constrained Parameter Space\label{results}}
In all of the following plots we take ${\mathrm {sign}}(\mu) > 0$, necessary to obtain the observed sign of $(g-2)_\mu$, and we take $A_0 = 0$ for simplicity, choosing to concentrate on the effects of varying $D^2$ and $m_{10}/m_{16}$. Note that we do not demand Yukawa unification as an additional constraint. For $\mu > 0$ we were unable to find any regions of parameter space consistent with Yukawa unification. Although in principle $m_{10}/m_{16}$ and $D^2$ can take on relatively large values we choose to explore the effects of fairly small deviations from universality. As a result there remains a close connection between the $(m_{1/2},m_0)$-plane of the CMSSM and the $(m_{1/2},m_{16})$-plane in the following analysis, i.e. for the most part, the sparticle spectrum will be similar in each case for corresponding points in the two planes, with possible important exceptions in the Higgs and chargino/neutralino sectors. Therefore we can still make a meaningful comparison between our ${\textrm{SO}}(10)$ scenario and the CMSSM. Our initial scan of the parameter space revealed that the most important effects only appear for large values of $\tan\beta$ and so we set $\tan\beta = 50$ throughout. For this case, we will show that even small deviations from the CMSSM can lead to large changes in the topography of the allowed regions in the $(m_{1/2},m_{16})$-plane.

\begin{figure}[tbp]
\begin{center}
\includegraphics[width=0.6\textwidth]{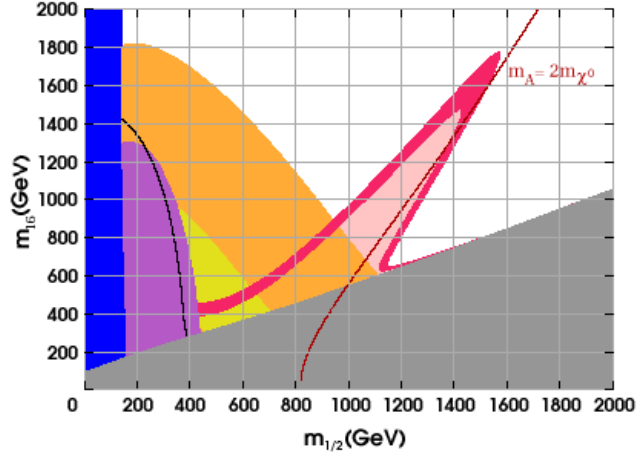}
\end{center}
\caption{{\em This plot shows contours in the $(m_{1/2},m_{16})$-plane for a variety of constraints. In this and subsequent plots we have $A_0 = 0$, $\mu > 0$ and $\tan\beta = 50$. In this figure we show the CMSSM parameter space: $m_{10}/m_{16} = 1$ and $\mathcal{D} = 0$. The blue (dark grey) strip and to the left of the black line at low $m_{1/2}$ are excluded by the LEP bounds on $m_{\tilde{\chi}^{\pm}}$ and $m_{h^0}$ respectively; in the grey triangular region at small $m_{16}$ the LSP is the $\tilde{\tau}$; the purple (medium grey) region extending out to $m_{1/2} \sim 500$ GeV is ruled out by $\mathcal{B}(b \rightarrow X_s \gamma)$; the orange (light grey) and yellow (very light grey) bands are the $(g-2)_{\mu}$ $1\sigma$ and $2\sigma$ favoured regions; the narrow crimson (darkish grey) curve satisfies the WMAP bounds; the dark red (dark grey) labelled line shows the exact position of the $A^0$-resonance; the light pink (very, very light grey) region is allowed by WMAP if there exists another source of CDM; finally, any white regions are ruled out by the WMAP upper bound.}}\label{D0M1}
\end{figure}

Fig.~\ref{D0M1} 
shows the $(m_{1/2},m_{16})$-plane in the standard CMSSM case. $D$-terms are set to zero and $m_{10} = m_{16}$. One can see the usual features: the region where $m_{1/2} \lesssim 150$ GeV is ruled out for any value of $m_{16}$ due to the LEP mass bound on the lightest chargino; a large triangular area where $m_{16} \simeq 0.5 m_{1/2}$ ruled out because the LSP is a stau; the quarter-egg-shaped region at small $m_{1/2}$ excluded by the lower bound on $\mathcal{B}(b\rightarrow X_s \gamma)$; the LEP lower bound on the lightest Higgs boson mass, valid for a Standard Model-like Higgs; the arcs representing the $1\sigma$ and $2\sigma$ favoured regions for the muon anomalous magnetic moment at small to moderate $m_{1/2}$ and $m_{16}$; the $A$-resonance around the region where $2m_{\tilde{\chi}^0_1} = m_{A^0}$ and rapid annihilation can occur via an S-channel $A^0$ (or, sub-dominantly, $H^0$) satisfying the upper bound on $\Omega_{{\textrm{CDM}}}h^2$; finally, there is the co-annihilation tail along the boundary demarking $m_{\tilde{\chi}_1^0} = m_{\tilde{\tau}_1}$. As usual, the region allowed by all the constraints, if we take them all seriously, remains a narrow strip at fairly low $m_{1/2}$ and $m_{16}$, mainly dictated by the highly stringent WMAP bounds on $\Omega_{{\textrm{CDM}}}h^2$. 

\begin{figure}[tbp]
\begin{center}
\includegraphics[width=0.6\textwidth]{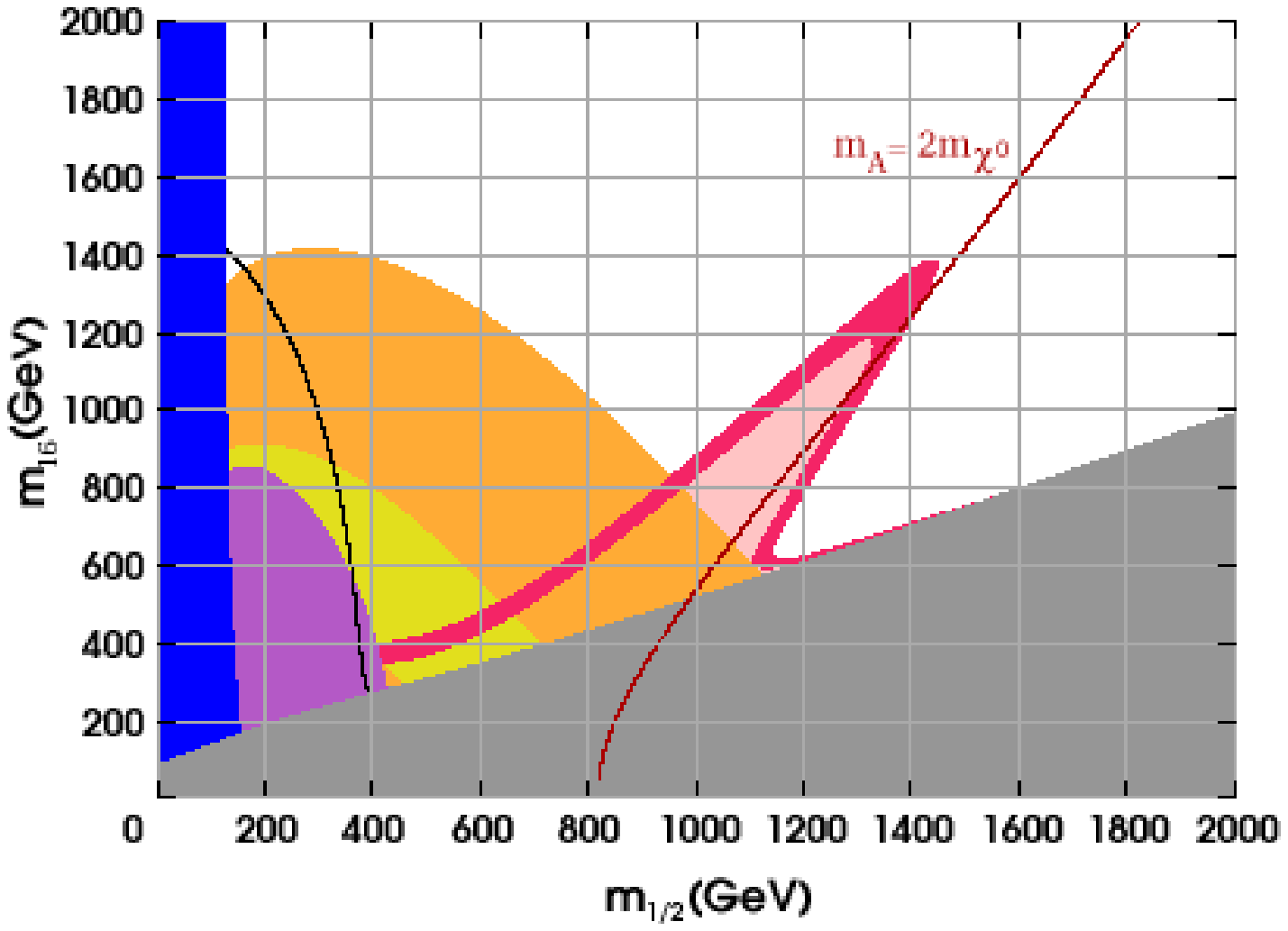}(a)
\includegraphics[width=0.6\textwidth]{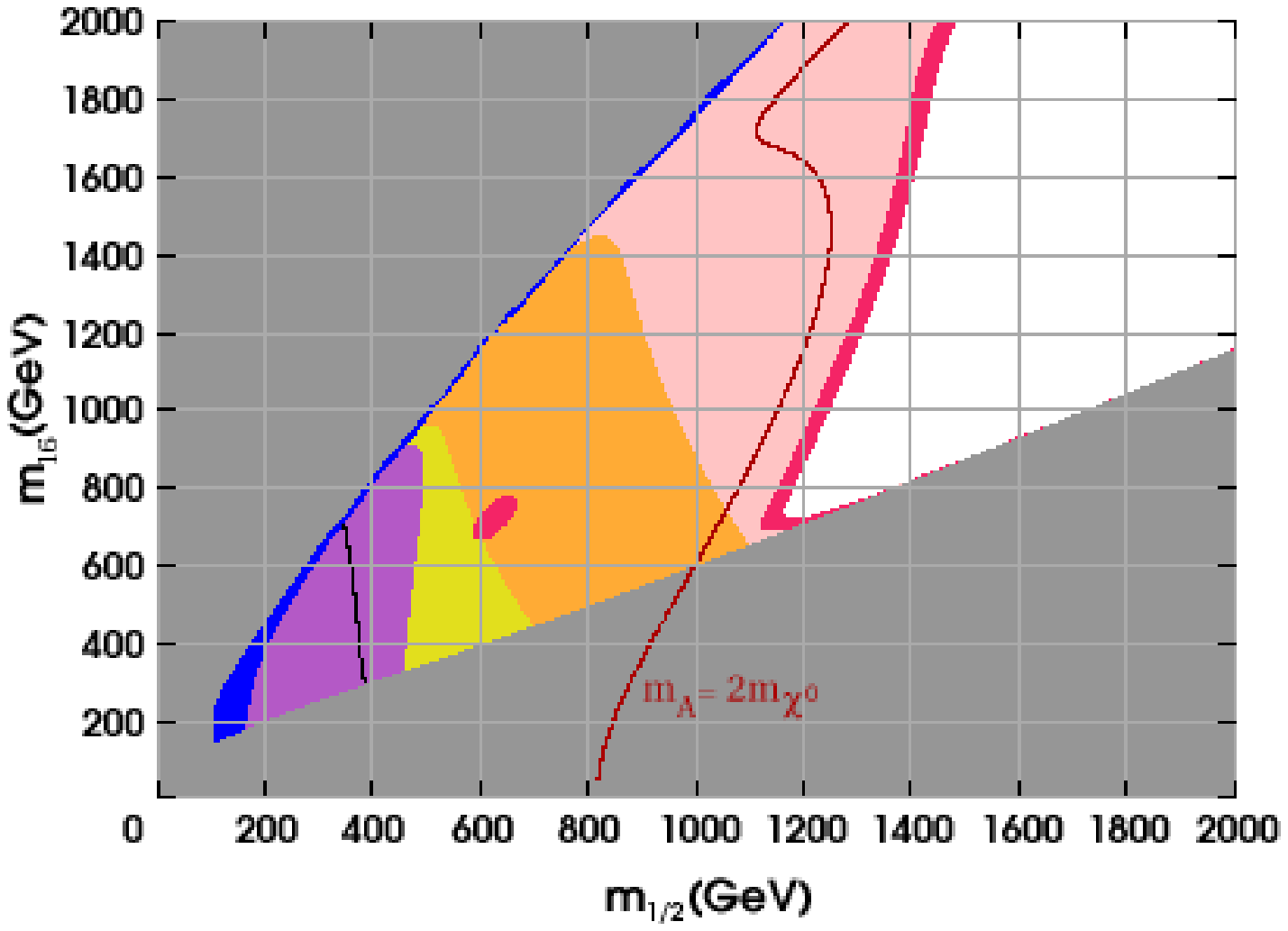}(b) 
\end{center}
\caption{{\em Same as Fig.~\ref{D0M1}, but showing the effects of varying $m_{10}/m_{16}$ only; $\mathcal{D} = 0$. In {\rm (a)} $m_{10}/m_{16} = 0.75$ and in {\rm (b)} $m_{10}/m_{16} = 1.25$. The new grey triangular region at small $m_{1/2}, $large $m_{16}$ in {\rm (b)} is ruled out by $|\mu|^2 < 0$ indicating that for this region there is no solution to EWSB.}}\label{DOMvar}
\end{figure}

Next, in Fig.~\ref{DOMvar} 
we look at the same plot, this time varying $m_{10}/m_{16}$ while keeping $\mathcal{D} = 0$. Fig.~\ref{DOMvar}(a) shows $m_{10}/m_{16} = 0.75$. First of all, due to the resulting increase in $\mu$, the \mbox{$\mathcal{B}(b\rightarrow X_s \gamma)$} exclusion region has shrunk, especially for $m_{16} \gg m_{1/2}$, as would be expected from our earlier analysis in Section~\ref{bsg}. This is because the chargino amplitude no longer over-cancels the Standard Model amplitude. The $(g-2)_\mu$ preferred regions follow a similar pattern (see Section~\ref{g2m} for details). Since $M_2$ is more or less unaltered and in this case dominates the mass of the lightest chargino, the region excluded by the LEP bound on the $\tilde{\chi}_{1}^{\pm}$ is almost unchanged. The stau LSP triangle is a little less restrictive since the $X_i$ factors in the soft mass RGEs (see Section~\ref{spectrum}) are smaller and the lighter right-handed stau is slightly heavier as a result. Due to overall heavier $A^0$ masses, the contours of equal $m_{A^0}$ move to the left and as a result of where they overlap with the contours of equal $m_{\chi_1^0}$ along the line $m_{A^0} = 2m_{\chi_1^0}$, the $A^0$ resonance rapid annihilation funnel is at a flatter angle as it emerges from the stau LSP boundary. The end result is a larger relic density and a smaller allowed region.

Looking now at  Fig.~\ref{DOMvar}(b) where $m_{10}/m_{16} = 1.25$, we see a huge change. First of all, a large region appears in the upper-left of the parameter space which is excluded by the constraint $|\mu|^2 > 0$ at the EWSB scale. Since $\mu$ is very small in the parameter space bordering this region, the lightest chargino becomes Higgsino-like and a thin strip appears along which the $m_{\chi_1^{\pm}} > 103$ GeV bound is violated. The $b\rightarrow X_s \gamma$ excluded region along with the zones of preferred $(g-2)_\mu$ are enhanced, noticeably so at higher $m_{16}$ towards the edge of the EWSB limit where $\mu$ is rapidly decreasing. Most interestingly, the enhancement of the $A^0$ resonance annihilation of neutralinos has opened up the whole of the parameter space preferred by $(g-2)_\mu$ and allowed by the other constraints, at least when one ignores the lower bound on $\Omega_{{\textrm{CDM}}}h^2$. There are two main reasons for this. The first is that the $A^0$ resonance, marked by the contour $m_{A^0} = 2m_{\chi_1^0}$, is steeper and exhibits a characteristic kink, thus bringing it to smaller values of $m_{1/2}$. The kink originates from a complex interplay in the $(m_{1/2},m_{16})$-plane near the EWSB boundary, between the difference $(m_{H_1}^2 - m_{H_2}^2)$ (dependent on a particular combination of the RGE factors $X_t$ and $X_b$ (discussed in Section~\ref{rge}) and which controls the evolution of $m_{A^0}$) and the absolute value of $m_{H_2}^2$ (dependent on $X_t$ alone and which determines $\mu$ or equivalently the Higgsino component and mass of the lightest neutralino). With the $A^0$ resonance at smaller $m_{1/2}$, it is more effective at reducing $\Omega_{{\textrm{CDM}}}h^2$ which is proportional to $m_{\tilde{\chi}^0_1}$ and therefore, for most of the parameter space (away from the EWSB boundary), roughly proportional to $m_{1/2}$. The second reason is that neutralino LSPs in the region of parameter space close to the boundary where $|\mu| \simeq 0$ have large Higgsino components (as can be inferred from Eq.~\ref{neutevec}). Since this region is at much lower $m_{16}$ and much closer to the $A^0$ resonance than in the CMSSM case the effects are very much greater, namely that the neutralino LSP's coupling to the $A^0$ and also its coupling to gauge bosons is greatly enhanced. Next to the $m_{\tilde{\chi}_1^{\pm}} = 103$ GeV boundary there is significant co-annihilation with charginos. All this conspires to produce a greatly diminished relic density over much of the parameter space, despite the fact that the GUT scale boundary conditions do not deviate massively from the universal case. 

\begin{figure}[tbp]
\begin{center}
\includegraphics[width=0.6\textwidth]{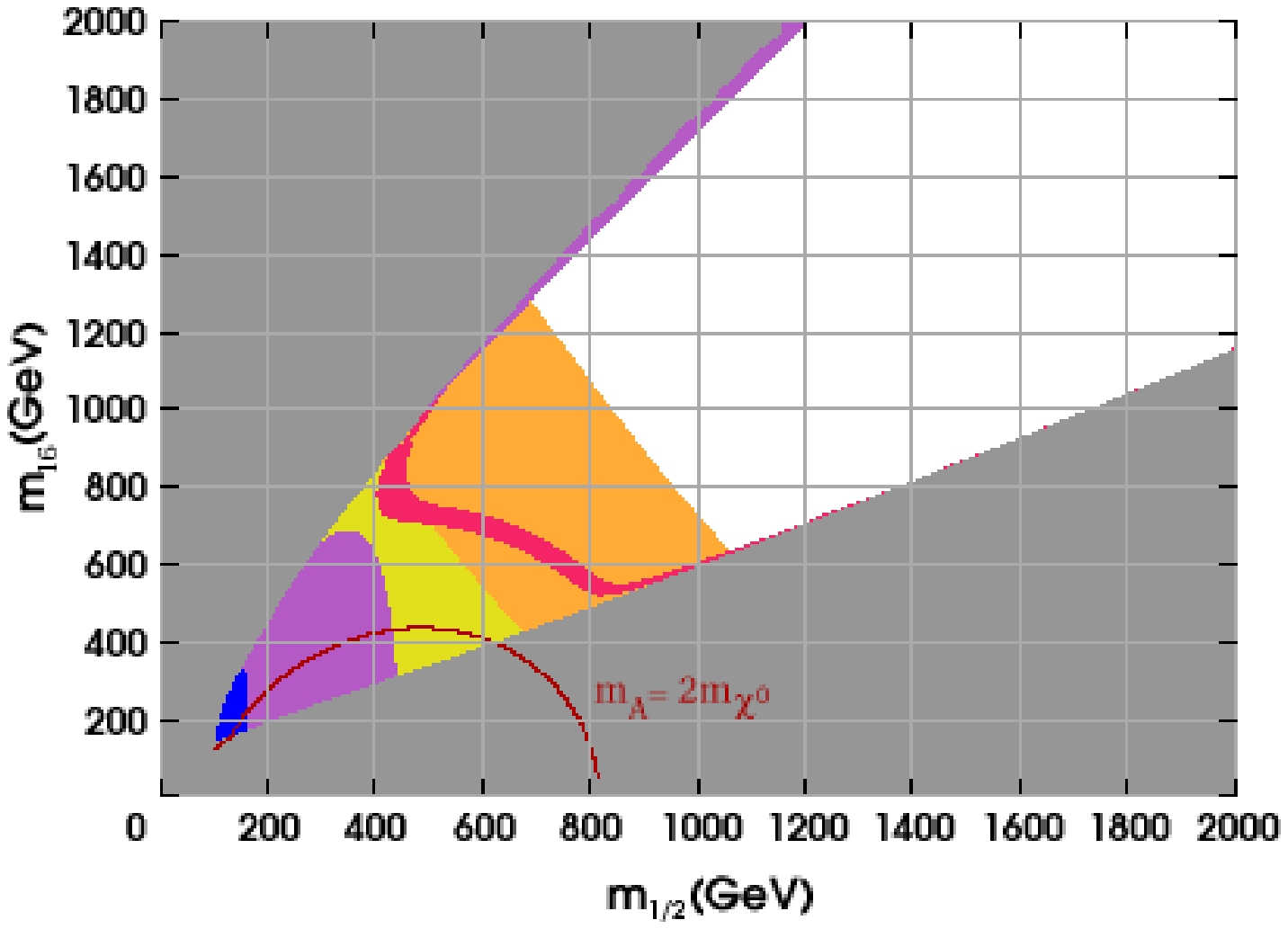}(a) 
\includegraphics[width=0.6\textwidth]{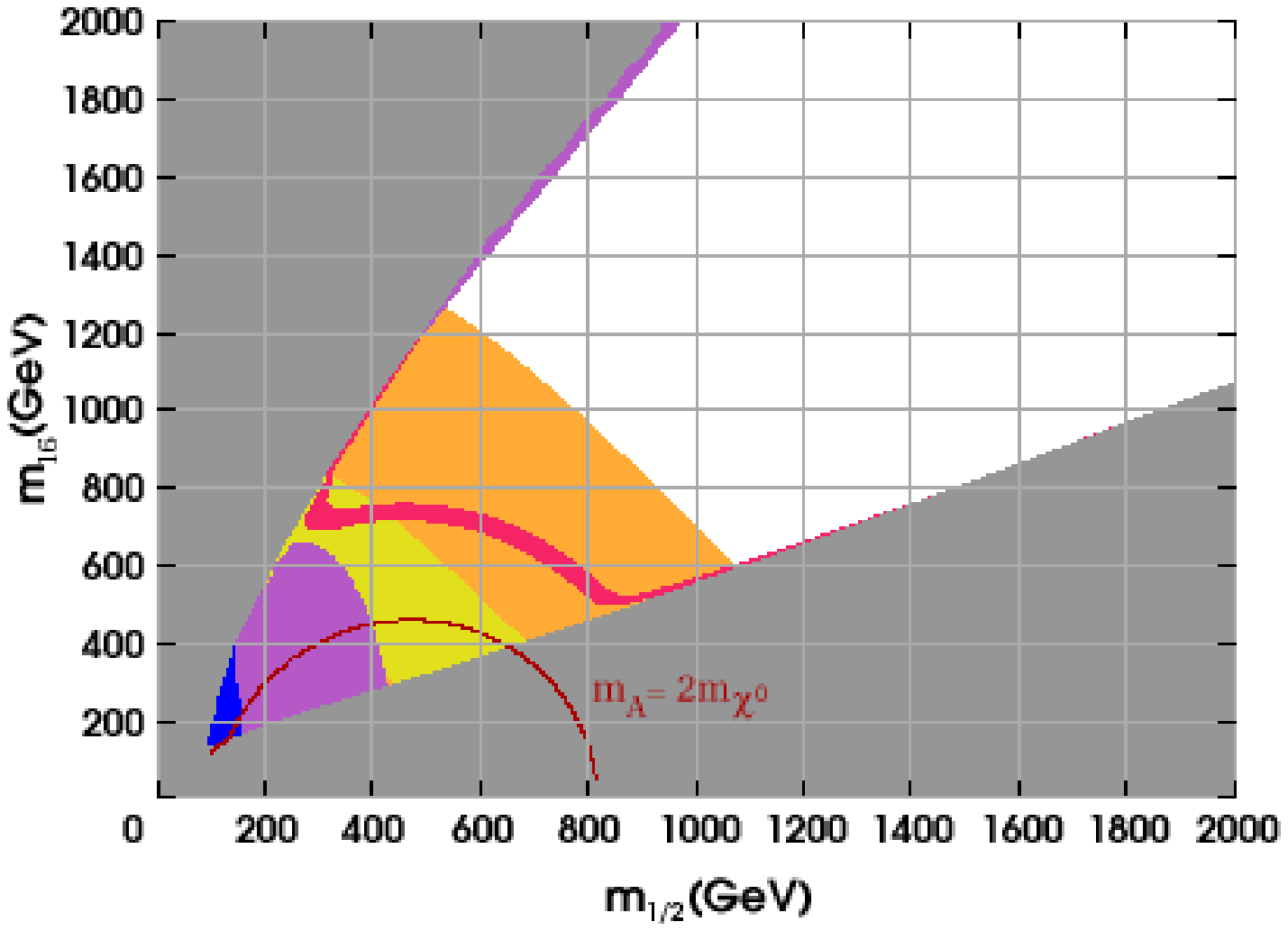}(b)
\includegraphics[width=0.6\textwidth]{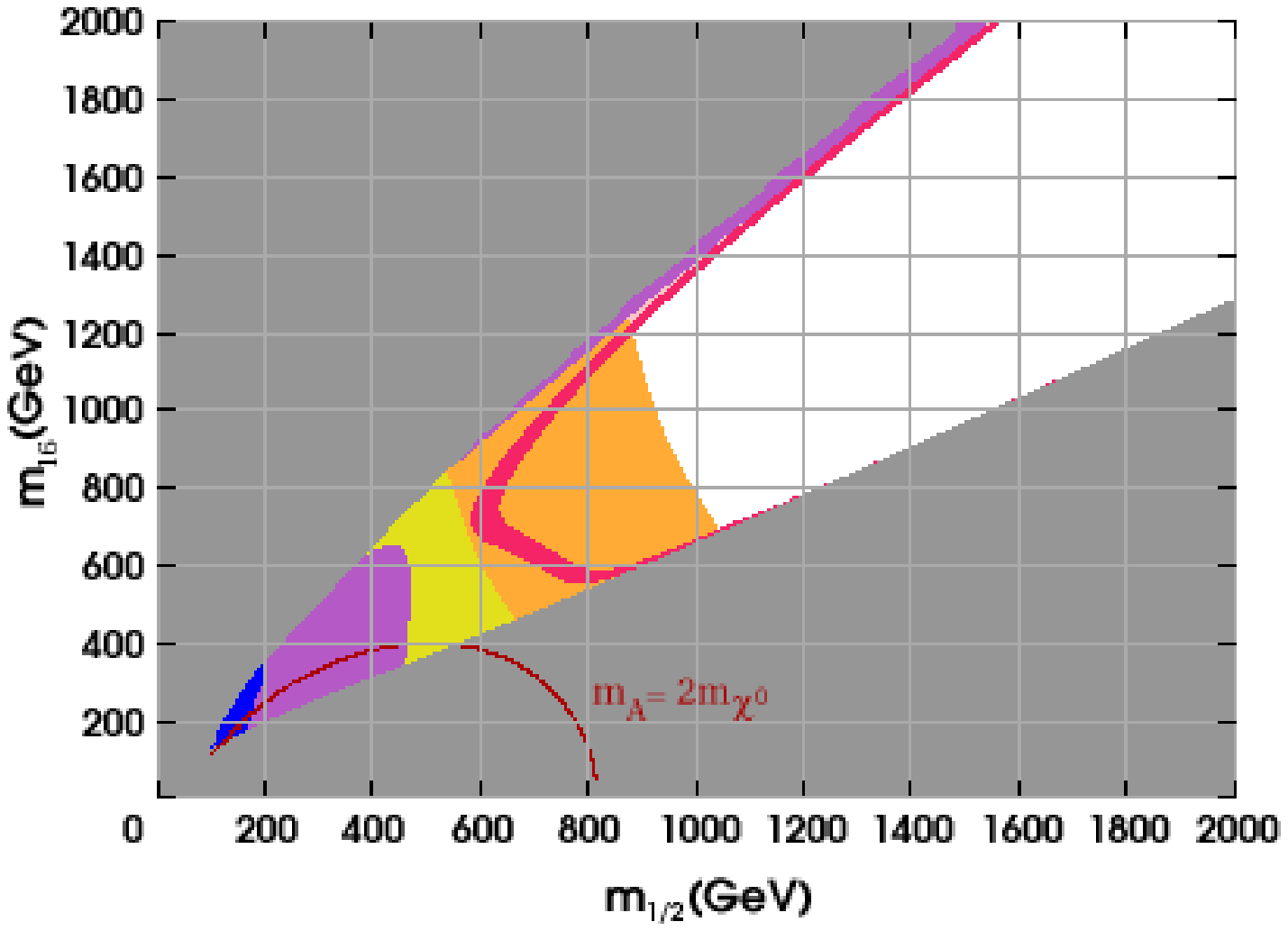}(c) 
\end{center}
\caption{{\em Same as Fig.~\ref{D0M1}, but with $\mathcal{D} = -0.4$. In {\rm (a)}, $m_{10}/m_{16} = 1$, in {\rm (b)} $m_{10}/m_{16} = 0.75$ and in {\rm (c)} $m_{10}/m_{16} = 1.25$. The grey triangular region at small $m_{1/2}$, large $m_{16}$ in all three graphs is ruled out this time {\rm not} because $|\mu|^2 < 0$, but because $m_{A^0}^2 < 0$, again indicating a failure to break EWSB properly. N.B. close to this region the Higgs masses are all very small.}}\label{Dm40}
\end{figure}

\begin{figure}[tbp]
\begin{center}
\includegraphics[width=0.6\textwidth]{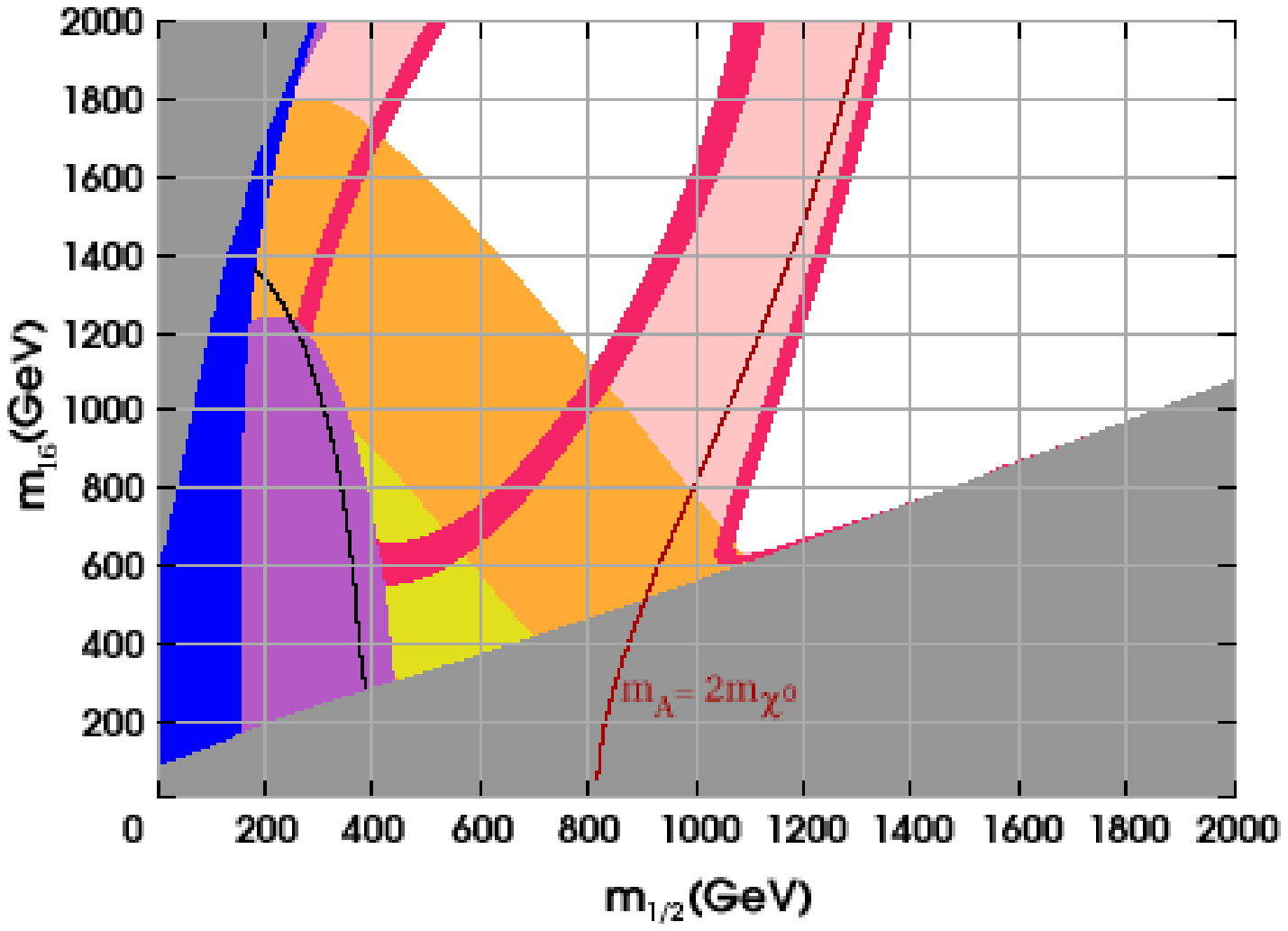}(a) 
\includegraphics[width=0.6\textwidth]{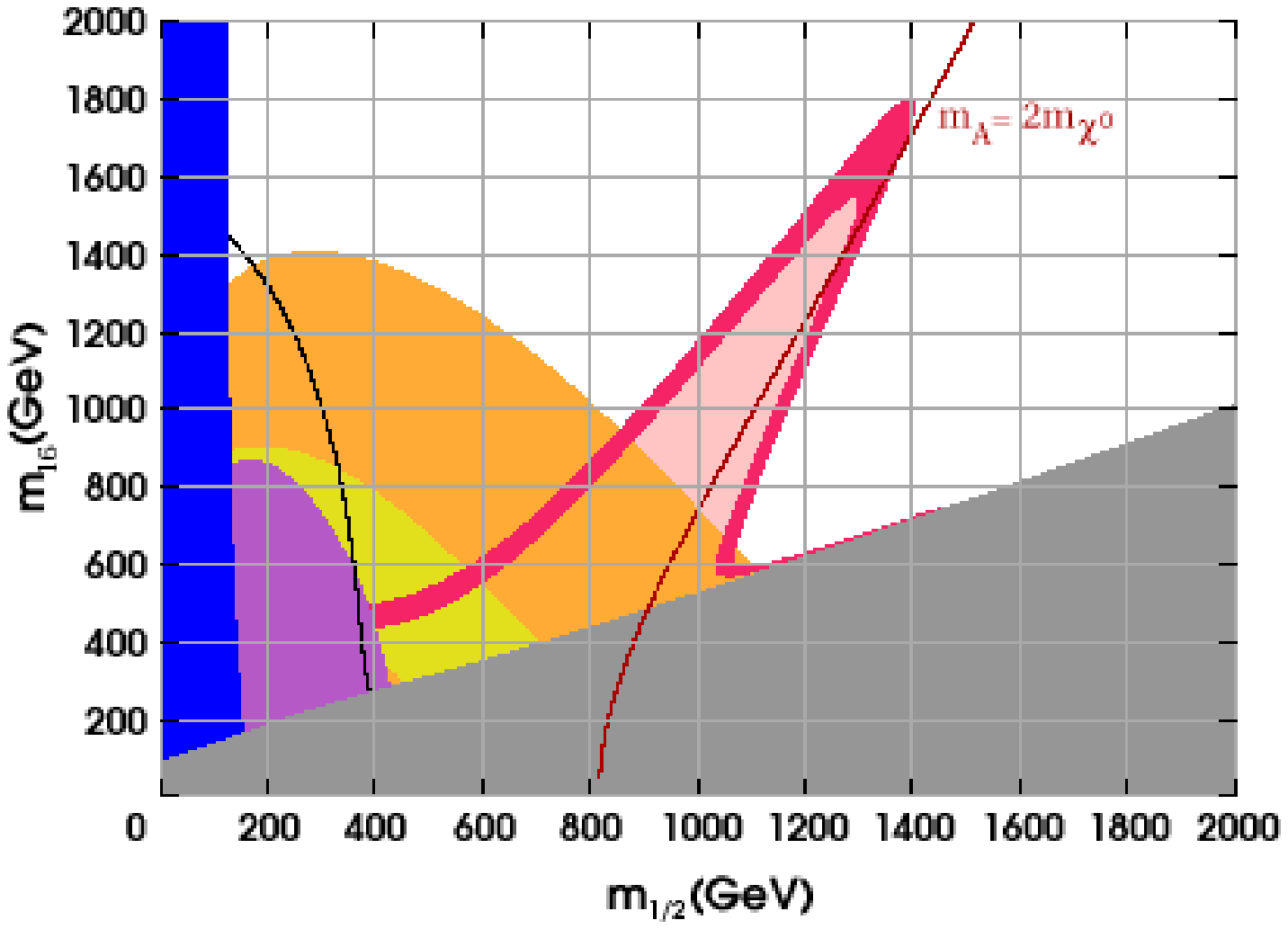}(b)
\includegraphics[width=0.6\textwidth]{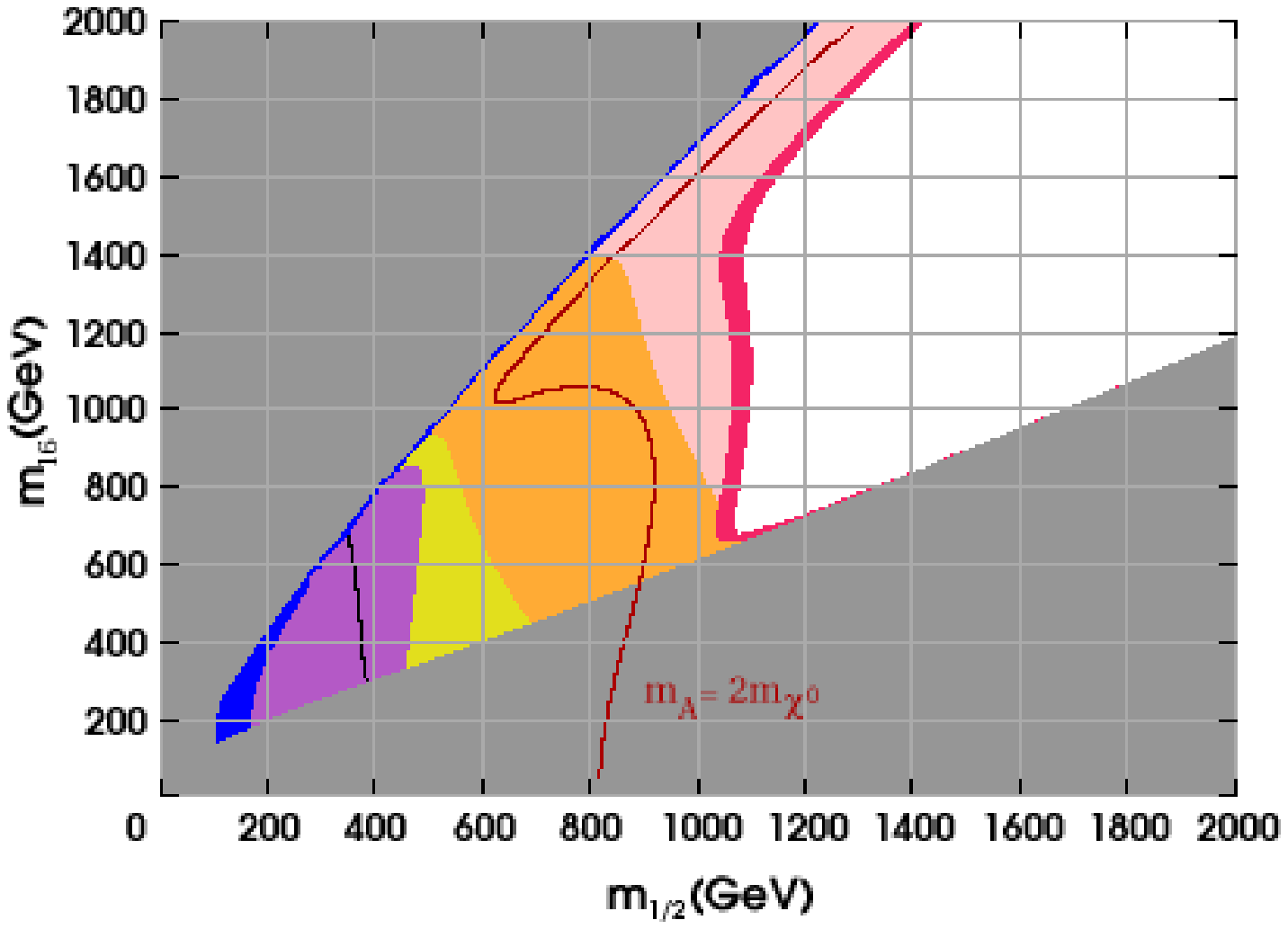}(c) 
\end{center}
\caption{{\em Same as Fig.~\ref{D0M1}, but with $\mathcal{D} = -0.2$. In {\rm (a)}, $m_{10}/m_{16} = 1$, in {\rm (b)} $m_{10}/m_{16} = 0.75$ and in {\rm (c)} $m_{10}/m_{16} = 1.25$. The grey triangular region at large $m_{16}$ in graphs {\rm (a)} and {\rm (c)} is ruled out because $|\mu|^2 < 0$ as for the case $\mathcal{D} = 0$.}}\label{Dm20}
\end{figure}

\begin{figure}[tbp]
\begin{center}
\includegraphics[width=0.6\textwidth]{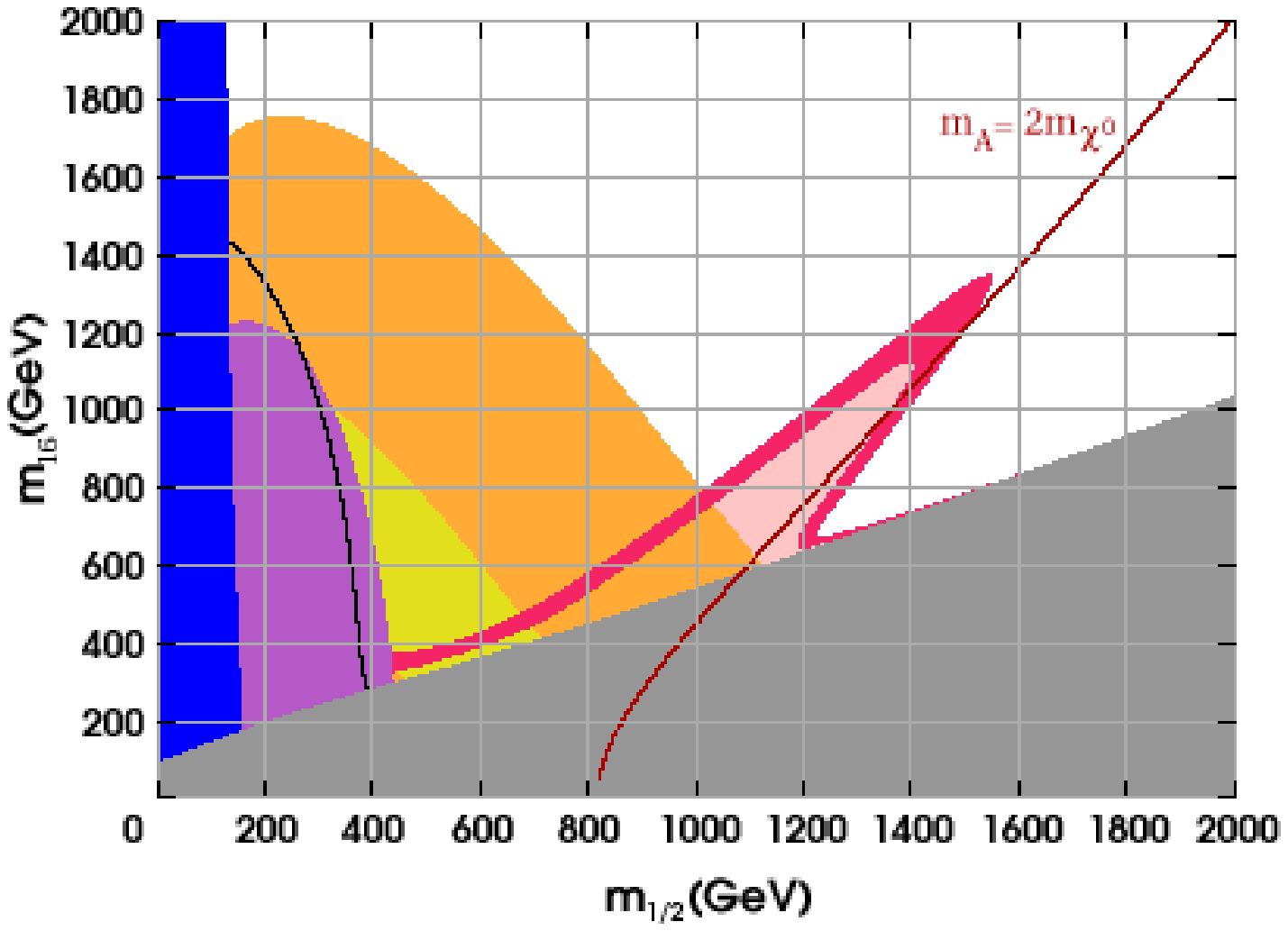}(a)
\includegraphics[width=0.6\textwidth]{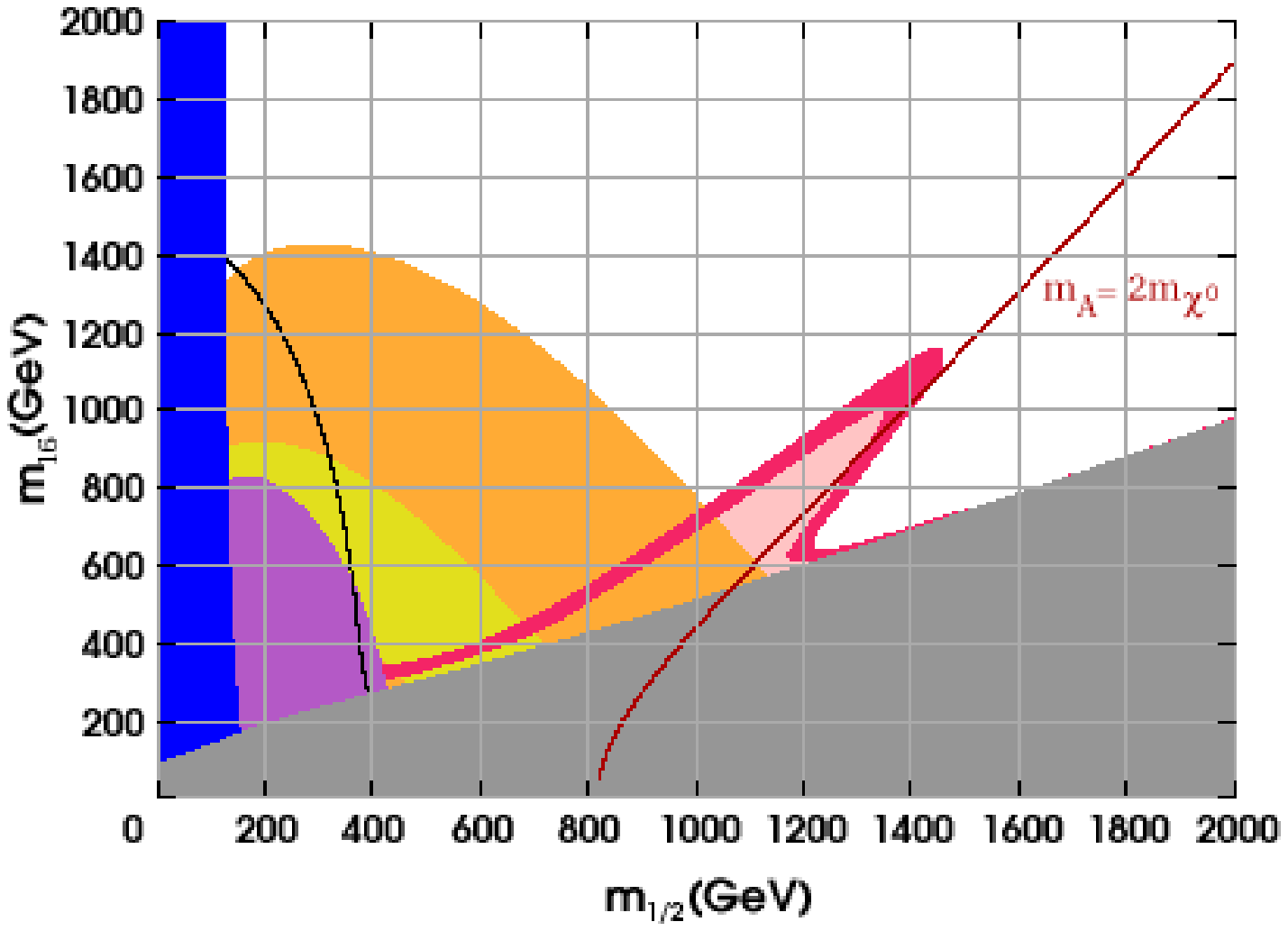}(b) 
\includegraphics[width=0.6\textwidth]{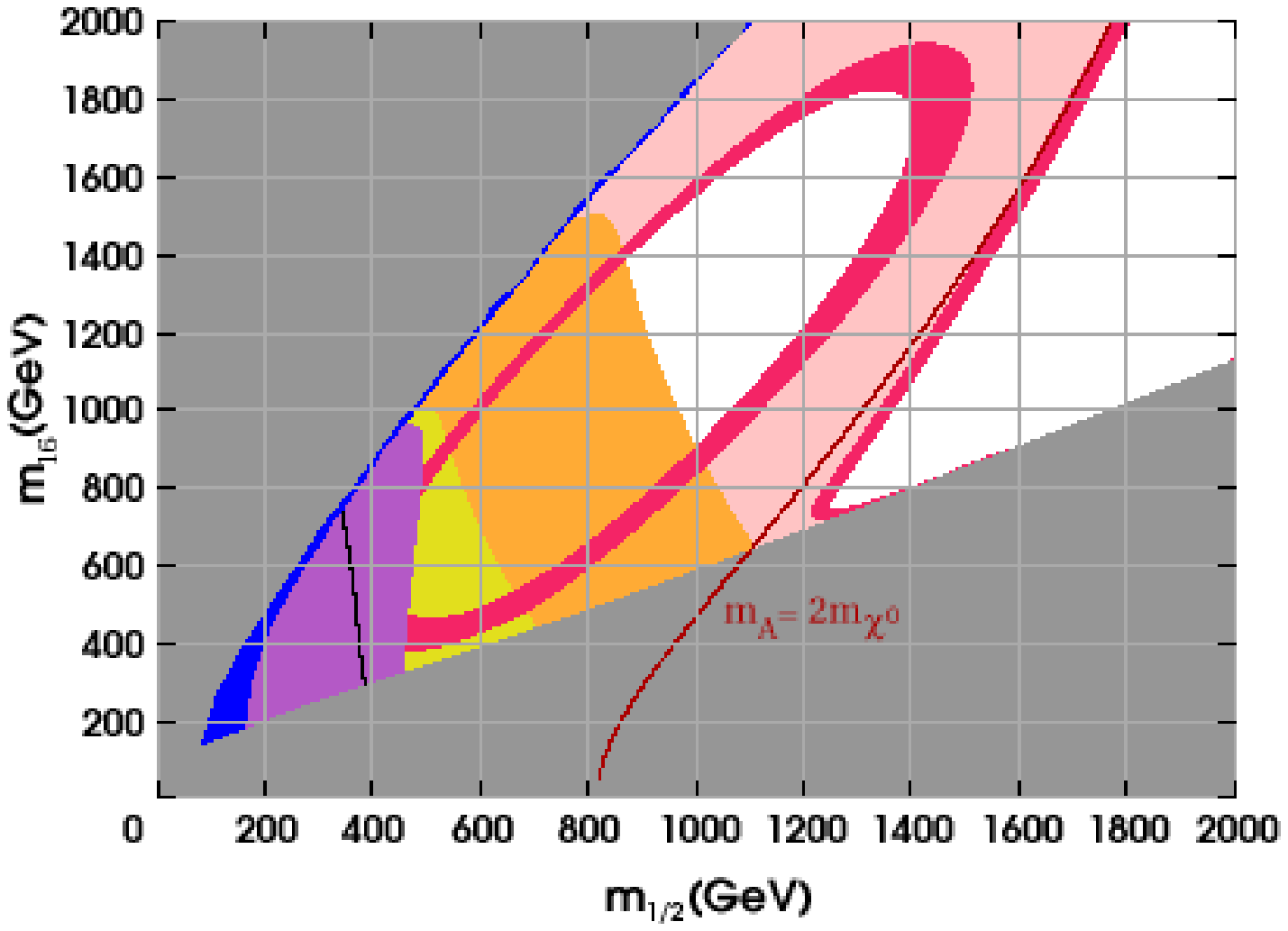}(c) 
\end{center}
\caption{{\em Same as Fig.~\ref{D0M1}, but with $\mathcal{D} = 0.2$. In {\rm (a)}, $m_{10}/m_{16} = 1$, in {\rm (b)} $m_{10}/m_{16} = 0.75$ and in {\rm (c)} $m_{10}/m_{16} = 1.25$. The grey triangular region at large $m_{16}$ in {\rm (c)} is ruled out because $|\mu|^2 < 0$ as for the case $\mathcal{D} = 0$.}}\label{D20}
\end{figure}

\begin{figure}[tbp]
\begin{center}
\includegraphics[width=0.6\textwidth]{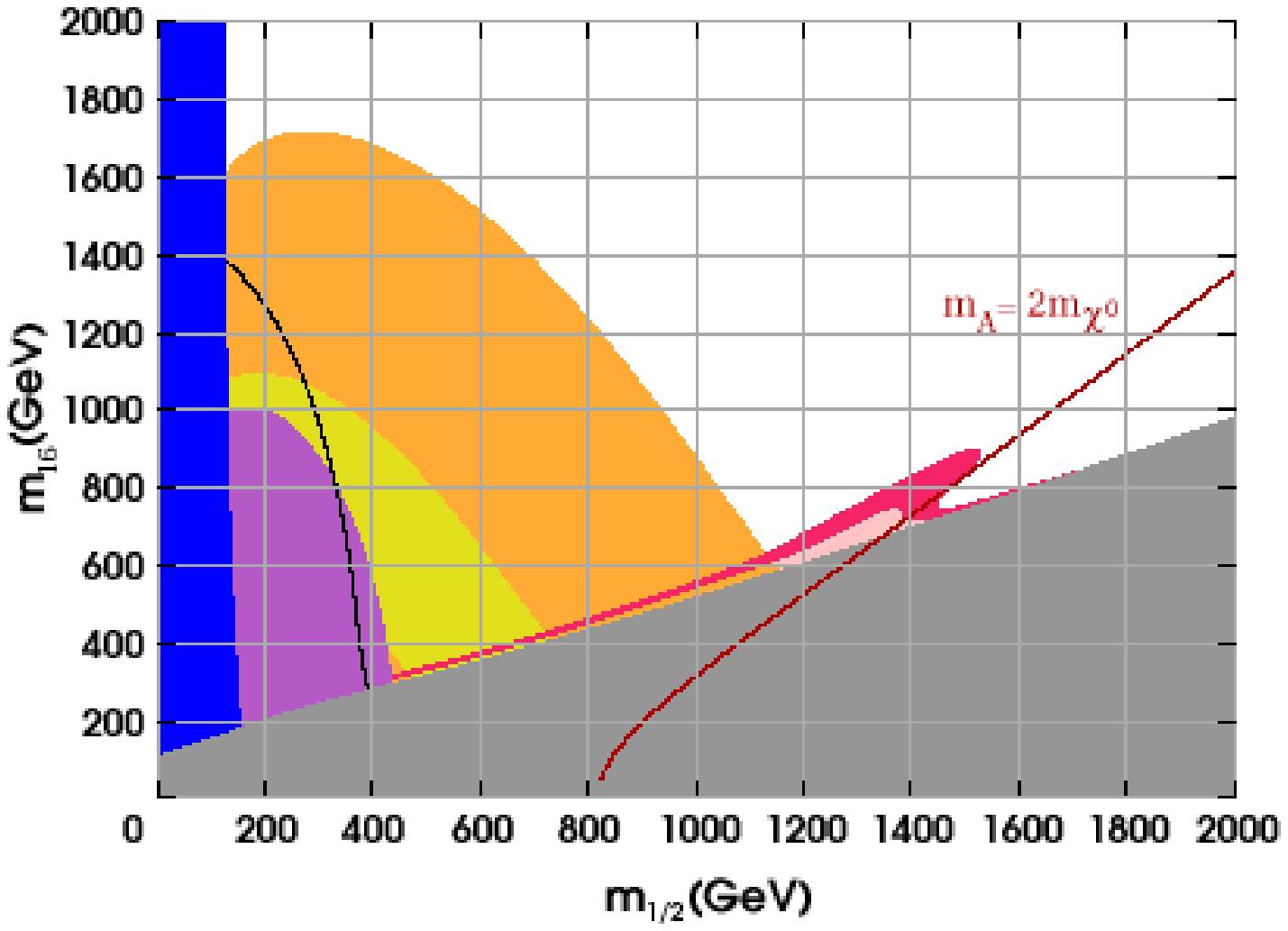}(a)
\includegraphics[width=0.6\textwidth]{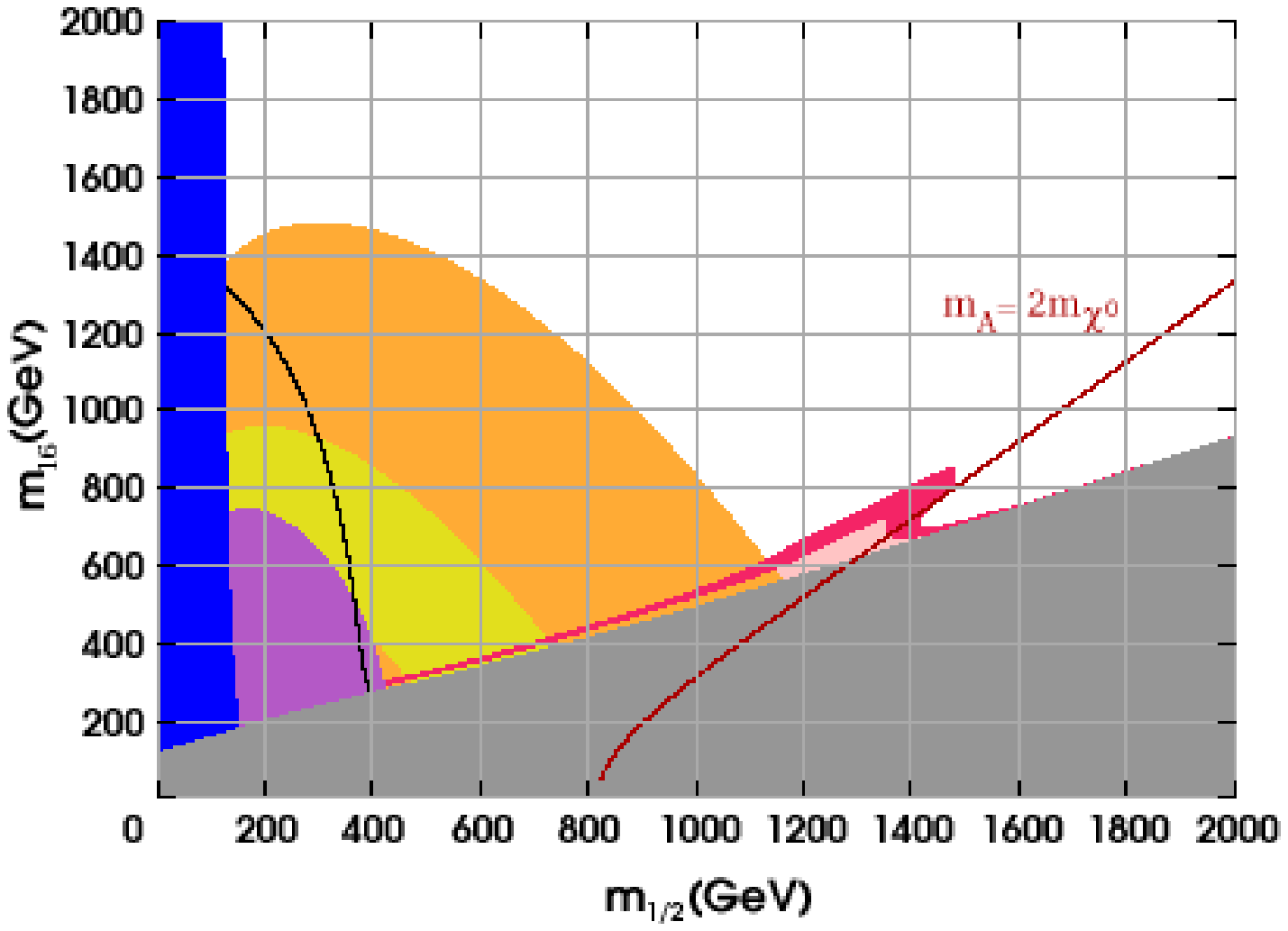}(b) 
\includegraphics[width=0.6\textwidth]{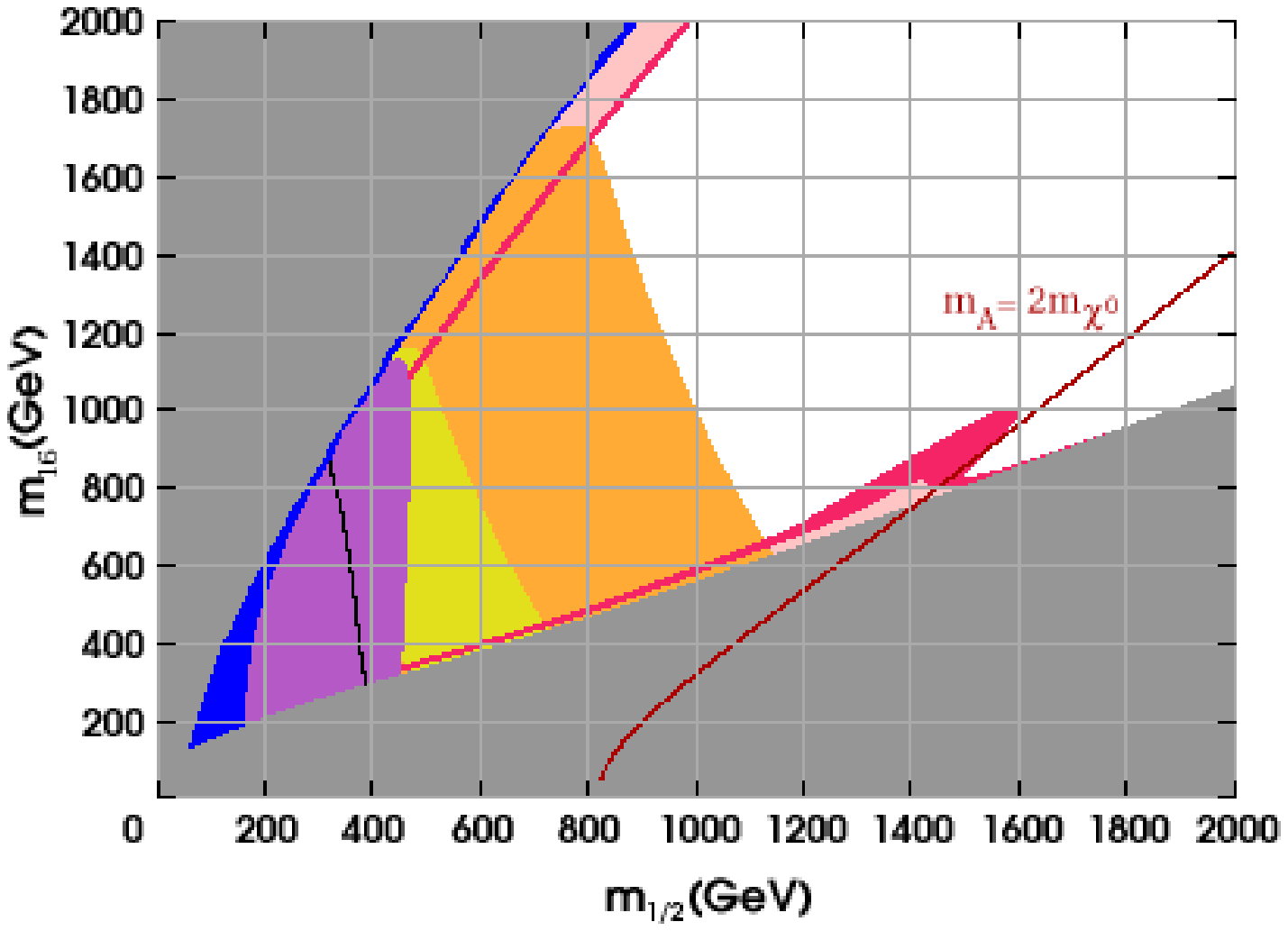}(c)
\end{center}
\caption{{\em Same as Fig.~\ref{D0M1}, but with $\mathcal{D} = 0.4$. In {\rm (a)}, $m_{10}/m_{16} = 1$, in {\rm (b)} $m_{10}/m_{16} = 0.75$ and in {\rm (c)} $m_{10}/m_{16} = 1.25$. The grey triangular region at large $m_{16}$ in {\rm (c)} is ruled out because $|\mu|^2 < 0$ as for the case $\mathcal{D} = 0$.}}\label{D40}
\end{figure}

In order to compare with Figs.~\ref{D0M1} and~\ref{DOMvar}, Figs.~\ref{Dm40},~\ref{Dm20},~\ref{D20} and~\ref{D40} show the $(m_{1/2},m_{16})$-plane for (a) $m_{10}/m_{16} = 1$, (b) $m_{10}/m_{16} = 0.75$ and (c) $m_{10}/m_{16} = 1.25$, this time in order of increasing $\mathcal{D}$. 

Fig.~\ref{Dm40} shows the case of $\mathcal{D} = -0.4$. All three graphs are fairly similar although very different from the CMSSM plot. Here, several curious things have happened. The area in which EWSB does not occur is due to $m_{A^0}^2 < 0$ this time. $(g-2)_\mu$ is slightly smaller than in the $\mathcal{D}=0$ scenario, as expected from Section~\ref{g2m}. Also $\mathcal{B}(b\rightarrow X_s\gamma)$ is substantially less restrictive at low $m_{16}$ and the excluded region drops off sharply next to the $m_{A^0}^2 = 0$ boundary where the charged Higgs diagram successfully cancels out the large chargino diagram. At larger values of $m_{16}$, still along the edge of $m_{A^0}^2 = 0$ boundary, there is a narrow strip excluded by the observed value for $\mathcal{B}(b\rightarrow X_s\gamma)$, this time because the Higgs component of the amplitude is overcompensating for the chargino component, interfering constructively with the Standard Model contribution and giving too large a value. For $\mathcal{D} = -0.4$, the stau LSP excluded region is at its largest due to the lighter right-handed stau receiving a negative contribution due to the negative $D$-term. The slope of this region is greatest for $m_{10}/m_{16} = 1.25$ where $X_\tau$ is at its largest, decreasing the stau mass even more. The A-resonance in this case has a completely different shape due to the fact that the Higgses are very light and beyond $m_{1/2} \simeq 800$ GeV and $m_{16} \simeq 450$ GeV, $m_{A^0} < 2m_{\chi^0_1}$ everywhere. The particulars of this situation produce a semi-circular curve in the bottom-left of the $(m_{1/2},m_{16})$-plane and the corresponding relic density curves as shown. An enhancement of the Higgsino component of the LSP results in a slightly different shape for Fig.~\ref{Dm40}(c). Note, however, that in each of (a), (b) and (c), there is no boundary where $\mu$ vanishes ($m_{A^0}$ always gets there first) and there are no regions where co-annihilations involving charginos play a significant r\^ole. 

Fig.~\ref{Dm20} shows similar plots, but with $\mathcal{D} = -0.2$. This time, the situation is more like the situation with $\mathcal{D} = 0$ in that the no EWSB region is due to $|\mu|^2 < 0$ rather than $m_{A^0}$ vanishing (although $m_{A^0}$ is very small close to the boundary compared to the CMSSM). For $m_{10}/m_{16} = 1$, Fig.~\ref{Dm20}(a), unlike in the CMSSM case, the $D$-terms help to keep $m_{H_2}^2$ above zero at the top-left-hand side of the plot and there is a slim region where $|\mu|^2 < 0$. Next to it, one can see at $m_{1/2} \simeq 300$ GeV, $m_{16} \simeq 2000$ GeV a tiny sliver of parameter space ruled out by the $\mathcal{B}(b\rightarrow X_s \gamma)$ upper bound due to the existence of a light charged Higgs. The main $\mathcal{B}(b\rightarrow X_s \gamma)$ exclusion and the $(g-2)_\mu$ preferred regions are basically indistinguishable from the CMSSM. Close to the no EWSB boundary at moderate to large $m_{16}$ there is again significant neutralino annihilation and there exists a region of acceptable relic density due to the large LSP Higgsino component. As one moves away from this region, increasing $m_{1/2}$, the proportion of Higgsino drops off and the relic density increases above the WMAP upper bound. Increasing $m_{1/2}$ further, one eventually reaches the A-resonance where again the WMAP bounds are satisfied. This results in a large, slanted {\em U}-shaped band where $\Omega_{{\textrm{CDM}}}h^2$ is perfectly adjusted to account for the WMAP observations. Fig.~\ref{Dm20}(b) is very similar to the case with zero $D$-terms, except that it has a larger A-resonance rapid-annihilation funnel. In Fig~\ref{Dm20}(c) the $m_{A^0} = 2m_{\tilde{\chi}^0_1}$ kink is more pronounced, tending towards the semi-circular arc of the Fig.~\ref{Dm40} plots.

We note that close to the boundary where radiative EWSB fails in Figs.~\ref{D0M1},~\ref{DOMvar}, and~\ref{Dm20} there is a sudden dip in the $(g-2)_\mu$ preferred region and the $b\rightarrow X_s \gamma$ excluded region where the Higgsino mass drops rapidly with $\mu$, becomes nearly degenerate with the wino, and large cancellations occur between the partial amplitudes corresponding to the two mass eigenstates. For the magnetic moment of the muon, although our approximate formulae no longer apply, one can infer this behaviour directly from the exact result Eqs.~\ref{g2exact} and~\ref{g2exact2} and inspection of the chargino mass matrix diagonalization equation, Eq.~\ref{chgmm}. When $\mu^2 \simeq M_2^2 \gg M_W^2$, then $\theta_L \simeq \theta_R \simeq \pi/4$ and $m_{\tilde{\chi}_1^{\pm}} \simeq m_{\tilde{\chi}_2^{\pm}} \simeq M_2 \simeq \mu$ and the contributions cancel. Of course, this cancellation is not exact, and, in any case, we have neglected the smuon-neutralino contributions. 

Increasing $\mathcal{D}$ to 0.2 and then to 0.4, we obtain the plots of Figs.~\ref{D20} and~\ref{D40}. Here, and referring back to Figs.~\ref{D0M1},~\ref{DOMvar},~\ref{Dm40} and~\ref{Dm20}, we will remark on some general trends: 
\begin{itemize}
\item{There is only a small change in $\mathcal{B}(b\rightarrow X_s \gamma)$ which is most apparent at large $m_{16}$ and small $m_{1/2}$, due to the increased sensitivity of $\mu$ and $m_{A^0}$ to the input parameters in this region. For $m_{10}/m_{16} = 1$ or 0.75, it is at a maximum for $\mathcal{D} \simeq 0$. This happens because, for large negative $D^2$, a large contribution from a propagating $H^{\pm}$ results in a partial cancellation of the chargino contribution, thus preventing its over-destruction of the Standard Model amplitude, while at large positive $D^2$, although the charged Higgs diagram is comparatively negligible, the chargino loop too is much smaller due to the increasing masses. Both of these cases lead to a decrease in the excluded region. On the other hand, for $m_{10}/m_{16} = 1.25$ as one increases $D^2$, the $\mathcal{B}(b\rightarrow X_s \gamma)$ exclusion zone continues to increase as the $|\mu|^2 = 0$ boundary recedes away to smaller $m_{1/2}$ and larger $m_{16}$ and the effect of the charged Higgs decreases.}
\item{For $(g-2)_{\mu}$, there are three distinct situations regarding its behaviour when $D^2$ is varied, corresponding to the different values for $m_{10}/m_{16}$. For $m_{10}/m_{16} = 1$, as $D^2$ grows, at high $m_{16}$ and small $m_{1/2}$, $(g-2)_{\mu}$ decreases. This is due to the greater effect that $D^2$ has on $\mu$ in this region as noted at the end of Section~\ref{g2m}. On the other hand, at points in the parameter space for which $m_{1/2} \sim m_{16}$, such as the $m_{1/2} = m_{16} = 500$ GeV point we analysed in detail in Section~\ref{g2m}, $(g-2)_{\mu}$ increases with $D^2$. The effect of increasing $\mathcal{D}$ at $m_{10}/m_{16} = 1$, therefore, is to make the muon anomalous magnetic moment preferred region a little lower and slightly broader. For $m_{10}/m_{16} = 0.75$, there is no region close to the boundary of the $(g-2)_{\mu}$ favoured zone where $\mu$ is affected to such a large extent by the $D$-terms since we are significantly further away from where EWSB fails than in the previous case. As a result, the $(g-2)_\mu$ preferred region increases in all parts of the nearby parameter space as one increases $D^2$. Finally, for $m_{10}/m_{16} = 1.25$, increasing $D^2$, causes the $|\mu|^2 = 0$ boundary to be pushed back, opening a larger area of parameter space. This also results in a decrease in $m_{\tilde{\nu}_{\mu}}^2$. Consequently, the amplitude is increased over the whole region. However, note the dip in $(g-2)_\mu$ as $\mu$ becomes small, as commented on earlier.}
\item{For each value of $m_{10}/m_{16}$, increasing $D^2$ results in a smaller $\tilde{\tau}$ LSP region. This is because the lighter, right-handed stau mass steadily increases relative to the lightest neutralino mass since $m^2_{\tilde{e}_R} = m_{16}^2 + g_{10}^2 D^2$.}   
\item{As one increases $D^2$, $m_{A^0}$ increases and the A-resonance makes a more acute angle with the $m_{1/2}$ axis, emerging from the stau LSP excluded region at a larger value of $m_{1/2}$. Consequently, the neutralino dark matter density is much increased and the region allowed by WMAP is substantially reduced in this part of parameter space. For $\mathcal{D} = 0.4$, in Fig.~\ref{D40}, the characteristic A-resonance peak had almost disappeared.}   
\end{itemize}
To make a final note on these plots, the relic density curves in the plot shown in Fig.~\ref{D20}(c) resemble those of Fig.~\ref{Dm20}(a). The difference is that the $|\mu|^2 < 0$ region is due to an increased $m_{10}/m_{16}$ in Fig.~\ref{D20}(c) rather than a negative $D^2$ in Fig.~\ref{Dm20}(a). 

\subsubsection*{A Cautionary Note}
It should be strongly emphasized here that the exact location and form of the relic density curves, the location of the EWSB excluded region and also of the A-resonance (on both of which the relic density strongly depends) are not only dependent on the variation of the mSugra parameters due to $D$-terms or Higgs-sfermion splitting or any other changes one may make, but also on the precise value of the top mass which governs the highly important parameter $X_t$ and thus the Higgs soft masses. The value used in this paper is $m_t^{{\textrm{POLE}}} = 174.3$ GeV, the value quoted by the PDG~\cite{Eidelman:2004wy}, which is now at odds with the recently released results from the D$\emptyset$ collaboration~\cite{Abazov:2004cs} (and see also~\cite{Abazov:2004tg}), which puts the top mass at $m_t^{{\textrm{POLE}}} = 178.0 \pm 4.3$ GeV. Even this $\sim 2\% $ change would be enough to significantly change the form of our plots. However, even had we used this updated value for $m_t$, there is still the question of the $4.3$ GeV error. Varying $m_t$ within this would again lead to very different results. Acknowledging this, we also point out that the position and shape of the WMAP preferred region, the A-resonance, and the $|\mu|^2 = 0$ region are also highly dependent on the exact value of $\tan\beta$. To keep things under control, we used a constant value of $\tan\beta = 50$ throughout this paper. However, by tweaking this value and the values we set for $D^2$ and $m_{10}/m_{16}$ to some small degree we would be able to account for any difference in the top mass to a good approximation. As such, we believe that while qualitatively correct, our results for the relic density, for the exact position of the boundary where EWSB fails and for the exact value of the $A^0$ mass, due to their strong dependence on $\tan\beta$ and $m_t$, and also for the values of $\mathcal{B}(b\rightarrow X_s \gamma)$ and $(g-2)_{\mu}$ in the region of $\mu \sim 0$ or $m_{A^0} \sim 0$ (although they are reasonably stable elsewhere) should not be taken as quantitatively correct to any degree of accuracy. This is not a statement of the inaccuracy of the (fairly well-tested and state-of-the-art) computer programs we used, merely a statement about the sensitivity of certain observables to the input parameters especially in the region of parameter space close to where the radiative EWSB mechanism fails. For this reason we have chosen to focus on the qualitative aspects of our results and their origins.

\section{Summary and Conclusions\label{summary}}
In this paper we have discussed the effects that small deviations from universality, in the form of ${\textrm{SO}}(10)$-inspired ${\textrm{U}}(1)_{\textrm{X}}$ $D$-terms and Higgs-sfermion soft mass splitting, have on the parameter space of the CMSSM at large $\tan\beta$, with $\mu > 0$ and $A_0 = 0$.

In the first part of the paper we reviewed the origin of such terms before going on to follow, by use of various approximations, how the additional parameters $\mathcal{D} \equiv {\mathrm{sign}}(D^2)\sqrt{|D^2|}/m_{16}$ and $m_{10}/m_{16}$ feed through the renormalization group equations, electroweak symmetry breaking conditions and sparticle mass matrices to effect changes in the sparticle mass spectrum and mixings. 

It was noted that the $D$-terms are closely connected to the CP-odd $A^0$ Higgs mass and that $m_{10}/m_{16}$ has a large effect on the value of $\mu$ at the electroweak scale. At large negative values of $D^2$ or large values of $m_{10}/m_{16}$ one can expect the radiative electroweak symmetry breaking mechanism to break down due to $m_{A^0}^2 < 0$ in the former case and $|\mu|^2 < 0$ in the latter and for this to be especially important for small $m_{1/2}$, large $m_{16}$. 

The $D$-terms affect the scalars at tree-level, especially the heavy Higgs $A^0$, $H^0$ and $H^{\pm}$, and the right-handed down-type squarks and left-handed sleptons, but largely cancel out of the RGEs. On the other hand, $m_{10}/m_{16}$ affects the Higgs scalar soft masses at tree-level and feeds through the renormalization group equations through large third family Yukawa couplings, noticeably affecting the third generation of squarks and sleptons, though most strongly influencing $\mu$ and the Higgsino-like charginos and neutralinos. We explained how, contrary to first impressions, increasing $m_{10}/m_{16}$ also increases $m_{A^0}$ as the indirect result of bottom mass threshold corrections. We also gave approximate results for the mixing angles in the chargino sector and the components of the lightest neutralino, the Higgsino components of which can be seen to grow with $\mu$.

We analysed the computed spectrum away from such regions at $m_{1/2} = m_{16} = 500$ GeV for $\mathcal{D} \in [-0.4, 0.4]$ and $m_{10}/m_{16} \in [0.75, 1.25]$ and found deviations in the sparticle masses compared to the universal case resulting from additive effects at the GUT scale coming from the $D$-terms, and from loop effects feeding through the RGEs as a result of the Higgs-sfermion splitting. These changes were found to be very small for those masses predominantly dependent on the gluino mass $M_3$, for example the gluino and the squarks of the first two families (except the right-handed down-type squarks which receive larger $D$-term corrections), but could be sizeable for other sparticles such as the third family sleptons and the Higgsino-like charginos and neutralinos. It was noted that close to the boundary where EWSB fails, and otherwise at regions with small $m_{1/2}$ and large $m_{16}$, the effects of $D$-terms and $m_{10}/m_{16}$ can be much more important. We analyzed a set of points in these more sensitive regions of parameter space and observed much larger deviations in the mass spectrum from the universal case, especially in the heavy Higgs masses, and the Higgsino-like charginos and neutralinos. 

In the next section we analysed how the $D$-terms and $m_{10}/m_{16}$ affect the dominant $\tan\beta$-enhanced amplitudes of quantities such as $(g-2)_\mu$, $\mathcal{B}(b\rightarrow X_s \gamma)$ and the SUSY threshold corrections to the bottom mass. In general the effects were found to be relatively small in much of the parameter space, but subtle, and in general larger in the small $m_{1/2}$, large $m_{16}$ region. 

For $(g-2)_\mu$ we gave a detailed quantitative approximation in Section~\ref{g2m}, assuming that $\mu^2 - M_2^2 \gg M_W^2$ and taking the large $\tan\beta$ limit. It was observed that there is a trade-off between the external factor and the loop functions present in the approximate formula, and in regions where the approximation is valid it can be quite finely balanced especially in the case of varying the $D$-terms, resulting in only small deviations from the CMSSM case. We found that the effect of increasing $D^2$ away from the sensitive regions close to $|\mu|^2 = 0$ and $m_{A^0}^2 = 0$ and to where $m_{1/2} \ll m_{16}$ was to increase the amplitude due to the decreasing muon sneutrino mass despite the contrary effect of the increasing Higgsino mass. For the particular point $m_{1/2} = m_{16} = 500$ GeV we gave numerical examples, verifying our approximation. In other circumstances, however, often the effect of increasing $D^2$ was the reverse, leading to a decrease in $(g-2)_\mu$. Increasing $m_{10}/m_{16}$ was always found to increase $(g-2)_\mu$ as a result of the decrease in the Higgsino mass, though a drop-off was observed close to where radiative EWSB fails as a result of cancellation between the partial amplitudes arising from the two chargino mass eigenstates. All such effects were, for the most part, found to be relatively small. 

In the case of $\mathcal{B}(b\rightarrow X_s \gamma)$ it was discovered that for both $m_{10}/m_{16} = 1$ and 0.75, the size of the excluded region of $\mathcal{B}(b\rightarrow X_s \gamma)$ peaks for $\mathcal{D} \simeq 0$. This is because for positive $D^2$ the Higgsino mass increases and therefore the chargino contribution to the Wilson coefficients decreases and there is no longer an over-cancellation of the Standard Model contribution, while for negative $D^2$ the charged Higgs contribution grows rapidly and partially cancels the chargino amplitude again saving the successful Standard Model result over a larger region of parameter space. Increasing $m_{10}/m_{16}$ to 1.25 in general increases the chargino contribution resulting in a larger excluded zone. Increasing $D^2$ in this case pushes back the no EWSB boundary, decreases the charged Higgs contribution and results in a larger forbidden region. 

The SUSY threshold corrections to the bottom mass were found to decrease with increasing $m_{10}/m_{16}$ and decreasing $D^2$, a result of increasing the chargino contribution with respect to the dominant gluino contribution which has the opposite sign. This trend is what is needed for Yukawa unification favoured by GUT models, but the corrections were found to be far too small to achieve this aim.     

Increasing $m_{10}/m_{16}$ and decreasing $D^2$ were found to increase the area excluded by the existence of a stau LSP due to an associated decrease in the mass of the lighter right-handed stau. However, the change is quite small. 

The main effect of varying $m_{10}/m_{16}$ and $D^2$ was to cause substantial alterations to the position and shape of the curves of neutralino relic density required to satisfy the WMAP bounds. This was found to be due to both large changes in the position and shape of the A-resonance region where  $m_{A^0} \simeq 2m_{\tilde{\chi}_1^0}$ resulting from large corrections to the $A^0$ Higgs and the lightest neutralino mass (in the region where it changes from mainly bino to mainly Higgsino), and a massively increased Higgsino component of the LSP close to the boundary where $\mu = 0$, strongly affecting its couplings to the $A^0$ Higgs and gauge bosons.

A cautionary note was made regarding the sensitivity of these results to the changing value of $m_t$ and also to $\tan\beta$. 

In conclusion we note that although corrections to universality like $D$-terms and Higgs-sfermion splitting are straightforward to implement at the input scale, by the time they filter through to the sparticle masses and observables their effects can be complex and often subtle. We have attempted to clarify exactly what these effects are and how they originate. The various observables used to constrain low energy supersymmetric models are susceptible to change significantly even under small corrections to the sfermion/Higgs universality assumed in the CMSSM, especially observables sensitive to $\mu$ and $m_{A^0}$ such as the neutralino relic density, and particularly in areas of parameter space close to where the EWSB mechanism breaks down. As a result a much larger region of parameter space than in the CMSSM becomes viable. However, from the point of view of a successful ${\textrm{SO}}(10)$ unified model, there is still the persistent problem of achieving Yukawa unification in the $\mu > 0$ case, as has been previously noted by other authors~\cite{Auto:2003ys,Blazek:2002ta}. We hope to address this point in our future work.  

%These changes varied from negligible in the case of those sparticle masses predominantly dependent on the gaugino masses $M_i$, including the first two family squarks except for the right-handed down and strange squarks which receive relatively larger $D$-term corrections, to a few percent to $\sim 10\%$ in the case of the right-handed down-type squarks of the first two generations and the stops and sbottoms, to $\sim 15\%$ in the case of the third family sleptons.   

\ack{The author would like to thank B.~C.~Allanach, R.~Dermisek, M.~Hirsch, C.~Hugonie, S.~Profumo, W.~Porod and especially G.~G.~Ross for useful discussions, and the E.C. for a Marie Curie Training Site grant as part of the European Network Project ``Physics Beyond the Standard Model (HPMT-CT-2000-00124)''}

\bibliography{so10soft}

\begin{table}
\begin{center}
\begin{tabular}{|l|rrrrrrr|}
\hline
{\bf Standard Points}& A & B & C & D & E & F  & G\\
\hline
Input Parameters & & &  &  &  &  & \\
\hline
$m_{1/2}$ & 500 & 500 & 500 & 500 & 500 & 500 & 500\\
$m_{16}$ & 500 & 500 & 500 & 500 & 500 & 500 & 500\\
$\mathcal{D}$ & 0 & 0 & 0 & -0.4 &  0.4 & -0.4 & 0.4\\
$m_{10}/m_{16}$  & 1.0 & 0.75& 1.25 & 1.0 & 1.0 & 1.25 & 0.75\\
\hline
\hline
$\mu(M_Z)$  & 592.9 & 647.7& 514.4 & 561.5 & 622.9 & 478.0 & 675.3\\
\hline
\hline
Sparticle Masses & & &  &  &  &  & \\
\hline
$m_{h^0}$ & 116.1 &116.1 & 116.1&116.1 & 116.1 & 116.1 & 116.1\\
$m_{H^0},m_{A^0}$  &  500.0 & 502.9& 488.9 & 403.5 & 578.8 & 388.4 & 581.3\\
$m_{H^{\pm}}$  & 506.1 & 509.9& 496.3 & 412.4 & 585.0 & 397.8 & 587.3\\
$m_{\tilde{u}_{L}}$ & 1153.4& 1153.2 & 1153.7&1144.2 & 1162.5 & 1144.6 & 1162.3\\
$m_{\tilde{u}_{R}}$ & 1121.1& 1120.9& 1121.5 & 1113.2 & 1128.9 & 1113.6 & 1128.7\\
$m_{\tilde{d}_{L}}$ & 1156.0& 1155.8& 1156.4 & 1146.9 & 1165.1 & 1147.2 & 1164.9\\
$m_{\tilde{d}_{R}}$ & 1111.7& 1111.4& 1112.0 & 1138.1 & 1084.5 & 1138.5 & 1084.3\\
$m_{\tilde{e}_{L}}$ & 599.5& 599.5& 599.4 & 649.8 & 544.5 & 649.9 & 544.5\\
$m_{\tilde{e}_{R}}$ & 533.9& 533.9& 534.0 & 510.5 & 556.5 & 510.5 & 556.4\\
$m_{\tilde{\nu}_{e}}$ &594.5 & 594.6& 594.5 & 645.3 & 539.0 & 645.3 & 539.1\\
$m_{\tilde{t}_{1}}$ & 855.9& 870.3& 837.0 & 844.0 & 867.7 & 824.8 & 881.8\\
$m_{\tilde{t}_{2}}$ & 1025.5& 1036.7& 1010.2 & 1013.7 & 1037.0 & 997.9 & 1048.0\\
$m_{\tilde{b}_{1}}$ & 950.1& 959.0& 937.9 & 951.8 & 938.0 & 938.6 & 946.8\\
$m_{\tilde{b}_{2}}$ & 1010.7& 1023.2& 992.9 & 1022.2 & 1008.5 & 1004.4 & 1020.7\\
$m_{\tilde{\tau}_{1}}$  &350.6 &369.5& 324.3 & 328.0 & 357.8 & 297.3 & 373.0\\
$m_{\tilde{\tau}_{2}}$  &555.1 &567.0& 539.8 & 603.5 & 511.8 & 591.2 & 527.5\\
$m_{\tilde{\nu}_{\tau}}$ & 531.9 & 540.3 & 521.0 & 588.7 & 468.0 & 579.1 & 477.7\\

$m_{\tilde{\chi}^{\pm}_{1}}$ & 393.0& 395.7& 386.1 & 391.0 & 394.3 & 380.7 & 396.4\\
$m_{\tilde{\chi}^{\pm}_{2}}$ & 632.7& 685.9& 559.1 & 602.7 & 661.8 & 527.3 & 713.1\\
$m_{\tilde{\chi}^{0}_{1}}$ & 207.8& 207.9& 206.9 & 207.5 & 207.5 & 206.7 & 207.8\\
$m_{\tilde{\chi}^{0}_{2}}$ & 393.1& 395.7& 386.3 & 391.1 & 394.4 & 381.0 & 396.4\\
$m_{\tilde{\chi}^{0}_{3}}$ & 619.7& 675.6& 539.4 & 587.5 & 650.4 & 502.1 & 703.8\\
$m_{\tilde{\chi}^{0}_{4}}$ & 632.3& 685.4& 559.0 & 602.3 & 661.3 & 527.3 & 712.5\\
$m_{\tilde{g}}$  & 1167.4&1167.8& 1166.9 & 1167.5 & 1167.2 & 1167.0 & 1167.6\\
\hline
\hline
$a_{\mu}^{{\textrm{SUSY}}}  \times 10^{9}$   & 2.18& 2.08& 2.34 & 2.04 & 2.34 & 2.20 & 2.24\\
$\mathcal{B}(b\rightarrow X_{s}\gamma)  \times 10^{4}$ & 2.73& 2.83& 2.57 & 2.74 & 2.75 & 2.58 & 2.84\\
$\Omega_{{\textrm{CDM}}}h^2$ & 0.151& 0.208 & 0.078 & 0.074 & 0.632 & 0.013 & 0.774\\
\hline
\end{tabular}
\end{center}
\caption{{\em Here we show the mass spectra (in GeV) for the standard points, $m_{1/2} = m_{16}= 500$ GeV. The first column, {\rm A},  is the CMSSM case, {\rm B-E} show the effects of changing one of $D^2$ or $m_{10}/m_{16}$, and {\rm F} and {\rm G} are the combination of $D^2$ and $m_{10}/m_{16}$ that contrast the most.}\label{benchtbl}}
\end{table}

\begin{table}
\begin{center}
\begin{tabular}{|l|rrrrrr|}
\hline
{\bf Points of Interest}& U & V & W & X & Y & Z\\
\hline
Input Parameters & & & & & & \\
\hline
$m_{1/2}$ & 500 & 500 & 813 & 813 & 200 & 200\\
$m_{16}$ & 990 & 990 & 1200 & 1200 & 1300 & 1300\\
$\mathcal{D}$ & 0 & 0 & 0 & -0.4 & 0 & 0.4\\
$m_{10}/m_{16}$  & 1.0 & 1.25& 1.0 & 1.25 & 1.0 & 1.0\\
\hline
\hline
$\mu(M_Z)$  & 581.6 & 103.3 & 882.2 & 299.2 & 293.5 & 562.7\\
\hline
\hline
Sparticle Masses & & & & & &\\
\hline
$m_{h^0}$ & 116.6 &116.8 & 119.3 & 87.2 & 113.9 & 114.2\\
$m_{A^0}$  &  614.2 & 436.3& 846.0 & 88.8 & 521.2 & 986.7\\
$m_{H^0}$  &  614.3 & 436.3& 846.0 & 120.2 & 521.1 & 986.7\\
$m_{H^{\pm}}$  & 620.1 & 445.0& 850.2 & 127.5 & 528.0 & 990.2\\
$m_{\tilde{u}_{L}}$ & 1419.4& 1422.0 & 2007.2 & 1979.9 & 1346.7 & 1399.1\\
$m_{\tilde{u}_{R}}$ & 1395.6& 1398.1& 1960.5 & 1937.4 & 1347.7 & 1391.2\\
$m_{\tilde{d}_{L}}$ & 1421.5& 1424.1& 2008.6 & 1981.4 & 1349.0 & 1401.3\\
$m_{\tilde{d}_{R}}$ & 1388.6& 1391.1& 1947.2 & 2033.5 & 1347.7 & 1185.5\\
$m_{\tilde{e}_{L}}$ & 1040.0& 1040.2& 1308.5 & 1439.1 & 1300.8 & 1123.8\\
$m_{\tilde{e}_{R}}$ & 1006.1& 1006.1& 1235.3 & 1177.6 & 1300.2 & 1363.1\\
$m_{\tilde{\nu}_{e}}$ &1037.2 & 1037.4& 1306.2 & 1431.1 & 1298.5 & 1121.2\\
$m_{\tilde{t}_{1}}$ & 992.1& 922.7& 1462.3 & 1357.2 & 801.9 & 882.3\\
$m_{\tilde{t}_{2}}$ & 1169.2& 1096.0& 1681.2 & 1569.0 & 949.2 & 1053.4\\
$m_{\tilde{b}_{1}}$ & 1123.0& 1057.2& 1634.8 & 1538.6 & 929.1 & 893.4\\
$m_{\tilde{b}_{2}}$ & 1192.1& 1087.4& 1697.3 & 1682.3 & 1036.9 & 1036.1\\
$m_{\tilde{\tau}_{1}}$  &721.0 &674.5& 883.0 & 740.2 & 950.0 & 929.8\\
$m_{\tilde{\tau}_{2}}$  &922.7 &897.9& 1166.7 & 1287.2 & 1141.9 & 1045.1\\
$m_{\tilde{\nu}_{\tau}}$ & 914.5 & 894.4 & 1158.9 & 1284.7 & 1138.5 & 933.2\\
$m_{\tilde{\chi}^{\pm}_{1}}$ & 396.8 & 105.2& 657.8 & 312.8 & 152.1 & 162.5\\
$m_{\tilde{\chi}^{\pm}_{2}}$ & 625.0 & 422.7& 938.9 & 676.3 & 333.9 & 598.1\\
$m_{\tilde{\chi}^{0}_{1}}$ & 209.2 & 92.6& 347.5 & 292.7 & 80.8 & 82.6\\
$m_{\tilde{\chi}^{0}_{2}}$ & 396.9 & 118.7& 657.9 & 323.1 & 152.4 & 162.5\\
$m_{\tilde{\chi}^{0}_{3}}$ & 611.3 & 218.9& 927.9 & 365.5 & 316.6 & 592.1\\
$m_{\tilde{\chi}^{0}_{4}}$ & 624.6 & 422.8& 938.8 & 676.3 & 332.0 & 596.1\\
$m_{\tilde{g}}$  & 1200.8 &1198.6& 1857.2 & 1855.0 & 563.1 & 563.0\\
\hline
\hline
$a_{\mu}^{{\textrm{SUSY}}}  \times 10^{9}$   & 1.17 & 1.40& 0.61 & 0.72 & 1.17 & 1.11\\
$\mathcal{B}(b\rightarrow X_{s}\gamma)  \times 10^{4}$ & 2.90& 2.77& 3.37 & 5.09 & 2.41 & 2.89\\
$\Omega_{{\textrm{CDM}}}h^2$ & 0.857 & 0.009 & 0.979 & 0.016 & 1.364 & 29.414\\
\hline
\end{tabular}
\end{center}
\caption{{\em This table shows the mass spectra (in GeV) for three pairs of contrasting points in the parameter space. Points {\rm U} and {\rm V} compare an unremarkable point in the CMSSM space {\rm (U)} with the same point but with $m_{10}/m_{16} = 1.25$ {\rm (V)} demonstrating a huge change in $\mu$. Points {\rm W} and {\rm X} compare another unremarkable point in the CMSSM parameter space {\rm (W)} with the same point but with $m_{10}/m_{16} = 1.25$ and $\mathcal{D} = -0.4$ {\rm (X)} where the Higgs bosons are extremely light. Finally, points {\rm Y} and {\rm Z} compare two points at large $m_{16}$ and small $m_{1/2}$. Point {\rm Y} is the CMSSM point, while point {\rm Z} has $\mathcal{D} = 0.4$. $\mu$ is roughly as sensitive to the $D$-terms as $m_{A^0}$ in this region of parameter space.}\label{Inttbl}}
\end{table}

\begin{table}
\begin{center}
\begin{tabular}{|l|rrr|}
\hline
{\bf Point}& A & D & E\\
\hline
Input Parameters & & &\\
\hline
$m_{1/2}$ & 500 & 500 & 500\\
$m_{16}$ & 500 & 500 & 500\\
$\mathcal{D}$ & 0 & -0.4 & 0.4\\
$m_{10}/m_{16}$  & 1.0 & 1.0 & 1.0\\
\hline
\hline
$\mu(M_Z)$  & 592.9 & 561.5 & 622.9\\
$m_{\tilde{L}_\mu}$ & 601.6 & 651.8 & 546.7\\
$M_2$ & 382.8 & 382.8 & 382.7\\
\hline
\hline
$\frac{M_2\mu}{m_{\tilde{L}_\mu}^2(\mu^2 - M_2^2)} \times 10^6$ & 3.06 & 3.00 & 3.30\\
$F(\frac{M_2^2}{m_{\tilde{L}_\mu}^2})$ & 1.25 & 1.39 & 1.11\\
$F(\frac{\mu^2}{m_{\tilde{L}_\mu}^2})$ & 0.68 & 0.83 & 0.55\\
$\delta F$ & 0.57 & 0.56 & 0.56\\
$\delta a_\mu^{\tilde{\chi}^{\pm}} \times 10^9$ (approximation) & 2.54 & 2.42 & 2.68\\
$\delta a_\mu^{\tilde{\chi}^{\pm}} \times 10^{-9}$ (actual value) & 2.35 & 2.24 & 2.48\\
$\delta a_{\mu} \times 10^{9}$ (actual value)  & 2.18 & 2.04 & 2.34\\
\hline
\end{tabular}
\end{center}
\caption{{\em This table compares the approximation to the calculated value of $\delta a_\mu$ for points A, D and E in Table~\ref{benchtbl}.\label{Ftbl}}}
\end{table}

\end{document}